%% file: experts.tex
\documentclass[10pt,a4paper]{article}
\usepackage{graphicx}
\RequirePackage{fix-cm}

\usepackage{caption}
\usepackage{multicol}
\usepackage{supertabular}
\usepackage[latin1]{inputenc}
\usepackage[round,authoryear]{natbib}
\usepackage[english]{babel}
\ifnum\pdfoutput>0
\fi
\usepackage{rotating}
\usepackage{upgreek}
\usepackage{algorithm}
\usepackage{algorithmicx}
\usepackage{algpseudocode}
\usepackage{stmaryrd}
\usepackage{url}
\usepackage{caption}
\usepackage[list=true]{subcaption}
\usepackage{color}
\usepackage{enumitem}

\usepackage[T1]{fontenc}
\usepackage{fancyhdr}   
\usepackage[hmargin=1.25in, vmargin=2.8cm]{geometry}

\input{sections/experts_macros_rev}

\graphicspath{{figures/BesselOneStep/}{figures/Tobit/}{figures/BesselIterate/}{figures/LG/}}

\title{Adaptive sequential Monte Carlo \\by means of mixture of experts}
\author{Julien Cornebise\footnote{\texttt{julien@cornebise.com}}
E. Moulines\footnote{D\'epartement Traitement du Signal et de l'Image, Telecom ParisTech}, 
J. Olsson\footnote{Centre for Mathematical Sciences, Lund University}}

\date{September 13th\textsuperscript{th}, 2012}

\makeatletter

\newenvironment{algorithmhere}
{\def\@captype{algorithm}\captionsetup{font=bf}
\par\noindent\hrulefill}
{\hrulefill\par}
\newenvironment{figurehere}
{\def\@captype{figure}}
{}
\makeatother

\begin{document}
\maketitle

\setlength{\columnsep}{20pt}
\begin{multicols}{2}
\textbf{Abstract:}
Appropriately designing the proposal kernel of particle filters is an issue
of significant importance, since a bad choice may lead to deterioration of
the particle sample and, consequently, waste of computational power. In
this paper we introduce a novel algorithm adaptively approximating the
so-called optimal proposal kernel by a mixture of integrated curved
exponential distributions with logistic weights. This family of
distributions, referred to as \emph{mixtures of experts}, is broad enough
to be used in the presence of multi-modality or strongly skewed
distributions. The mixtures are fitted, via online-EM
methods, to the optimal kernel through minimisation of the Kullback-Leibler
divergence between the auxiliary target and instrumental distributions of
the particle filter. At each iteration of the particle filter, the
algorithm is required to solve only a single optimisation problem for the
whole particle sample, yielding an algorithm with only linear complexity. In addition, we illustrate in a simulation study how the method can be successfully applied to optimal filtering in nonlinear
state-space models.



\input{sections/introduction_rev}

\input{sections/notation_rev}

\input{sections/preliminaries_rev}

\input{sections/framework_rev}
\input{sections/algorithms_rev}

\input{sections/examples_rev}

\input{sections/conclusion_rev}

\section*{Acknowledgement}

We thank the anonymous referees for insightful comments that significantly improved the presentation of the paper. 

\appendix

\input{sections/proofs_rev}

\bibliographystyle{plainnat}
\bibliography{experts}
\end{multicols}

\end{document}

%% file: sections/experts_macros_rev.tex
\usepackage{graphicx}
\usepackage{amsmath,amssymb,bbm}
\usepackage{ifthen}
\usepackage{twoopt}
\usepackage{color}
\usepackage{xargs}

\newcommand{\cset}{\boldsymbol{\Gamma}}
\newcommand{\colv}[2]{{#1}_{|#2}}
\newcommand{\globmax}{\bar{\boldsymbol{\theta}}}
\newcommand{\jtarg}{\tilde{\mu}}
\newcommand{\jinit}{\mu}
\newcommand{\jvect}[1]{\mathbf{#1}}
\newcommand{\jvecth}[1]{\hat{\mathbf{#1}}}
\newcommand{\meanfd}[1]{\mathbf{h}(#1)}
\newcommandx\n[1][1=]{
\ifthenelse{\equal{#1}{}}
{N}
{N_{#1}}
}
\newcommand{\nablapar}{\nabla_{\hspace{-0.4mm} \boldsymbol{\theta}}}

\newcommand{\normconst}[1]{\hat{c}_{#1}}
\newcommandx\param[1][1=]{
\ifthenelse{\equal{#1}{}}
{\boldsymbol{\theta}}
{\boldsymbol{\theta}_{#1}}
}
\newcommand{\pset}{\boldsymbol{\Pi}}
\newcommand{\pstat}[2]{
\ifthenelse{\equal{#1}{}}
{\bar{\mathbf{p}}(#2)}
{\bar{p}_{#1}(#2)}
}
\newcommand{\pstatest}[2]{
\ifthenelse{\equal{#1}{}}
{\tilde{\mathbf{p}}(#2)}
{\tilde{p}_{#1}(#2)}
}
\newcommand{\set}[1]{\mathsf{#1}}
\newcommand{\sdim}[1]{
\ifthenelse{\equal{#1}{1}}
{s}
{s'}
}
\newcommand{\sset}{\boldsymbol{\Sigma}}
\newcommand{\step}[1]{\lambda_{#1}}
\newcommand{\sstatest}{\tilde{\jvect{s}}}
\newcommandx\sumwgts[2][1=]{
\ifthenelse{\equal{#1}{}}
{\tilde{\Omega}_{#2}}
{\tilde{\Omega}_{#2}^{(#1)}}
}
\newcommand{\trace}{\mathrm{Tr}}
\newcommand{\trans}{\intercal}
\newcommand{\tstat}{\bar{\mathbf{t}}}
\newcommand{\tstatest}{\tilde{\jvect{t}}}
\newcommand{\vstat}{\bar{\mathbf{v}}}
\newcommand{\vstatest}{\tilde{\jvect{v}}}

\newcommand{\indic}{\mathbbm 1}

\newcommand{\rset}{\ensuremath{\mathbb{R}}}

\newcommand{\dimpsp}{p}
\newcommand{\dimpsptd}{\tilde p}

\newcommand{\Amat}{\Lambda}

\newcommand{\E}{\mathbb{E}}

\newcommand{\1}{\indic}

\numberwithin{equation}{section}

\newcommand{\eqsp}{}
\newlength{\vertsepeq} 

\newcommand*{\ifopt}[2]{\ifthenelse{\equal{#1}{}}{}{#2}} 
\newcommand*{\ifbothopt}[3]{\ifopt{#1}{\ifopt{#2}{#3}}}
\newcommand*{\optexp}[1]{\ifopt{#1}{^{#1}}} 
\newcommand*{\optpexp}[1]{\ifopt{#1}{^{(#1)}}} 
\newcommand*{\optsub}[1]{\ifopt{#1}{_{#1}}} 

\newcommand{\eqdef}{:=}
\newcommand*{\equp}{\equiv} 

\newcommand{\R}{\mathbb{R}}

\newcommand*{\extd}[1]{\overline{#1}}
\newcommand{\ud}{d}
\newcommand{\mcb}[1]%
	{\ensuremath{\mathcal{F}_{\mathrm{b}}\left(#1\right)}} 
\newcommand{\argmax}{\operatornamewithlimits{arg\,max}}
\newcommand{\argmin}{\operatornamewithlimits{arg\,min}}

\newcommand*{\prmat}{\,}

\newcommand*{\normdens}[4][]{\mathcal{N}\optsub{#1}\left(#2; #3, #4\right)}

\newcommand{\iid}{\text{i.i.d.}}
\newcommand{\ie}{i.e.}
\newcommand{\eg}{e.g.}

\newcommand{\stsp}[1][]{{\mathsf{X}\optpexp{#1}}} 

\newcommand{\obsp}[1][]{{\mathsf{Y}\optpexp{#1}}} 

\newcommand{\filtpost}[2][]{\ensuremath{\phi_{\ifopt{#1}{#1,}#2}}}	

\newcommand*{\hk}{Q}
\newcommand{\hkdens}[1][]{q\optpexp{#1}}	
\newcommand{\olik}[1][]{g\optsub{#1}}	

\newcommand{\ukdens}[1][]{l\optsub{#1}}

\newcommand{\uknormdens}[1][]{l^\ast\optsub{#1}}

\newcommand{\propdens}[1][]{r\optsub{#1}}

\newcommand{\plim}{\ensuremath{\stackrel{\mathbb{P}}{\longrightarrow}}}

\newcommandx\entropy[1][1=]{
\ifthenelse{\equal{#1}{}}
{\ensuremath{\mathcal{E}}}
{\ensuremath{\mathcal{E}_{#1}}}
}

\newcommand{\propjoint}{\pi^\ast}
\newcommand{\targjoint}{\mu^\ast}
\newcommand*{\auxtarg}{\bar{\mu}}

\newcommand{\prior}{\pi_0}

\newcommand{\auxinstr}[1][]{\pi_{\ifasymptotic N \fi\ifopt{#1}{\ifasymptotic , \fi#1}}}
\newcommand{\auxinstrparam}[1]{\pi_{\ifasymptotic N,\fi #1}}
\newcommand*{\mixparamvec}[1][]{\boldsymbol\theta\optexp{#1}} 

\newcommand*{\symp}{{\mathbf{x}}}	
\newcommand*{\symP}{{\mathbf{X}}}	
\newcommand*{\symw}{\omega} 
\newcommand*{\symW}{\Omega} 
\newcommand*{\symi}{i}	
\newcommand*{\symI}{I}	
\newcommand*{\symj}{j}	
\newcommand*{\symJ}{J}	
\newcommand*{\symadj}{a}	
\newcommand*{\symAdj}{a}	
\newcommand*{\td}[1]{\tilde{#1}} 
\newcommand*{\optim}[1]{{#1}^\ast} 
\newif\ifasymptotic
\asymptoticfalse
\newcommand*{\psp}[1][]{\ensuremath{\boldsymbol{\Xi}\optsub{#1}}}		
\newcommand*{\psptd}{\ensuremath{\boldsymbol{\td{\Xi}}}}		

\newcommand*{\ppa}{\symp}	
\newcommand*{\ppr}{\td{\symp}} 
\newcommand*{\ppava}{\symP}	
\newcommand*{\pprva}{\td{\symP}} 
\newcommand*{\comp}{\symj}
\newcommand*{\compva}{\symJ}
\newcommand*{\indv}{\symi}
\newcommand*{\indva}{\symI}
\newcommand*{\base}[4]{
	\ensuremath{%
	{#1}
	\ifopt{#2#3}{^%
	{
		\ifopt{#2}{#2}
		\ifbothopt{#2}{#3}{,}
		\ifopt{#3}{[#3]}
	}
	}
	\ifasymptotic
	_{N\ifopt{#4}{,}#4}
	\else
	\ifopt{#4}{_{#4}}
	\fi
	}
}
\newcommand*{\baseneverN}[4]{%
	{\asymptoticfalse\base{#1}{#2}{#3}{#4}} 
}

\newcommandtwoopt*{\parti}[3][][]{\base{\symP}{#1}{#2}{#3}}
\newcommandtwoopt*{\partiva}[3][][]{\base{\symP}{#1}{#2}{#3}}
\newcommandtwoopt*{\partiextd}[3][][]{\base{\extd{\symP}}{#1}{#2}{#3}}
\newcommandtwoopt*{\wgt}[3][][]{\base{\symw}{#1}{#2}{#3}}

\newcommandtwoopt*{\wgtsum}[2][][]{\base{\symW}{#1}{#2}{}}
\newcommandtwoopt*{\ind}[3][][]{\base{\symI}{#1}{#2}{#3}}
\newcommandtwoopt*{\indTSS}[3][][]{\base{J}{#1}{#2}{#3}}
\newcommandtwoopt*{\partitd}[3][][]{\base{\td{\symP}}{#1}{#2}{#3}}
\newcommandtwoopt*{\partibar}[3][][]{\base{\bar{\symP}}{#1}{#2}{#3}}
\newcommandtwoopt*{\wgttd}[3][][]{\base{\td{\symw}}{#1}{#2}{#3}}
\newcommandtwoopt*{\wgtbar}[3][][]{\base{\bar{\symw}}{#1}{#2}{#3}}
\newcommandtwoopt*{\wgtsumtd}[2][][]{\base{\td{\symW}}{#1}{#2}{}}
\newcommandtwoopt*{\wgtsumbar}[2][][]{\base{\bar{\symW}}{#1}{#2}{}}

\newcommandtwoopt*{\partihat}[3][][]{\base{\hat{\symP}}{#1}{#2}{#3}}
\newcommandtwoopt*{\adj}[3][][]{\base{\symadj}{#1}{#2}{#3}}
\newcommand*{\adjfunc}[1][]{\baseneverN{\symAdj}{#1}{}{}} 
\newcommandtwoopt*{\adjfuncpath}[2][][]{\symAdj^{\mathrm{p}\ifopt{#1}{,(#1)}\ifopt{#2}{,[#2]}}}
\newcommandtwoopt*{\adjopt}[3][][]{\base{\symadj^{\ast}}{#1}{#2}{#3}}
\newcommand*{\adjfuncopt}[1][]{\baseneverN{\optim{\symAdj}}{#1}{}{}}

\newcommand{\wgtfunc}{w}

\newcommand{\KL}{\ensuremath{d_{\mathrm{KL}}}}

\newcounter{hyp}
	{\refstepcounter{hyp}\it\begin{itemize}\item[{\bf(A\arabic{hyp})}] \label{hyp:#1}}%
			{\end{itemize}}

\newcommand*{\rmd}{d}
\newcommand*{\propdensparam}[1]{\propdens_{#1}}

\newcommand*{\cA}{a}	
\newcommand*{\cB}{\mathbf{b}}	
\newcommand*{\uu}{\mathbf{u}}	
\newcommand*{\uuva}{\mathbf{U}}	
\newcommand*{\uusp}{{\boldsymbol{\Upsilon}}}	
\newcommand*{\curvparj}[1]{{\boldsymbol\eta_{#1}}}	
\newcommand*{\curvpar}[1][]{{\boldsymbol\eta\optsub{#1}}}	
\newcommand*{\curvparvec}[1][]{\underline{\boldsymbol\eta}\optexp{#1}}	

\newcommand*{\intcurv}[3]{\rho(#1,#2;#3)}
\newcommand*{\intcurvj}[3][j]{\intcurv{#2}{#3}{\curvpar_{#1}}} 
\newcommand*{\curv}[4]{\bar{\rho}(#1,(#2,#3);#4)}
\newcommand*{\curvexpestat}[2]{\bar{\mathbf{s}}_{#1}(#2)}  

\newcommand*{\auxinstrextd}[1]{\bar{\pi}_{#1}}

\newcommand*{\completelike}[1]{\auxinstrextd{#1}}

\newcommand*{\dimosp}{p}
\newcommand*{\dimosptd}{p'}
\newcommand*{\ds}{\ensuremath{d_s}}			

\newcommand*{\Expect}[2][]{\E\ifopt{#1}{_{#1}}\left[#2\right]}

\newcommand*{\normlaw}[3][]{\mathcal{N}\optsub{#1}\left(#2, #3\right)}
\newcommand*{\vectornorm}[1]{\left\|#1\right\|}

\newcommand*{\nablaop}[1]{\nabla_{\hspace{-0.4mm} #1}}

\newcommand*{\detm}[1]{\left\vert #1\right\vert}

\newcommand*{\mahalanobis}[3]{\delta( #1,#2,#3)} 

\newcommand{\dimo}{p}

\newcommand*{\mixvar}[2][]{\boldsymbol{\Upsigma}_{#2} \optexp{#1}}	
\newcommand*{\mixwgt}[2][]{\alpha_{#2}\optexp{#1}}	

\newcommand*{\mixmean}[2][]{\boldsymbol{\mu}_{#2}\optexp{#1}}

\newcommand*{\mixbeta}[1]{\boldsymbol\beta_{#1}}	
\newcommand*{\mixbetavec}[1][]{\underline{\boldsymbol\beta}\optexp{#1}}

\newcommand*{\mixparamopt}{\mixparamvec^\ast}	
\newcommand*{\mixparamsp}{\Theta} 
\newcommand*{\mixwgtcond}[4]{\completelike{#1}(#2 | #3, #4)} 
\newcommand*{\suffstat}{\mathbf{s}} 

\newcommand*{\tdens}[5][]{t\optsub{#1}(#2; #3, #4, #5)}
\newcommand*{\gammadens}[3]{\gamma\left(#1; #2, #3\right)}
\newcommand*{\df}{\nu}

\def\clap#1{\hbox to 0pt{\hss#1\hss}}

\def\mathclap{\mathpalette\mathclapinternal}

\def\mathclapinternal#1#2{%
\clap{$\mathsurround=0pt#1{#2}$}}

\catcode`\@=11
\def\indep{\setbox0\hbox{$\m@th\perp$}\mathrel{\hbox
         to1.25\wd0{\copy0\hss\box0}}}
\def\indep{\setbox0\hbox{$\m@th\perp$}\ifmmode\mathrel\fi{\hbox
         to1.25\wd0{\copy0\hss\box0}}}
\catcode`\@=12

\newtheorem{proposition}{Proposition}[section]
\newtheorem{example}{Example}[section]
\newtheorem{assumption}{Assumption}[section]
\newtheorem{remark}{Remark}[section]

\usepackage{ifmtarg}
\makeatletter
\newenvironment{Proposition}[1][]{%
  \begin{proposition}\@ifnotmtarg{#1}{\textbf{(#1)}}%
  }{%
  \end{proposition}%
}
\newenvironment{Example}[1][]{%
  \begin{example}\@ifnotmtarg{#1}{\textbf{(#1)}}%
  }{%
  \end{example}%
}
\newenvironment{Assumption}[1][]{%
  \begin{assumption}\@ifnotmtarg{#1}{\textbf{(#1)}}%
  }{%
  \end{assumption}%
}
\newenvironment{Remark}[1][]{%
  \begin{remark}\@ifnotmtarg{#1}{\textbf{(#1)}}%
  }{%
  \end{remark}%
}
\makeatother


%% file: sections/introduction_rev.tex
\section{Introduction}
\label{sec:introduction}

During the last decade, sequential Monte Carlo (SMC)  methods have
developed from being a tool for solving specific
problems, such as the optimal filtering problem in general state-space
models or simulation of rare events \citep[these topics are abundantly
covered in][and the references therein]{liu:2001,doucet:defreitas:gordon:2001,ristic:arulampalam:gordon:2004,cappe:moulines:ryden:2005,delmoral:doucet:jasra:2006}. The high flexibility of these methods makes it possible to use them for  efficiently executing key steps of composite algorithms tackling very complex problems, e.g., as in particle Markov chain Monte Carlo (MCMC) methods for joint inference on static and dynamic variables \citep[see][]{andrieu:doucet:holenstein:2009}.

However, while there has been abundant work on adaptation of MCMC methods, from the early algorithms of \cite{haario:2001} to the mathematical
study of disminishing adaptation established in
\cite{andrieu:2006} and \cite{roberts:2009}, adaptive methods for SMC
are still in their infancy. This article, which sets out from a
theoretical framework for adaptive SMC presented in
\cite{cornebise:moulines:olsson:2008}, is part of the effort to fill this
gap---which is required to achieve maximum efficiency and
user-friendliness. A mathematically less detailed pre-version of the present paper appeared as part of the Ph.~D. dissertation by \cite{cornebise:2009}. 

Such adaptive methods are especially suited to cope with the rising complexity of \emph{state-space models} (SSMs) in recent applications. While simple unidimensional filtering problems will rarely justify even a slight computational overhead of adaptive methods over plain SMC methods (such as the \emph{bootstrap particle filter}; see Section~\ref{sec:SMC}), the situation is reversed for many state-of-the-art SSMs, where forward simulation may require costly physical emulators or the evaluation of computationally expensive transition densities. This occurs, e.g., as soon as the model involves differential equations and numeric solutions of the same. Recent examples range from health sciences \citep[see e.g.][for a study of the epidemiology of human papillomavirus infections]{korotsil:2012} to environmental sciences \citep[see e.g. ][who studied data assimilation for atmospheric dispersion of hazardous airborne biological material]{robins:2009}. 

In the SMC framework, the objective is to approximate a \emph{sequence}
of target densities, defined on a sequence of state spaces as follows. Given a density $\jinit$ on some state space $\psp$, the next density $\jtarg$ in the sequence is defined by
\begin{equation}
    \jtarg(\ppr) \eqdef \frac{\int \jinit(\ppa) \ukdens(\ppa, \ppr) \, \ud \ppa}{\iint \jinit(\ppa)
    \ukdens(\ppa, \ppr) \, \ud \ppa \, \ud \ppr} \eqsp, \quad \ppr \in \psptd \eqsp, 
    \label{eq:SISrecursion}
\end{equation}
where $\psptd$ is some other state space and $\ukdens : \psp \times \psptd \rightarrow \R_+$ is an un-normalised transition density. Here the dimensions of $\psptd$ and $\psp$ typically differ. This framework encompasses, inter alia, the classical problem of \emph{optimal filtering} in SSMs recalled in Section~\ref{sec:examples} (see
in particular Equation~\eqref{eq:filteringRecursion}). 

SMC methods approximate online the densities generated by \eqref{eq:SISrecursion} by a recursively updated set of random \emph{particles} $\{ \parti{i} \}_{i = 1}^N$ and associated \emph{importance weights} $\{ \wgt{i} \}_{i = 1}^N$. At each iteration of the algorithm, the weighted particle sample $\{ (\parti{i}, \wgt{i}) \}_{i = 1}^N$ approximates the corresponding target density $\jinit$ in the sense that  
$ \sum_{i = 1}^N \wgt{i} f(\parti{i}) / \wgtsum$, with $\wgtsum \eqdef \sum_{\ell = 1}^N \wgt{\ell}$, 
approximates $\E_{\jinit}[f(\mathbf{X})] \eqdef \int f(\ppa) \jinit(\ppa) \, \ud \ppa$ (this will be our generic notation for expectations) for all bounded measurable functions $f$ on the corresponding state space $\psp$. All SMC methods have two main operations in common: a \emph{mutation step} and a \emph{selection step}. In the former, the particles are randomly scattered in the state space by means of a \emph{proposal} transition density $\propdens : \psp \times \psptd \rightarrow \R_+$; more specifically, the mutation operation generates a set of new, $\psptd$-valued particles by moving the particles according to the transition density $r$ and assigning each mutated particle an updated importance weight proportional to $\ukdens / \propdens$; consequently, each weight reflects the relevance of the corresponding particle as measured by the likelihood ratio. In general, the mutated particles are conditionally independent given the ancestor particles. In the selection step, the particles are duplicated or eliminated with probabilities depending on the importance weights. This is typically done by multinomially resampling the particles with probabilities proportional to the corresponding weights. In the \emph{auxiliary particle filter} \citep[proposed by][]{pitt:shephard:1999} the particles are resampled according to weights that are modified by a factor determined by an \emph{adjustment multiplier function} $a : \psp \rightarrow \R_+$. Introducing adjustment multipliers makes it possible to, by using some convenient lookahead strategy, increase the weight of those particles that will most likely contribute significantly to the approximation of the subsequent target distribution after having been selected. In this way computational power is directed toward zones of the state space where the mass of the subsequent target density is located. 

The choice of the adjustment weight function
$\adjfunc$ and the proposal density $\propdens$ affects significantly the quality of the generated
sample, and in this paper we focus on adaptive design of the latter. Letting $r(\ppa, \ppr) = \ukdens(\ppa, \ppr) / \int \ukdens(\ppa, \ppr) \, \rmd \ppr$ is appealing as the particles in this case are propagated according to a dynamics that is closely related to that of the recursion \eqref{eq:SISrecursion}. Indeed, by combining this choice of proposal kernel with the adjustment multiplier function $\adjfunc(\ppa) = \int \ukdens(\ppa, \ppr) \, \rmd \ppr$ yields perfectly uniform importance weights. In this case the SMC algorithm is referred to as \emph{fully adapted}. However, this so-called \emph{optimal} proposal kernel and adjustment multiplier weight function can be expressed on closed-form only in a few cases and one is in general referred to approximations of these quantities. 

There is a large literature dealing with the problem of approximating the optimal proposal kernel. When using SMC for optimal filtering in SSMs, \cite{doucet:godsill:andrieu:2000} suggest approximating each function $\ukdens(\ppa, \cdot)$ by a Gaussian density whose mean and covariance are obtained using the \emph{extended Kalman filter}. A similar approach based on the \emph{unscented Kalman
filter} was proposed by \cite{vandermerwe:doucet:defreitas:wan:2000} \citep[see also][]{vandermerwe:wan:2003}. 
This technique presupposes that the model under consideration can be well approximated by a nonlinear Gaussian SSM with linear measurement equation; however, this far from always the case, and careless linearisation (\eg\ in cases where the local likelihood is multimodal) may boomerang and cause severe deterioration of the particle sample.
\citet[Section~7.2.2.4]{cappe:moulines:ryden:2005} make a form of Laplace approximation of $\ukdens(\ppa, \cdot)$ by simply locating the mode of the function in question and centering a Gaussian density or the density of a student's $t$-distribution around the same. This technique, which goes back to \cite{pitt:shephard:1999}, is appropriate when the function is log-concave (or strongly unimodal); nevertheless, as the mode of each $\ukdens(\ppa, \cdot)$ is a function of $\ppa$, this approach involves solving one optimisation problem per particle. Thus, in spite of this intense recent activity in the field, the state-of-the-art algorithms have met only mitigated success as they implicitly assume that the functions $\ukdens(\ppa, \cdot)$ have a single mode.

A common practice \citep[see e.g.][]{oh:berger:1993} for
designing proposal distributions in standard (non-sequential) importance sampling is to consider a parametric family of proposal distributions and then identify a parameter that minimizes some measure of discrepancy between the target and the proposal distributions. Common choices are the widely used squared \emph{coefficient of variation} of the importance weights or the negated \emph{Shannon entropy} \citep[see \eg][p.~235, for definitions of these quantities]{cappe:moulines:ryden:2005} of the same. Both are maximal when the sample is completely degenerated, i.e. when one single particle carries all the weight, and minimal when all the importance weights are equal. The coefficient of variation often appears in a transformed form as the \emph{effective sample size} \citep[we refer again to][for a definition]{cappe:moulines:ryden:2005}. The same quantities have been widely used also within the framework of SMC,
but until recently only for triggering selection and not as
a tool for adaptive design of the instrumental distribution.
The key result of \citet[Theorems 1 and 2]{cornebise:moulines:olsson:2008} was to relate the
Shannon entropy to the \emph{Kullback-Leibler divergence} (KLD; see Equation~\eqref{eq:def:KLD})
between the instrumental and proposal distributions of the particle filter.
More specifically, when selection is carried through using multinomial resampling, each particle cloud update (selection followed by resampling) can be viewed as standard importance sampling where a weighted mixture  of $N$ stratas---the $i$th stratum being proportional to $\wgt{i} \ukdens(\parti{i}, \cdot)$---approximating $\jtarg$ is targeted using a mixture proposal $\pi$ comprising $N$ stratas proportional to $\{ \wgt{i} \propdens(\parti{i}, \cdot) \}_{i = 1}^N$. As the number of particles tends to
infinity, one may prove that the KLD between these two mixtures tends to the KLD between two distributions having densities $\targjoint(\ppa, \ppr) \propto \ukdens(\ppa, \ppr)  \jinit(\ppa)$ and $\propjoint(\ppa, \ppr)  \propto \adjfunc(\ppa) \propdens(\ppa, \ppr) \jinit(\ppa)$, which can be viewed as ``asymptotic'' target and proposal distributions on the product space $\psp \times \psptd$. In addition, one may prove that the Shannon entropy has the same limit. (\cite{cornebise:moulines:olsson:2008} established a similar relation between the coefficient of
variation of the particle weights and the \emph{chi-square divergence} between the same distributions.) This gives a sound theoretical support for using the Shannon entropy of the importance weights for measuring the quality of the particle sample. In addition, it suggests that the KLD between the target mixture and $\pi$ could be used in lieu of the Shannon entropy for all purposes, especially adaptation. As a matter of fact, in the context of adaptive design of SMC methods, the KLD is \citep[as pointed out by][Section 2.3]{cornebise:moulines:olsson:2008} highly practical since it decouples the adaptation of the adjustment weight function $\adjfunc$ and that of the proposal kernel $\propdens$; see the next section.  

Henceforth, in the present article we extend and implement fully the methodology indicated and sketched by \citet{cornebise:moulines:olsson:2008} and select the auxiliary proposal
distribution $\pi$ from a family $\{\auxinstrparam{\mixparamvec} ; \mixparamvec \in \mixparamsp\}$
of candidate distributions by picking $\mixparamopt$ such that the KLD between the target mixture and $\auxinstrparam{\mixparamopt}$ is minimal and letting $\pi = \auxinstrparam{\mixparamopt}$. On the one hand, the chosen parametric family should be flexible enough to approximate complex transition kernels; on the other hand, sampling from $\auxinstrparam{\mixparamvec}$ should be easy. Finally, the parameterisation should be done in such a way that the problem of estimating the parameters is as simple as possible. In this article, we suggest modeling the proposal $\auxinstrparam{\mixparamvec}$ as a mixture of integrated curved exponential distributions. By letting the mixture weights of the chosen proposal depend on the ancestor particles $\{ \ppava_i \}_{i = 1}^N$, we allow for partitioning of the input space into regions corresponding to a specialised kernel. Each component of the mixture belongs to a family of integrated curved exponential distributions, whose two most known members are the multivariate Gaussian distribution and the Student's $t$-distribution. Also the parameters of the mixture depend on the ancestors $\{ \ppava_i \}_{i = 1}^N$. This parameterisation of the proposal distribution is closely related to the (hierarchical) \emph{mixture of experts} appearing in the machine learning community and described in \citep{jordan:jacobs:1994, jordan:xu:1995}. The flexibility of our approach allows for efficient approximation of the optimal kernel for a large class of intricate (nonlinear non-Gaussian) models. Unlike typical EM-type parameter estimation procedures, which are in general run iteratively until stabilisation, our adaptive SMC algorithm only requires a decrease of the KLD, not an exact minimum. As illustrated in our examples in Section~\ref{sec:examples}, such a gain typically occurs already at early iterations of the algorithm (a characteristic of EM-type algorithms), impliying the need of only very few EM iterations and therefore a minimal computational overhead.

The paper is organized as follows. In the next section we introduce some matrix notation and list the most common notation used throughout the paper. Section~\ref{sec:preliminaries} describes, in a nutshell, auxiliary SMC methods as well as a KLD-based approach to adaptation of these methods that is used in the paper. Section~\ref{sec:framework} recalls mixtures of experts and Section~\ref{sec:algo} treats optimisation of the mixture parameters by means of stochastic approximation methods; the latter section is the core of the paper and contains our main results. In Section~\ref{sec:examples} we illustrate the efficiency of the method on several
simulation-based examples and Section~\ref{sec:proofs} contains some proofs.

%% file: sections/notation_rev.tex
\section{Notation}
\label{sec:notations}
\subsection{Matrix notation}
All vectors and matrices are typeset in boldface. Vectors are \emph{column vectors} unless precised differently. The $i$th column vector of a matrix $\bf A$ is denoted by $\colv{\mathbf{A}}{i}$. We denote by $\trace(\bf A)$ the trace of a matrix $\bf A$ and for any matrices $\bf A$ and $\bf B$ of dimensions $m \times n$  and $p \times q$, respectively, we denote their \emph{direct sum} by
\begin{equation} \label{eq:def:direct:sum}
\bf A \varoplus \bf B 
\eqdef
\begin{pmatrix} 
\bf A & \bf 0_{p \times q} \\ 
\bf 0_{m \times n}& \bf B 
\end{pmatrix} \eqsp, 
\end{equation}
which is such that $(\bf A \varoplus \bf B)^\trans = \bf A^\trans \varoplus \bf
B^\trans$ and, for any two  matrices $\bf C$ and $\bf D$ of
compatible dimensions, $(\bf A \varoplus \bf B)  (\bf C \varoplus \bf
D)^\trans = (\bf A \bf C^\trans) \varoplus (\bf B \bf D^\trans)$, where $\trans$ denotes the transpose. 

\subsection{List of notation}

The following quantities are defined in the corresponding equations. \\

\let\mcnewpage=\newpage
\makeatletter
\renewcommand\newpage{%
        \if@firstcolumn
                \hrule width\linewidth height0pt
                \columnbreak
        \else
                \mcnewpage
        \fi
}
\makeatother
\begin{center}
\tablehead{\hline \hline Quantity & Equation\\\hline}
\begin{supertabular}{l l} 
$\adjfuncopt$ & \eqref{eq:def:optimal:weight} \\
$\mixwgt{j}$ & \eqref{eq:definition-mixwgt} \\
$\KL$ & \eqref{eq:def:KLD} \\
$\curvparvec$ & \eqref{eq:def:param} \\
$\filtpost{k}$ & \eqref{eq:def:filter} \\
$G$ & \eqref{eq:obs:eqn} \\
$\olik$ & \eqref{eq:obs:eqn} \\ 
$\jvect{h}$ & \eqref{eq:def:mean:field} \\
$\kappa$ & \eqref{eq:def:kappa} \\
$l$ & \eqref{eq:def:l} \\
$\uknormdens$ & \eqref{eq:def:optimal:kernel} \\
$\jtarg$ & \eqref{eq:SISrecursion} \\
$\bar{\jvect{p}}$ & \eqref{eq:def:pstat:vector} \\
$\tilde{\jvect{p}}$ & \eqref{eq:def:IS:pstat:vector} \\
$\auxinstrparam{\mixparamvec}$ & \eqref{eq:define:auxinstrdensparam} \\
$\completelike{\mixparamvec}( \, )$ & \eqref{eq:completelike} \\ 
$\completelike{\mixparamvec}(\, | \,)$ & \eqref{eq:def:responsibilities} \\ 
$\propdensparam{\mixparamvec}$ & \eqref{eq:definition-propdensparam} \\
$\varoplus$ & \eqref{eq:def:direct:sum} \\
$\hk$ & \eqref{eq:state:eqn} \\
$\hkdens$ & \eqref{eq:state:eqn} \\ 
$\rho$ & \eqref{eq:definition-propdensparam} \\
$\bar{\rho}( \, )$ & \eqref{eq:curv} \\
$\bar{\rho}(\, | \,)$ & \eqref{eq:def:cond:u} \\
$\bar{\jvect{s}}$ & \eqref{eq:define:curvexpestat:vector} \\
$\sstatest$ & \eqref{eq:define:IS:curvexpestat:vector} \\
$\tstat$ & \eqref{eq:def:grad:hessian:exp:stat} \\
$\tstatest$ & \eqref{eq:def:IS:v:t:stat} \\
$\mixparamvec$ & \eqref{eq:def:param} \\
$\globmax$ & \eqref{eq:def:glob:max} \\ 
$\vstat$ & \eqref{eq:def:grad:hessian:exp:stat} \\
$\vstatest$ & \eqref{eq:def:IS:v:t:stat} \\
$\wgtfunc_{\mixparamvec}$ & \eqref{eq:definine:wgttd} \\
\end{supertabular}
\end{center}

%% file: sections/preliminaries_rev.tex
\section{Preliminaries}
\label{sec:preliminaries}

\subsection{Auxiliary SMC methods}
\label{sec:SMC}

In order to precisely describe one iteration of the SMC algorithm, let $\jinit$ be a target probability density function over a
space $\psp$  and suppose that we have at hand a weighted sample $\{ (\parti{i}, \wgt{i}) \}_{i = 1}^N$ such that $\sum_{i = 1}^N \wgt{i} f(\parti{i}) / \wgtsum$, with $\wgtsum = \sum_{i = 1}^N \wgt{i}$, estimates $\int f(\ppa) \jinit(\ppa) \, \rmd \ppa$ for any $\jinit$-integrable function $f$. As a notational convention, we use capitals to denote random variables and lower case for function arguments.

We now wish to transform (by moving the particles
and adjusting accordingly the importance weights) the sample $\{ (\parti{i}, \wgt{i}) \}_{i = 1}^N$
into a new weighted particle sample approximating the probability density $\jtarg$ defined on $\psptd$ through \eqref{eq:SISrecursion}. Having access to $\{ (\parti{i}, \wgt{i}) \}_{i = 1}^N$,
an approximation of $\jtarg$ is naturally obtained by plugging the weighted
empirical measure associated with this weighted sample into
\eqref{eq:SISrecursion}, yielding the mixture density
\begin{align} 
\jtarg(\ppr) & \approx \frac{\sum_{i = 1}^N \wgt{i} \ukdens(\parti{i},
\ppr)}{\sum_{j = 1}^N \wgt{j} \int \ukdens(\parti{j}, \ppr) \, \rmd \ppr}
\notag \\
& = \sum_{i = 1}^N \left( \frac{\wgt{i} \adjfuncopt(\parti{i})}{\sum_{j = 1}^N \wgt{j}  \adjfuncopt(\parti{j})} \uknormdens(\parti{i}, \ppr) \right) \eqsp,  
\label{eq:non-aux:target}
\end{align}
where we have introduced the normalised transition density 
\begin{equation} \label{eq:def:optimal:kernel}
	\uknormdens(\ppa, \ppr) \eqdef \frac{\ukdens(\ppa, \ppr)}{\int
	\ukdens(\ppa, \ppr) \, \rmd \ppr}  
\end{equation}
and the partition function  
\begin{equation} \label{eq:def:optimal:weight}
	\adjfuncopt(\ppa) \eqdef \int \ukdens(\ppa, \ppr) \, \rmd \ppr 
\end{equation}
of $\ukdens(\ppa, \cdot)$, and ideally an updated particle sample would be obtained by drawing new particles $\{ \partitd{i} \}_{i = 1}^N$ independently from \eqref{eq:non-aux:target}. There are however two problems with this
approach: firstly, in general, the integral $\int \ukdens(\cdot, \ppr) \,
\rmd \ppr$ lacks closed-form expression, ruling out direct computation of
the mixture weights; secondly, even if the integral was known in
closed form, the resulting algorithm would have an $\mathcal{O}(N^2)$ computational complexity due to the normalization of the mixture weights. To cope with these issues, we will, as proposed by \cite{pitt:shephard:1999}, aim instead at  sampling from an \emph{auxiliary target distribution} having density 
\begin{equation} 
	\auxtarg(\indv, \ppr) \eqdef \frac{\wgt{\indv} \adjfuncopt(\parti{\indv})}{\sum_{j = 1}^N \wgt{j} \adjfuncopt(\parti{j})} \uknormdens(\parti{\indv},\ppr) \eqsp, \label{eq:def:auxtarg}
\end{equation}
over the product space $\{1, \ldots, N\} \times \psptd$ of indices and particle positions. To
sample $\auxtarg$ we take an importance sampling
approach consisting in drawing independent pairs $\{(\ind[]{i}, \partitd{i})\}_{i = 1}^N$ of indices and positions from the proposal distribution
\begin{equation} \label{eq:define:auxinstr}
	\auxinstr(\indv,\ppr) \eqdef \frac{\wgt{\indv} \adjfunc(\parti{\indv})}
    {\sum_{j=1}^{N} \wgt{j} \adjfunc(\parti{j})} \propdens (\parti{\indv}, \ppr)
\end{equation}
over the same extended space and assigning each draw
$(\ind[]{i}, \partitd{i})$ the importance weight $\wgttd[]{i} \eqdef w(\ind[]{i}, \partitd{i})$, where
\begin{equation*}
	w(\indv, \ppr) \eqdef \frac{\ukdens(\parti{\indv},\ppr)}{\adjfunc(\parti{\indv})
	\propdens(\parti{\indv},\ppr)} \propto \frac{\auxtarg(\indv,\ppr)}{\auxinstr(\indv,\ppr)} \eqsp.
\end{equation*}
Here $\propdens$ and $\adjfunc$ are the proposal transition density resp. adjustment multiplier weight function mentioned in the introduction. As for all importance sampling, it is crucial that the target distribution $\auxtarg$ is absolutely continuous with respect to the proposal distribution $\auxinstr$. We thus require that the adjustment multiplier weight function is positive and that the proposal transition density is such that for all $\ppa \in \psp$,
$$
\propdens(\ppa, \ppr) = 0 \Rightarrow \ukdens(\ppa, \ppr) = 0 \eqsp, 
$$
\ie\ the support of $\ukdens(\ppa, \ppr)$ in included in that of $\propdens(\ppa, \ppr)$.  Finally, we discard the indices $\ind[]{i}$
and keep $\{ (\partitd[]{i}, \wgttd[]{i}) \}_{i = 1}^N$
as an approximation of $\jtarg$. 

%

This scheme, which is traditionally referred
to as the auxiliary particle filter \citep{pitt:shephard:1999}, encompasses the simpler
framework of the bootstrap particle filter proposed originally by
\cite{gordon:salmond:smith:1993}, which simply amounts to setting
$\adjfunc = \1_{\psp}$ (implying a gain of simplicity at the
price of flexibility). Moreover, SMC schemes where resampling is performed
only at random times can similarly be cast into the setting of the
auxiliary particle filter by composing the kernels involved in several
consecutive steps of the of recursion \eqref{eq:SISrecursion}  \citep[see
e.g.][Chapter 4, for details]{cornebise:2009}. A theoretical analysis of bootstrap-type SMC methods with random resampling schedule is also given by \citet{delmoral:doucet:jasra:2012}. Therefore, any methodology
built for the auxiliary particle filter also applies to these simpler
frameworks without modification.

\subsection{A foundation of SMC adaptation}
\label{sec:adaptive:SMC}


We now describe a foundation of SMC adaptation underpinning the coming developments. Recall that the KLD between two distributions $\mu$ and $\nu$, defined on the same state space $\psp$, is defined as
\begin{equation} \label{eq:def:KLD}
	\KL(\mu \| \nu) \eqdef \E_\mu \left[ \log 
	\frac{\rmd \mu}{\rmd \nu}(\mathbf{X}) \right] \eqsp,
\end{equation}
provided that $\mu$ is absolutely continuous with respect to $\nu$. Here $\E_\mu$ denotes the expectation under $\mu$. Casting the KLD into the framework of SMC methods and replacing $\mu$ and $\nu$ by the auxiliary target distribution $\auxtarg$ and the proposal $\auxinstr$, respectively, yields
\begin{multline}
	\KL(\auxtarg \| \auxinstr)
	 = \Expect[\auxtarg]{\log \frac{\auxtarg(\ind[],\pprva)}{\auxinstr(\ind,\pprva)}} \eqsp, \label{eq:KLD:define} 
\end{multline}
which decouples as
\begin{multline}
	\KL(\auxtarg \| \auxinstr) 
	 = \underbrace{	\Expect[\auxtarg]{
	\log 	\frac{\uknormdens(\parti{\indva},\pprva)}{\propdens(\parti{\indva},\pprva)}}
	\vphantom{\left[\frac{\sum_i^N}{\sum_i^N}\right]}
	}_{\mathclap{\text{\normalsize Depends only on $\propdens$}}} \\ 
	\quad +
	\underbrace{\Expect[\auxtarg]{\log
			\frac{\wgt{\indva} \adjfuncopt(\parti{\indva}) / \sum_{j=1}^N \wgt{j} \adjfuncopt(\parti{j})}
				{\wgt{\indva} \adjfunc(\parti{\indva}) / \sum_{j=1}^N \wgt{j}
				\adjfunc(\parti{j})}}
				}_{\text{\normalsize Depends only on $\adjfunc$}} \eqsp.
				\label{eq:KLD:decouples}
\end{multline}
As clear from the definition \eqref{eq:def:auxtarg}, the measure $\auxtarg$ is random as it depends on the ancestor particles $\{ \parti{i} \}_{i = 1}^N$; the locations of the latter should thus be viewed as fixed within the brackets of the expectations in \eqref{eq:KLD:define} and \eqref{eq:KLD:decouples}. The first term in \eqref{eq:KLD:decouples} corresponds to the discrepancy induced by mutating the particles $\{ \partitd{i} \}_{i = 1}^N$ using a suboptimal proposal kernel, and the second term corresponds
to the discrepancy induced by sampling the ancestor indices $\{ \ind{i} \}_{i = 1}^N$ according
to suboptimal adjustment weights. Moreover, 
\begin{multline*} 
	\KL\left(\auxtarg\|\auxinstr\right)  = - \Expect[\auxtarg]{\log
	\propdens(\parti{\indva},\pprva)} \\
	- \Expect[\auxtarg]{\log \frac{\adjfunc(\parti{\indva})}{\sum_{j=1}^N \wgt{j}
	\adjfunc(\parti{j})}} + c
	\label{eq:KLD:first:term}
\end{multline*}
where the additive term $c$ involves the optimal quantities $\adjfuncopt$
and $\uknormdens$ only and is irrelevant for the optimisation problem. Equality up to a constant will in the following be denoted by $\equp$.
\emph{Restricting ourselves to adaptation of the proposal kernel}, we obtain the simple expression
\begin{equation} \label{eq:final:KLD}
\KL\left(\auxtarg\|\auxinstr\right) \equp - \E_{\auxtarg} \left[ \log \propdens(\parti{\indva}, \pprva) \right] \eqsp. 
\end{equation}
In the next section we present a method for minimising the KLD \eqref{eq:final:KLD} over a family $\{\auxinstrparam{\mixparamvec} ; \mixparamvec \in \mixparamsp\}$ of instrumental distributions where the proposal transition density of each member belongs to parametrised family of mixtures of experts. Formally, we will solve 
\begin{equation} \label{eq:opt:problem}
\mixparamopt \eqdef \argmin_{\mixparamvec \in \mixparamsp}
\KL(\auxtarg \| \auxinstrparam{\mixparamvec})
\end{equation}
and take $\pi = \auxinstrparam{\mixparamopt}$. Although the right hand side of \eqref{eq:final:KLD} is most often not directly computable,
such a quantity can be approximated \emph{on-the-fly} using the weighted sample already generated;
the resulting algorithms (Algorithms~\ref{alg:main:1} and \ref{alg:main:2}; see Section~\ref{sec:algo}) closely resembles the \emph{cross-entropy method} \citep[see][]{rubinstein:kroese:2004} method.

%% file: sections/framework_rev.tex
\section{Mixture of experts}
\label{sec:framework}
In this contribution, we consider $\psp = \rset^{\dimpsp}$ and $\psptd = \rset^{\dimpsptd}$ and let the proposal kernel have density (with respect to Lebesgue measure)
\begin{equation}
\label{eq:definition-propdensparam}
	\propdensparam{\mixparamvec}(\ppa,\ppr) \eqdef
	\sum_{j = 1}^d \mixwgt{j}(\ppa; \mixbetavec) \rho(\ppa,\ppr ; \boldsymbol{\eta}_j) \eqsp,
\end{equation}
where $\{ \mixwgt{j} \}_{j = 1}^d$ are nonnegative \emph{weighting
functions} summing into unity and $\rho$ is a Markov transition kernel from $\psp$ to $\psptd$. The weighting functions and the Markov transition density are parametrised by parameters $\mixbetavec$ and $\boldsymbol{\eta}$, respectively, and the $j$th stratum of the mixture is characterised by a certain choice $\boldsymbol{\eta}_j$ of the latter. The integer $d$ is fixed. 
Denote  
\begin{equation} \label{eq:def:param}
\begin{split}
\curvparvec &\eqdef \left\{\curvpar[1], \ldots, \curvpar[d]\right\} \eqsp, \\
\mixparamvec &\eqdef \left\{ \mixbetavec, \curvparvec \right\} \eqsp.
\end{split}
\end{equation}
Putting \eqref{eq:definition-propdensparam} into the auxiliary particle filter framework in Section~\ref{sec:SMC}, the associated
proposal distribution is given by  
\begin{multline}
	 \auxinstrparam{\mixparamvec}(\indv, \ppr) =\\
	 \frac{\wgt{\indv} \adjfunc(\parti{\indv})}{\sum_{\ell = 1}^N \wgt{\ell} \adjfunc(\parti{\ell})} \sum_{j=1}^d\mixwgt{j}(\parti{\indv}, \mixbetavec) \rho(\parti{\indv},\ppr ; \boldsymbol{\eta}_j) \eqsp.
	\label{eq:define:auxinstrdensparam}
\end{multline}
We then assign the importance weight $\wgttd[]{i}
= \wgtfunc_{\mixparamvec}(\ind[]{i}, \partitd[]{i})$ to each draw $(\ind[]{i}, \partitd[]{i})$ from $\auxinstrparam{\mixparamvec}$, where 
\begin{align} \label{eq:definine:wgttd}
\wgtfunc_{\mixparamvec}(\indv, \ppr) &= \dfrac{\ukdens(\parti{\indv},\ppr)}{\adjfunc(\parti[]{\indv}) \sum_{j = 1}^d \mixwgt{j}(\parti[]{i},\mixbetavec) \rho(\parti{\indv},\ppr ; \boldsymbol{\eta}_j)} \notag \\
&\propto \frac{\auxtarg(\indv,\ppr)}{\auxinstrparam{\mixparamvec}(\indv,\ppr)} \eqsp.
\end{align}
The set $\{ \mixwgt{j} \}_{j = 1}^d$ 
of weight functions typically partitions the input space $\psp$ into sub-regions with smooth transitions by assigning a vector of mixture weights to each point of the input space. In each sub-region, one proposal kernel will eventually dominate the mixture, specialising the whole proposal $\propdensparam{\mixparamvec}$ on a per-region basis. As in \cite{jordan:xu:1995}, we consider \emph{logistic} weight functions, \ie\
$\mixbetavec = \{\mixbeta{1}, \ldots, \mixbeta{d-1}\}$ and
\begin{equation} \label{eq:definition-mixwgt}
\mixwgt{j}(\ppa, \mixbetavec) \eqdef 
\begin{cases} 
\dfrac{\exp(\mixbeta{j}^\trans\prmat\extd{\ppa})}{1 + \sum_{\ell = 1}^{d - 1} \exp(\mixbeta{\ell}^\trans \prmat \extd{\ppa})} \eqsp, \quad j \neq d \eqsp, \\
\dfrac{1}{1 + \sum_{\ell = 1}^{d - 1} \exp(\mixbeta{\ell}^\trans \prmat \extd{\ppa})} \eqsp, \quad j = d \eqsp, 
\end{cases}
\end{equation}
where $\extd{\ppa} \eqdef (\ppa^\trans \ 1)^\trans$ and the $\mixbeta{j}$'s are vectors in $\R^{p+1}$. It is sometimes 
of interest to resort to simpler mixtures whose weights do not depend
on $\ppa$ by letting, for a set $\mixbetavec = (\beta_j)_{j = 1}^d$ nonnegative scalars summing into unity, $\mixwgt{j}(\ppa,\mixbetavec) \eqdef \beta_j$, $1 \leq j \leq d$, independently of $\ppa$ and hence without partitioning the input space. This model for the transition kernel is then similar to the \emph{switching regression model} \citep{quandt:ramsey:1972}.


The original mixture of experts proposed by \cite{jordan:jacobs:1994} was based on Gaussian strata $\rho$, which is indeed is the most straightforward choice. However the same algorithm can, at no additional cost, be extended to the broader class of densities of \emph{integrated curved exponential form}, described in Assumption~\ref{ass:integrated:curved:exponential:form} below, which still allows for the same trade-off between flexibility, ease of sampling, and convenient estimation.

\begin{Assumption} \label{ass:integrated:curved:exponential:form}
There exist a state space $\uusp \subseteq \rset^u$ 
and a Markov transition kernel from $\psp$ to $\psptd \times \uusp$ having a density of form   
\begin{multline} \label{eq:curv}
	 \curv{\ppa}{\ppr}{\uu}{\curvpar} \eqdef \gamma(\uu) h(\ppa, \ppr, \uu)\\
	 \times \exp \left( -\cA(\curvpar) + \trace(\cB(\curvpar)^\trans \suffstat(\ppa, \ppr, \uu)) \right) \eqsp, 
\end{multline}
where the functions $\gamma$, $h$,  $\cA$ and $\suffstat$, $\cB$ are real-valued resp. $\rset^{\sdim{1} \times \sdim{2}}$-valued, such that 
\begin{equation}
	\intcurv{\ppa}{\ppr}{\curvpar} = \int_{\uusp} \curv{\ppa}{\ppr}{\uu}{\curvpar} \, \ud \uu \eqsp.
	\label{eq:intcurv}
\end{equation} 
\end{Assumption}
In other words, $\intcurv{\ppa}{\ppr}{\curvpar}$ is the density of the marginal of a curved exponential distribution. In \eqref{eq:curv}, $\suffstat$ is a vector of sufficient statistics.

In the next section we present a method fitting efficiently the proposal $\auxinstrparam{\mixparamvec}$ to the target $\auxtarg$ by solving \eqref{eq:opt:problem}. We will illustrate the details for the specific case of multidimensional Gaussian distributions and multidimensional student's $t$-distributions. It should however be kept in mind that the algorithm is valid for any member of the family of integrated curved exponential distributions as long as Assumption~\ref{hyp:easycompute} below is fulfilled.

%% file: sections/algorithms_rev.tex
\section{Main results and algorithms}
\label{sec:algo}

In the following we discuss how the intricate optimisation problem \eqref{eq:opt:problem} can be cast into the framework of \emph{latent data problems}. Parameter estimation for mixtures of experts is most often carried through using the EM algorithm \citep[see][Section 8.4.6]{mclachlan:krishnan:2008}, and also in this paper a recursive EM-type algorithm will be used for finding close to optimal mixture parameters. More precisely, we develop an algorithm that is closely related to the \emph{online EM algorithm} for latent data models proposed by \cite{cappe:moulines:2009}. From a practical point of view, a key feature of the adaptive SMC approach described in Section~\ref{sec:adaptive:SMC} is that it aims at decreasing the KLD rather than obtaining a fully converged parameter estimate: very few EM iterations are therefore needed for obtaining a significant computational gain. This will be illustrated further in the simulation study of Section 6.

\subsection{Main assumptions and definitions}
\label{sec:main:assumptions:and:definitions}

In order to describe clearly the algorithm, \emph{we limit ourselves initially to the case of constant mixture weights}, \ie\ we let $\mixwgt{j}(\ppa,\mixbetavec) \eqdef \beta_j$, $1 \leq j \leq d$, for a set $\mixbetavec = \{ \beta_j \}_{j = 1}^d$ of nonnegative scalars summing into unity.  The more general choice \eqref{eq:definition-mixwgt} will be treated in Section~\ref{sec:algorithm:logistic:weights}.

We now augment the pair $(\indva, \pprva)$ with the index $\compva$ of the mixture component as well as the auxiliary, $\uusp$-valued variable $\uuva$ of the curved exponential family. More specifically, we introduce an extended random variable $(\indva, \compva, \pprva, \uuva)$ having mixed-type distribution with probability function 
\begin{multline} \label{eq:completelike}
	\completelike{\mixparamvec}(\indv, \comp, \ppr, \uu) \\ \eqdef \frac{\wgt{\indv} \adjfunc(\parti{\indv})}{\sum_{\ell = 1}^N \wgt{\ell} \adjfunc(\parti{\ell})} \mixwgt{\comp}(\parti{\indv},\mixbetavec) \curv{\parti{\indv}}{\ppr}{\uu}{\curvparj{\comp}}
\end{multline}
on the product space $\{1, \dots, N\} \times \{1,\dots,d\} \times \psptd \times \uusp$.
It is easily checked that $\auxinstrparam{\mixparamvec}$ is the marginal of $\auxinstrextd{\mixparamvec}$ in $\indva$ and $\pprva$. The following assumption is critical for convenient estimation. Although surprising at first glance, it simply requires that the state space extension used for handling integrated curved exponential distributions can be easily reverted for the sufficient statistics: the expectation of the sufficient statistics under the  conditional distribution of $\jvect{U}$ given $\ppava_I = \ppa$ and $\pprva = \ppr$, having density
\begin{equation} \label{eq:def:cond:u}
\bar{\rho}(\uu | \ppa, \ppr ; \boldsymbol{\eta}) \eqdef \frac{\curv{\ppa}{\ppr}{\uu}{\curvpar}}{\intcurv{\ppa}{\ppr}{\curvpar}} \eqsp, 
\end{equation}
should be available in closed form. Since the sufficient statistics of exponential families often have a simple form, this assumption is most often satisfied; e.g. it is satisfied for the Student's $t$-distribution in Example~\ref{sec:algo:kernels:student}. 
\begin{Assumption}
\label{hyp:easycompute}
For all $\boldsymbol{\eta}$ and $(\ppa, \ppr) \in \psp \times \psptd$, the expectation 
\begin{equation} \label{eq:Sclosedform}
	 \int \suffstat(\ppa, \ppr, \uu) \bar{\rho}(\uu | \ppa, \ppr ; \boldsymbol{\eta}) \, \ud \mathbf{u} 
\end{equation}
is available in closed form. 
\end{Assumption}

Under Assumption~\ref{hyp:easycompute}, let for $j \in \{1, \ldots, d\}$, 
\begin{equation} \label{eq:define:curvexpestat}
\begin{split} 
\pstat{j}{\mixparamvec} &\eqdef \Expect[\auxtarg]{\mixwgtcond{\mixparamvec}{j}{\indva}{\pprva}} \eqsp, \\
\curvexpestat{j}{\mixparamvec} &\eqdef \E_{\auxtarg} \left[
\mixwgtcond{\mixparamvec}{j}{\indva}{\pprva} 
\vphantom{\int \suffstat(\parti{\indva}, \pprva, \uu) \bar{\rho}(\uu | \parti{\indva}, \pprva ; \curvparj{j}) \, \ud \uu} 
\right.  \\
& \hspace{5mm} \left. \times \int \suffstat(\parti{\indva}, \pprva, \uu) \bar{\rho}(\uu |
	 \parti{\indva}, \pprva ; \curvparj{j}) \, \ud \uu \right] \eqsp,
\end{split} 
\end{equation}
where we have defined the \emph{responsibilities}
\begin{equation} \label{eq:def:responsibilities}
\begin{split}
\mixwgtcond{\mixparamvec}{j}{i}{\ppr} &\eqdef \frac{\int \completelike{\mixparamvec}(\indv, \comp, \ppr, \uu) \, \ud \uu}{\sum_{\ell = 1}^d \int \completelike{\mixparamvec}(\indv, \ell, \ppr, \uu) \, \ud \uu} \\
&= \frac{\beta_j \intcurv{\ppava_i}{\ppr}{\curvpar_j}}{\sum_{\ell = 1}^d \beta_\ell \intcurv{\ppava_i}{\ppr}{\curvpar_\ell}} \eqsp. 
\end{split}
\end{equation}
Note that the quantities defined in \eqref{eq:define:curvexpestat} depend implicitly on the ancestor sample. In addition, we collect these statistics in  
the matrix
\begin{equation} \label{eq:define:curvexpestat:vector}
\curvexpestat{}{\mixparamvec} \eqdef \curvexpestat{1}{\mixparamvec}  \varoplus \curvexpestat{2}{\mixparamvec} \varoplus \cdots \varoplus \curvexpestat{d}{\mixparamvec} 
\end{equation}
and the vector
\begin{equation} \label{eq:def:pstat:vector}
\pstat{}{\mixparamvec} \eqdef (\pstat{1}{\mixparamvec}, \pstat{2}{\mixparamvec}, \ldots, \pstat{d}{\mixparamvec})^\trans \eqsp. 
\end{equation}

We impose the following additional assumption. 
\begin{Assumption} \label{ass:existence:maximum}
There are subsets $\pset \subseteq \rset_+^d$ and $\sset \subseteq \rset^{\sdim{1} \times \sdim{2}}$ such that 
for all $\{ \jvect{s}_j \}_{j = 1}^d \in \sset^d$ and $\jvect{p} 
\in \pset$ 
, the mapping
\begin{equation} \label{eq:def:l}
\begin{split}
\lefteqn{\mixparamvec \mapsto l(\jvect{s}, \jvect{p} ; \mixparamvec)}\\
&\eqdef - \left( \cA(\curvpar_1), \ldots, \cA(\curvpar_d) \right) \jvect{p}\\ 
&\hphantom{=}+ \left( \log \beta_1, \ldots, \log \beta_d \right) \jvect{p}\\ 
&\hphantom{=}+ \sum_{j = 1}^d \trace(\cB(\curvpar_j)^\trans \mathbf{s}_j) 
\end{split}
\end{equation}
has \emph{unique} global maximum denoted by $\globmax(\jvect{s}, \jvect{p})$. In particular,
\begin{equation} \label{eq:def:glob:max}
\globmax(\jvect{s}, \jvect{p}) \eqdef \argmax_{\mixparamvec \in \Theta} l(\jvect{s}, \jvect{p} ; \mixparamvec) \eqsp. 
\end{equation}
\end{Assumption}

\begin{Remark}[Decoupling, optimal strata weights] \label{eq:rem:decoupling}
It should be noted that due to the additive form of $l$, the optimisation problem \eqref{eq:def:glob:max} can be split into the two separate subproblems
\begin{enumerate}
\item 
$$
\begin{cases}
\argmax_{\mixbetavec} \left( \log \beta_1, \ldots, \log \beta_d \right) \jvect{p} \\
\mbox{subject to } \sum_{j = 1}^d \beta_j = 1
\end{cases}
$$
\item 
\begin{multline*}
\argmax_{\curvparvec} \left( \left( \cA(\curvpar_1), \ldots, \cA(\curvpar_d) \right) \jvect{p} \vphantom{\sum_{j = 1}^d \trace \left( \cB(\curvpar_j)^\trans \mathbf{s}_j \right)} \right. \\ 
+ \left. \sum_{j = 1}^d \trace \left( \cB(\curvpar_j)^\trans \mathbf{s}_j \right) \right) \eqsp, 
\end{multline*}
\end{enumerate}
corresponding to maximisation over mixture weights and strata parameters, respectively. In the framework considered so far, where the mixture weights are assumed to be constant, the first of these problems (1) has, for all $\jvect{p} \in \rset_+^d$ the solution 
$$
\bar{\beta}_j = \frac{p_j}{\sum_{\ell = 1}^d p_\ell} 
$$
for all $j$. 
\end{Remark}

In the following two examples we specify the solution to the second problem (2) in Remark~\ref{eq:rem:decoupling} for the two---fundamental---special cases of \emph{multivariate Gaussian distributions} and \emph{multivariate Student's} $t$\emph{-distributions}; it should however be kept in mind that the method is valid for any integrated curved exponential distribution satisfying Assumption~\ref{hyp:easycompute}.

\begin{Example}[Updating formulas for Gaussian strata] \label{ex:gaussian:strata}
In a first example we let each stratum be the density of a linear Gaussian regression parameterised by
$\curvparj{} = (\mixmean{}, \mixvar{})$. The original hierarchical mixture of experts was based on plain Gaussian distributions and a possibly deep hierarchy of experts, incurring a computational overhead possibly larger than affordable for an algorithm used for adaptiation. Dropping the hierarchical approach is a possible way to reduce this overhead at the price of reduced flexibility. Here we compensate somewhat for this loss by using a linear regression within each expert. Thus, given $J = j$, the new particle location
$\pprva$ has $\dimosptd$-dimensional Gaussian distribution with
mean $\mixmean{j} \partiextd{\indva}$, where the \emph{regression matrix}
$\mixmean{j}$ is of size $\dimosptd \times (\dimosp+1)$, and symmetric, positive definite \emph{covariance matrix} $\mixvar{j}$ of size $\dimosptd \times \dimosptd$. In this case, the sufficient statistics are $\suffstat(\ppa, \ppr, \uu) = (\ppr \ppr^\trans) \varoplus
(\extd{\ppa} \extd{\ppa}^\trans) \varoplus (\ppr \prmat
\extd{\ppa}^\trans)$, 
leading to the expected responsibilities
\begin{equation*}
\pstat{j}{\mixparamvec} \eqdef 
\Expect[\auxtarg]{\mixwgtcond{\mixparamvec}{j}{I}{\pprva} } 
\end{equation*} 
and the expected sufficient statistics
$\curvexpestat{j}{\mixparamvec} \eqdef \curvexpestat{j,1}{\mixparamvec} \varoplus \curvexpestat{j,2}{\mixparamvec} \varoplus \curvexpestat{j,3}{\mixparamvec}$ with
\begin{equation*}
\begin{split}
\curvexpestat{j,1}{\mixparamvec} & \eqdef 
\Expect[\auxtarg]{\mixwgtcond{\mixparamvec}{j}{I}{\pprva} \, \pprva \pprva^\trans}
\eqsp, \\
\curvexpestat{j,2}{\mixparamvec} &  \eqdef 
\Expect[\auxtarg]{\mixwgtcond{\mixparamvec}{j}{I}{\pprva} \,
\partiextd{I} \partiextd{I}^\trans} \eqsp, \\
\curvexpestat{j,3}{\mixparamvec} & \eqdef 
\Expect[\auxtarg]{\mixwgtcond{\mixparamvec}{j}{I}{\pprva} \, \pprva \prmat \partiextd{I}^\trans} \eqsp.
\label{eq:define:mixsuffstats:gauss}
\end{split}
\end{equation*}

In addition, the functions $\cA$ and $\cB$ in \eqref{eq:curv} are in this case given by
\begin{equation} \label{eq:a:b:gaussian:case}
\begin{split}
\cA(\curvpar) &= \frac{1}{2} \log \detm{\mixvar{}}  \eqsp, \\
\cB(\curvpar) &= \left( - \frac{1}{2} \mixvar{}^{-1} \right) \varoplus \left( -\frac{1}{2} \mixmean{} \mixvar{}^{-1} \mixmean{}^\trans \right) \\
&\hphantom{=} \varoplus \left(- \mixvar{}^{-1} \mixmean{} \right) \eqsp.
\end{split}
\end{equation}
Consequently, the argument $\curvpar_j$ that maximizes $l(\jvect{s}, \jvect{p} ; \mixparamvec)$ for a given sufficient statistics $\suffstat_j$ is given by
\begin{equation} \label{eq:EM:solution:gaussian}
\begin{split}
\bar{\mixmean{}}_j(\suffstat_j) &\eqdef \suffstat_{j,3} \suffstat_{j,2}^{-1} \eqsp,\notag\\
\bar{\mixvar{}}_j(p_j,\suffstat_j) & \eqdef p_j^{-1} \left( \suffstat_{j,1} - \suffstat_{j,3} \suffstat_{j,2}^{-1} \suffstat_{j,3}^\trans \right) \eqsp.
\end{split}
\end{equation}
\begin{Remark}[Pooling the covariances] \label{rk:pooling}
In practice, a robustified version of the algorithm above can be obtained by \emph{pooling} the covariances \citep[see e.g.][]{rayens:greene:1991}. This means that a common covariance matrix is used for all the components of the mixture, \ie\ $\mixvar{j} = \mixvar{}$ for all $j$. By doing this, the well-known problem of mixture models with strata components degenerating to Dirac masses is avoided. 
It is straightforward to enforce such a restriction in the optimisation procedure above, leading to
\begin{equation} \label{eq:EM:solution:gaussian}
\bar{\mixvar{}}_j(p_j,\suffstat_j) \eqdef d^{-1} \sum_{\ell = 1}^d
\left( \suffstat_{\ell, 1} - \suffstat_{\ell, 3} \suffstat_{\ell, 2}^{-1} \suffstat_{\ell, 3}^\trans \right)
\end{equation}
for all $j$.
\end{Remark}
\end{Example}

\begin{Example}[Updating formulas for Student's $t$-distribution-based strata]
\label{sec:algo:kernels:student}

A common grip in importance sampling \citep[see, for instance,][]{oh:berger:1993}
is to replace, in order to allow for more efficient exploration of the state space, Gaussian distributions by Student's $t$-distributions. 
Therefore, as a more robust alternative to the approach taken in Example~\ref{ex:gaussian:strata}, one may use instead a $\dimosptd$-dimensional Student's $t$-distribution with $\nu$ degrees of freedom, \ie\ 
\begin{equation} \label{eq:students:t:stratum}
\intcurvj{\ppa}{\ppr} = \tdens[\dimosptd]{\ppr}{\mixmean{j} \extd{\ppa}}{\mixvar{j}}{\df} \eqsp. 
\end{equation}
\begin{Remark}[Fixing the degrees of freedom]
The number $\df$ of degrees of freedom of the
Student's $t$-distributions is fixed, typically to
$\df \in \{3,4\}$, beforehand and is common to all strata. A similar choice has
been made by, among others, \citet[Section 7]{peel:mclachlan:2000} and \cite{cappe:douc:guillin:marin:robert:2008}
as it allows for closed form optimisation in the $M$-step.
\end{Remark}
The choice \eqref{eq:students:t:stratum} can be cast into the framework of
Section~\ref{sec:framework}, with $\Upsilon = \rset$, thanks to the \emph{Gaussian}-\emph{Gamma
decomposition} of multivariate Student's $t$-distributions used by
\citet[Section 2]{liu:rubin:1995} and \citet[Section 3]{peel:mclachlan:2000}:
\begin{multline*}
\tdens{\ppr}{\mixmean{} \extd{\ppa}}{\mixvar{}}{\df}\\ =\int_0^{\infty} \normdens{\ppr}{\mixmean{} \extd{\ppa}}{\mixvar{} u^{-1}} \gammadens{u}{\frac{\df}{2}}{\frac{\df}{2}} \ud u \eqsp,
\end{multline*}
where $\gammadens{u}{a}{b} \eqdef b^a u^{a - 1} \exp( -b u) / \Gamma(a)$ is the density of the Gamma distribution with shape parameter $a$ and scale parameter $b$. Hence, the multivariate Student's $t$-distribution is an integrated curved exponential distribution~\eqref{eq:curv} with $\gamma(u) = \gammadens{u}{\df / 2}{\df / 2}$, $h(\ppa,\ppr, u) = (2 \pi)^{-\dimo/2}$, sufficient statistics $\suffstat(\ppa, \ppr,
u) = (u \ppr \ppr^\trans ) \varoplus (u \extd{\ppa}
\extd{\ppa}^\trans) \varoplus (u \ppr \extd{\ppa}^\trans)$, and the same $\cA$ and $\cB$ as in the Gaussian case \eqref{eq:a:b:gaussian:case}. Since the Gamma distribution is conjugate prior with respect to precision for the Gaussian distribution with known mean, the expectation $\eta$ of $U$ under $\bar{\rho}(u | \ppa, \ppr ; \curvpar)$ can in this case be expressed as 
\begin{equation} \label{eq:def:cond:mean:u}
\begin{split}
\eta(\ppr, \mixmean{} \extd{\ppa}, \mixvar{}) &\eqdef \int_0^\infty u \bar{\rho}(u | \ppa, \ppr ; \curvpar) \, \ud u \\
&= \frac{\df + \dimosptd}{\df + \mahalanobis{\ppr}{\mixmean{} \extd{\ppa}}{\mixvar{}}}\eqsp,
\end{split}
\end{equation}
where $\mahalanobis{\ppr}{\mixmean{} \extd{\ppa}}{\mixvar{}} = (\ppr - \mixmean{} \extd{\ppa})^\trans \mixvar{}^{-1} (\ppr - \mixmean{} \extd{\ppa})$ is the \emph{Mahalanobis distance} with covariance matrix $\mixvar{}$. This leads to the expected responsibilities
$$
	\pstat{j}{\mixparamvec} \eqdef \Expect[\auxtarg]{\mixwgtcond{\mixparamvec}{j}{I}{\pprva} }
$$
and the expected sufficient statistics $\curvexpestat{j}{\mixparamvec} = \curvexpestat{j,1}{\mixparamvec} \varoplus \curvexpestat{j,2}{\mixparamvec} \varoplus \curvexpestat{j,3}{\mixparamvec}$ with
\begin{equation} \label{eq:define:mixsuffstats:tstud}
	\begin{split}		
		\curvexpestat{j,1}{\mixparamvec} &\eqdef 
		\E_{\auxtarg}\left[\mixwgtcond{\mixparamvec}{j}{I}{\pprva} \eta(\pprva, \mixmean{} \extd{\ppava}, \mixvar{}) \pprva \pprva^\trans\right] \eqsp, \\
		\curvexpestat{j,2}{\mixparamvec} &\eqdef 
		\E_{\auxtarg}\left[\mixwgtcond{\mixparamvec}{j}{I}{\pprva} \eta(\pprva,
		\mixmean{} \extd{\ppava}, \mixvar{}) \partiextd{I}\partiextd{I}^\trans\right]
		\eqsp, \\
	 	\curvexpestat{j,3}{\mixparamvec} &\eqdef 
	 	\E_{\auxtarg}\left[\mixwgtcond{\mixparamvec}{j}{I}{\pprva} \eta(\pprva, \mixmean{} \extd{\ppava}, \mixvar{}) \pprva \partiextd{I}^\trans \right] \eqsp.
	\end{split}
\end{equation}
Since the functions $\cA$ and $\cB$ are unchanged from the Gaussian case, the same equations \eqref{eq:EM:solution:gaussian} can be applied for updating the parameter $\curvparvec$ based on the expected sufficient statistics \eqref{eq:define:mixsuffstats:tstud}. Moreover, covariance pooling can be achieved in exactly same way as in Remark~\ref{rk:pooling}.
\end{Example}

\subsection{Main algorithm}
\label{sec:main:algorithm}

Under the assumptions above we define the \emph{mean field}
\begin{equation} \label{eq:def:mean:field}
	\meanfd{\jvect{s}, \jvect{p}} \eqdef \curvexpestat{}{\globmax(\jvect{s}, \jvect{p})}  \varoplus \pstat{}{\globmax(\jvect{s}, \jvect{p})} - \jvect{s} \varoplus \jvect{p} \eqsp. 
\end{equation}
We now have the following result (proved in Section~\ref{sec:proofs}), which serves as a basis for the method proposed in the following. 

\begin{Proposition} \label{prop:key:result}
	Under Assumptions~\ref{ass:integrated:curved:exponential:form}--\ref{ass:existence:maximum}, the following holds true. If $(\jvect{s}^*, \jvect{p}^*)$ is a root of the mean field $\jvect{h}$, defined in \eqref{eq:def:mean:field}, then $\globmax(\jvect{s}^*, \jvect{p}^*)$ is a stationary point of $\mixparamvec \mapsto \nablapar \KL(\auxtarg \| \auxinstrparam{\mixparamvec})$. Conversely, if $\mixparamvec^*$ is a stationary point of the latter mapping, then $(\jvect{s}^*, \jvect{p}^*)$, with 
	\begin{equation} \label{eq:def:root:candidate}
		\begin{split}
			\jvect{s}^* &= \curvexpestat{}{\mixparamvec^*} \eqsp, \\
			\jvect{p}^* &= \pstat{}{\mixparamvec^*} \eqsp, 
		\end{split}
	\end{equation}
	is a root of $\jvect{h}$. 
\end{Proposition}

From Proposition~\ref{prop:key:result} it is clear that the optimal parameter $\mixparamvec^*$ solving the optimisation problem \eqref{eq:opt:problem} over the parameter space $\Theta$ can, under the assumptions stated above, be obtained by solving the equation $\jvect{h}(\jvect{s}, \jvect{p}) = \jvect{0}$ in the space $\sset^d \times \pset$ of sufficient statistics. Nevertheless, solving the latter equation obstructed by the fact that the expected sufficient statistics    \eqref{eq:define:curvexpestat}, and consequently the mean field itself, cannot be expressed on closed form (as we do not know the target distribution $\auxtarg$). Thus, following the online EM approach taken by \cite{cappe:moulines:2009}, we aim at finding this root using stochastic approximation. More specifically, we apply the classical \emph{Robbins-Monro procedure}
\begin{equation} \label{eq:Robbins:Monro}
	\jvecth{s}_{\ell + 1} \varoplus \jvecth{p}_{\ell + 1} =  \jvecth{s}_\ell \varoplus \jvecth{p}_\ell + 
	\step{\ell + 1} \tilde{\jvect{h}}(\jvecth{s}_\ell, \jvecth{p}_\ell) \eqsp, 
\end{equation}
where $\{ \step{\ell} \}_{\ell \geq 1}$ is a decreasing sequence of positive step sizes such that 
\begin{equation} \label{eq:step:size:criterion}
	\sum_{\ell = 1}^\infty \step{\ell} = \infty \eqsp, \quad
	\sum_{\ell = 1}^\infty \step{\ell}^2 < \infty 
\end{equation}
and $\tilde{\jvect{h}}(\jvecth{s}_\ell, \jvecth{p}_\ell)$ is a noisy observation of $\meanfd{\jvecth{s}_\ell, \jvecth{p}_\ell}$. We form these noisy observations by means importance sampling in the following way: at iteration $\ell + 1$ (i.e. when computing $\jvecth{s}_{\ell + 1}$ and $\jvecth{p}_{\ell + 1}$ given $\jvecth{s}_\ell$ and $\jvecth{p}_\ell$) we estimate 
$$
	\meanfd{\jvecth{s}_\ell, \jvecth{p}_\ell} = \curvexpestat{}{\globmax(\jvecth{s}_\ell, \jvecth{p}_\ell)}  \varoplus \pstat{}{\globmax(\jvecth{s}_\ell, \jvecth{p}_\ell)} - \jvecth{s}_\ell \varoplus \jvecth{p}_{\ell}
$$
by, first, setting $\param[\ell] \eqdef \globmax(\jvecth{s}_\ell, \jvecth{p}_\ell)$, second, estimating $\curvexpestat{}{\param[\ell]} $ and $\pstat{}{\param[\ell]}$ by drawing particles and indices $\{ (\ind[\ell]{i}, \partitd[\ell]{i}) \}_{i = 1}^{\n[\ell]}$ from $\auxinstrparam{\param[\ell]}$ (the currently best fitted proposal distribution), computing the associated weights $\{ \wgttd[\ell]{i} \}_{i = 1}^{\n[\ell]}$, and forming the importance sampling estimates
\begin{equation} \label{eq:def:IS:suff:stat}
	\begin{split}
		\pstatest{j}{{\param[\ell]}} &\eqdef \sum_{i = 1}^{\n[\ell]} \wgttd[\ell]{i} 
		\mixwgtcond{{\param[\ell]}}{j}{\ind[\ell]{i}}{\partitd[\ell]{i}} \eqsp, \\
		\sstatest_j(\param[\ell])&\eqdef \sum_{i = 1}^{\n[\ell]} \wgttd[\ell]{i} 
		\mixwgtcond{{\param[\ell]}}{j}{\ind[\ell]{i}}{\partitd[\ell]{i}} \\
		&\hspace{-7mm} \times \int \suffstat(\parti{\ind[\ell]{i}}, \partitd[\ell]{i}, \uu) \bar{\rho}(\uu | \parti{\ind[\ell]{i}}, \partitd[\ell]{i} ; \boldsymbol{\eta}_j^{\ell}) \, \ud \uu \eqsp. 
	\end{split}
\end{equation}
Note that we let the Monte Carlo sample size $\n[\ell]$, which can be relatively small compared to the sample size $\n$ of the particle filter, increase with the iteration index $\ell$. As before, we collect the statistics defined in \eqref{eq:def:IS:suff:stat} in a matrix
\begin{equation} \label{eq:define:IS:curvexpestat:vector}
	\sstatest(\param[\ell]) \eqdef \left( \sstatest_1(\param[\ell]) \ \sstatest_2(\param[\ell]) \cdots \sstatest_d(\param[\ell]) \right) 
\end{equation}
and a vector
\begin{equation} \label{eq:def:IS:pstat:vector}
	\pstatest{}{{\param[\ell]}} \eqdef (\pstatest{1}{\param[\ell]}, \pstatest{2}{\param[\ell]}, \ldots, \pstatest{d}{\param[\ell]})^\trans \eqsp. 
\end{equation}
Note that the importance sampling estimates defined in \eqref{eq:def:IS:suff:stat} lack normalisation $\sum_{i = 1}^{\n[\ell]} \wgttd[\ell]{i}$. The role of the latter is to estimate the normalising constant $\iint  \jinit(\ppa) \ukdens(\ppa, \ppr) \, \ud \ppa \, \ud \ppr$ of the target distribution defined in \eqref{eq:SISrecursion}, as it holds that
\begin{equation*} \label{eq:norm:const:estimator}
	\frac{1}{N} \sum_{i = 1}^{\n} \wgttd{i} \plim \iint \jinit(\ppa) \ukdens(\ppa, \ppr)
	\, \ud \ppa \, \ud \ppr 
\end{equation*}
as $\n \rightarrow \infty$ \citep[see e.g.][]{douc:moulines:olsson:2008}. Here $\plim$ denotes convergence in probability. However, as $\n[\ell]$ is supposed to be considerably smaller that $\n$, which is necessary for obtaining a computationally efficient algorithm, this estimator suffers from large variance in our case. Thus, in order to robustify the estimator we combine the Robbins-Monro procedure \eqref{eq:Robbins:Monro} with a similar procedure 
\begin{equation} \label{eq:robbins:monro:norm:const}
\normconst{\ell + 1} = (1 - \step{\ell + 1}) \normconst{\ell} + \frac{\step{\ell + 1}}{\n[\ell]} \sum_{i = 1}^{\n[\ell]} \wgttd[\ell]{i}  
\end{equation}
for the normalising constant. The recursion \eqref{eq:robbins:monro:norm:const} is typically initialised with 
$$
\normconst{0} = \frac{1}{\n[0]} \sum_{i = 1}^{\n[0]} \wgttd[0]{i} \eqsp. 
$$
Obviously, \eqref{eq:robbins:monro:norm:const} does not guarantee that the normalised weights $\{ \wgttd[\ell]{i} / \normconst{\ell + 1} \n[\ell] \}_{i = 1}^{\n{\ell}}$ sum to unity, and our approach thus leads to approximation of a probability measure with a finite measure. However, this does not impede the convergence of the stochastic approximation algorithm. Needless to say, this does not effect the properties of the particle approximation as the weights of the final particle sample obtained using the adapted kernel are renormalised in the usual manner and thus sum to unity. The final weighted empirical measure associated with the particle cloud is therefore still a probability distribution.

So, the Robbins-Monro recursion \eqref{eq:Robbins:Monro} is executed for, say, $L$ iterations \emph{before every updating step of the SMC algorithm} or only at time steps decided by the user, and the final parameter fit $\param[L]$ defines the instrumental distribution $\auxinstrparam{\mixparamopt}$ used for propagating the particles. Before formulating the final algorithm, working in practice, the following assumption is needed to formulate a practically useful algorithm. 
\begin{Assumption} \label{ass:closed:conv:comb}
There is a set $\cset \subseteq \rset_+$ such that for all $N \geq 1$, $\{ \partitd{i} \}_{i = 1}^N \in \tilde{\boldsymbol{\Xi}}^N$, $\{ \ind{i} \}_{i = 1}^N \in \{ 1, \ldots N \}^N$, $\param \in \Theta$, $c \in \cset$, $\jvect{s} \in \sset^d$, $\jvect{p} \in \pset$, and $\lambda \in (0, 1)$, it holds that 
\begin{equation*}
\begin{split}
(1 - \lambda) c &+ \frac{\lambda}{\n} \sum_{i = 1}^{\n} \wgttd[\ell]{i} \in \cset \eqsp, \\
(1 - \lambda) \jvect{s} &+ \frac{\lambda}{c \n} \sstatest(\param) \in \sset^d \eqsp, \\
(1 - \lambda) \jvect{p} &+ \frac{\lambda}{c \n} \pstatest{}{{\param}} \in \pset \eqsp, \\
\end{split}
\end{equation*}
where the quantities $\sstatest(\param)$ and $\pstatest{}{{\param}}$ are defined through \eqref{eq:define:IS:curvexpestat:vector} and \eqref{eq:def:IS:pstat:vector}, respectively. 
\end{Assumption}
Under Assumption~\ref{ass:closed:conv:comb} our main algorithm, combining the two procedures \eqref{eq:Robbins:Monro} and \eqref{eq:robbins:monro:norm:const}, is well defined and goes as follows. 
\begin{algorithmhere}
	\caption{}
	\label{alg:main:1}
	\begin{algorithmic}[1]
	\Require $\{ (\parti{i}, \wgt{i}) \}_{i = 1}^{\n}$, $\param[0]$, $\normconst{0}$, $\jvecth{s}_0$, $\jvecth{p}_0$
	\For{$\ell = 0 \to L - 1$}
		\For{$i = 1 \to \n[\ell]$}
			\State draw $(\ind[\ell]{i}, \partitd[\ell]{i}) \sim \auxinstrparam{\param[\ell]}$
			\State set $\wgttd[\ell]{i} \gets \wgtfunc_{\param[\ell]}(\ind[\ell]{i}, \partitd[\ell]{i})$
		\EndFor
		\State compute $\sstatest(\param[\ell])$ and $\pstatest{}{{\param[\ell]}}$ as in \eqref{eq:def:IS:suff:stat}
		\State set
		$$
			\normconst{\ell + 1} \gets (1 - \step{\ell + 1}) \normconst{\ell} + \frac{\step{\ell + 1}}{\n[\ell]} \sum_{i = 1}^{\n[\ell]} \wgttd[\ell]{i} 
		$$
		\State set
		\begin{multline*}
			\jvecth{s}_{\ell + 1} \varoplus \jvecth{p}_{\ell + 1} \gets (1 - \step{\ell + 1})
			\jvecth{s}_\ell \varoplus \jvecth{p}_{\ell} \\
			+ \frac{\step{\ell + 1}}{\normconst{\ell + 1} \n[\ell]}
			\sstatest(\param[\ell]) \varoplus \pstatest{}{{\param[\ell]}} 
		\end{multline*}
		\State set $\param[\ell + 1] \gets \globmax(\jvecth{s}_{\ell + 1}, \jvecth{p}_{\ell + 1})$
	\EndFor
	\end{algorithmic}
\end{algorithmhere}

\subsection{Algorithm formulation for logistic weights}
\label{sec:algorithm:logistic:weights}

We now extend the case of constant mixture weights to the considerably more complicated case of logistic weights \eqref{eq:definition-mixwgt}. In this case the gradient $\nablapar \KL(\auxtarg \| \auxinstrparam{\mixparamvec})$, the crucial quantity in the proof of Proposition~\ref{prop:key:result} (given in Section~\ref{sec:proofs}), is given by 
\begin{equation*} 
\begin{split}
\lefteqn{\nablapar \KL(\auxtarg \| \auxinstrparam{\mixparamvec})} \\
&= \E_{\auxtarg} \left[ \nablapar \log \auxinstrparam{\mixparamvec}(\indva, \pprva) \right] \\
&= - \left( \nablapar \cA(\curvpar_1) \cdots \nablapar \cA(\curvpar_d) \right) \pstat{}{\mixparamvec} \\
&\hphantom{=} +  \sum_{j = 1}^d \sum_{m = 1}^{\sdim{2}} \nablapar \colv{\cB}{m}(\curvpar_j)^\trans \curvexpestat{j|m}{\mixparamvec} +  \nablapar\kappa(\mixparamvec) \eqsp, 
\end{split}
\end{equation*}
where 
\begin{equation} \label{eq:def:kappa}
\kappa(\mixparamvec) \eqdef \E_{\auxtarg} \left[ \sum_{j = 1}^d \log \mixwgt{j}(\ppava_I, \mixbetavec) \mixwgtcond{\mixparamvec}{j}{I}{\pprva} \right]
\eqsp,
\end{equation}
with responsibilities $\mixwgtcond{\mixparamvec}{j}{\indv}{\ppr}$ given by, in this case,
\begin{equation*} \label{eq:define:mixwgtcond:generic}
	\mixwgtcond{\mixparamvec}{j}{\indv}{\ppr} \eqdef \frac{\mixwgt{j} (\parti{\indv},\mixbetavec) \intcurvj{\parti{\indv}}{\ppr}}
	{\sum_{\ell = 1}^d \mixwgt{\ell}(\parti{\indv}, \mixbetavec) \intcurvj[\ell]{\parti{\indv}}{\ppr}} 
\end{equation*}
and the vectors $\pstat{}{\mixparamvec}$ and $\curvexpestat{}{\mixparamvec}$ being defined as in \eqref{eq:def:pstat:vector}. However, since we are no longer within the framework of exponential families, we will not be able to find a closed-form zero of $\nablapar \kappa(\mixparamvec)$ (e.g. a closed-form global maximum of $\kappa(\mixparamvec)$) when $\pstat{}{\mixparamvec}$ and $\curvexpestat{}{\mixparamvec}$ are replaced by stochastic approximation iterates $\jvecth{p}_\ell$ and $\jvecth{s}_\ell$, respectively. On the other hand, for all $\param[\ell] \in \Theta$, the mapping 
\begin{multline*}
\mixbetavec \mapsto \bar{\kappa}(\mixbetavec ; \param[\ell]) \\ \eqdef \E_{\auxtarg} \left[ \sum_{j = 1}^d \log \mixwgt{j}(\ppava_I, \mixbetavec) \mixwgtcond{\param[\ell]}{j}{I}{\pprva} \right]
\end{multline*}
is \emph{concave} (this stems directly from the multinomial logistic regression by computing conditional expectations) and can thus be approximated well by a second degree polynomial (in $\mixbetavec$). More specifically, consider the gradient $\nablaop{\mixbetavec}  \bar{\kappa}(\mixbetavec ; \param[\ell])$, whose $j$th block ($j \in \{1, \ldots, d - 1 \}$) is given by
\begin{multline} \label{eq:gradient:block}
\nablaop{\mixbeta{j}} \bar{\kappa}(\mixbetavec ; \param[\ell]) \\
=\Expect[\auxtarg]{\left( \mixwgtcond{\param[\ell]}{j}{I}{\pprva} - \mixwgt{j}(\parti{\indva}, \mixbetavec) \right) \partiextd{\indva}} \eqsp, 
\end{multline}
and the Hessian $\nablaop{\mixbetavec} \nablaop{\mixbetavec} \bar{\kappa}(\mixbetavec ; \param[\ell])$, having the matrix 
\begin{multline} \label{eq:Hessian:block}
\nablaop{\mixbeta{j}} \nablaop{\mixbeta{j'}}  \bar{\kappa}(\mixbetavec ; \param[\ell]) =  \E_{\auxtarg} \left[ \mixwgt{j}(\parti{\indva},\mixbetavec) \left( \mixwgt{j'}(\parti{\indva},\mixbetavec) \vphantom{\frac{ \indic_{\{j = j'\}}}{1 + \sum_{\ell = 1}^{d - 1} \exp (\mixbeta{\ell}^\trans \prmat \partiextd{\indva})}} \right. \right.  \\
\left. \left. - \frac{ \indic_{j = j'}}{1 + \sum_{m = 1}^{d - 1} \exp (\mixbeta{m}^\trans \prmat \partiextd{\indva})} \right) \partiextd{\indva}  \partiextd{\indva}^\trans  \right] 
\end{multline}
as block $(j, j') \in \{1, \ldots, d - 1\}^2$. Note that the Hessian above does not depend on $\param[\ell]$.

Using these definitions, the function $\bar{\kappa}(\mixbetavec ; \param[\ell])$ can be approximated by means of a second order Taylor expansion around the previous parameter estimate $\mixbetavec[\ell]$ according to 
\begin{multline} \label{eq:Taylor:expansion}
\bar{\kappa}(\mixbetavec ; \param[\ell]) \approx \bar{\kappa}(\mixbetavec[\ell] ; \param[\ell]) + (\mixbetavec - \mixbetavec[\ell]) \cdot \nablaop{\mixbetavec} \bar{\kappa}(\mixbetavec ; \param[\ell]) \big|_{\mixbetavec = \mixbetavec[\ell]} \\+ \frac{1}{2} (\mixbetavec - \mixbetavec[\ell]) \cdot [(\mixbetavec - \mixbetavec[\ell]) \cdot \nablaop{\mixbetavec} \nablaop{\mixbetavec} \bar{\kappa}(\mixbetavec ; \param[\ell]) \big|_{\mixbetavec = \mixbetavec[\ell]}] \eqsp. 
\end{multline}
In the light of \eqref{eq:Taylor:expansion}, we approximate the gradient and Hessian quantities (which again lack closed-form expressions due to the expectation over $\auxtarg$) using stochastic approximation. For this purpose, we set
\[
\begin{split}
\tstat_j(\mixparamvec) &\eqdef \nablaop{\mixbeta{j}} \bar{\kappa}(\mixbetavec ; \param) \eqsp, \\
\vstat_{j, j'}(\mixparamvec) &\eqdef \nablaop{\mixbeta{j}} \nablaop{\mixbeta{j'}}  \bar{\kappa}(\mixbetavec ; \param) \eqsp,
 \end{split}
\]
and define
\begin{equation} \label{eq:def:grad:hessian:exp:stat}
\begin{split}
\tstat(\mixparamvec) &\eqdef (\tstat_1(\mixparamvec) \ \cdots \ \tstat_{d - 1}(\mixparamvec)) \eqsp, \\
\vstat(\mixparamvec) &\eqdef 
\left(
\begin{array}{ccc}
\vstat_{1,1}(\mixparamvec) & \cdots  &\vstat_{1,d - 1}(\mixparamvec) \\
\vdots         & \ddots  & \vdots \\
\vstat_{d - 1, 1}(\mixparamvec) & \cdots & \vstat_{d - 1, d - 1}(\mixparamvec) \\
\end{array}
\right) \eqsp, 
\end{split}
\end{equation}
and find, using again the Robbins-Monro procedure, a root to the extended mean field
\begin{multline*} \label{eq:def:ext:mean:field}
\meanfd{\jvect{s}, \jvect{p}, \jvect{v}, \jvect{t}} 
\eqdef 
\curvexpestat{}{\globmax(\jvect{s}, \jvect{p}, \jvect{v}, \jvect{t})} \varoplus \pstat{}{\globmax(\jvect{s}, \jvect{p}, \jvect{v}, \jvect{t})} \\ 
\varoplus 
\tstat(\globmax(\jvect{s}, \jvect{p}, \jvect{v}, \jvect{t})) \varoplus
\vstat(\globmax(\jvect{s}, \jvect{p}, \jvect{v}, \jvect{t}))
-
\jvect{s} \varoplus
\jvect{p} \varoplus
\jvect{v} \varoplus
\jvect{t} \eqsp. 
\end{multline*}
In this case, the mapping $\globmax$ updates the strata parameters in analogy with the case of constant mixture weights (Section~\ref{sec:main:assumptions:and:definitions}) while the $\mixbetavec$ parameter gets updated according to the Newton-Raphson-type formula
\begin{equation*} \label{eq:nrdir:define}
\mixbetavec[\ell + 1] \eqdef \mixbetavec[\ell] - \jvect{v}^{-1} \jvect{t} \eqsp. 
\end{equation*}
The full procedure is summarised below, where we have defined $\pstatest{}{{\param[\ell]}}$ as in \eqref{eq:def:IS:pstat:vector} and 
\begin{equation} \label{eq:def:IS:v:t:stat}
\begin{split}
\tstatest_j(\param[\ell])&\eqdef \sum_{i = 1}^{\n[\ell]} \wgttd[\ell]{i} 
\mixwgtcond{{\param[\ell]}}{j}{\ind[\ell]{i}}{\partitd[\ell]{i}} \\
&\hspace{-7mm} \times \int \tstat(\parti{\ind[\ell]{i}}, \partitd[\ell]{i}, \uu) \bar{\rho}(\uu | \parti{\ind[\ell]{i}}, \partitd[\ell]{i} ; \boldsymbol{\eta}_j^{\ell}) \, \ud \uu \eqsp, \\
\vstatest_j(\param[\ell])&\eqdef \sum_{i = 1}^{\n[\ell]} \wgttd[\ell]{i} 
\mixwgtcond{{\param[\ell]}}{j}{\ind[\ell]{i}}{\partitd[\ell]{i}} \\
&\hspace{-7mm} \times \int \vstat(\parti{\ind[\ell]{i}}, \partitd[\ell]{i}, \uu) \bar{\rho}(\uu | \parti{\ind[\ell]{i}}, \partitd[\ell]{i} ; \boldsymbol{\eta}_j^{\ell}) \, \ud \uu \eqsp.
\end{split}
\end{equation}

\begin{algorithmhere}
\caption{}
\label{alg:main:2}
\begin{algorithmic}[1]
\Require $\{ (\parti{i}, \wgt{i}) \}_{i = 1}^{\n}$, $\param[0]$, $\normconst{0}$, $\jvecth{s}_0$, $\jvecth{p}_0$, $\jvecth{t}_0$, $\jvecth{v}_0$
\For{$\ell = 0 \to L - 1$}
\For{$i = 1 \to \n[\ell]$}
\State draw $(\ind[\ell]{i}, \partitd[\ell]{i}) \sim \auxinstrparam{\param[\ell]}$
\State set $\wgttd[\ell]{i} \gets \wgtfunc_{\param[\ell]}(\ind[\ell]{i}, \partitd[\ell]{i})$
\EndFor
\State compute $\sstatest(\param[\ell])$ and $\pstatest{}{{\param[\ell]}}$ through  \eqref{eq:def:IS:suff:stat}
\State compute $\tstatest(\param[\ell])$ and $\vstatest{}(\param[\ell])$ through  \eqref{eq:def:IS:v:t:stat}
\State set
$$
\normconst{\ell + 1} \gets (1 - \step{\ell + 1}) \normconst{\ell} + 
\frac{\step{\ell + 1}}{\n[\ell]} \sum_{i = 1}^{\n[\ell]} \wgttd[\ell]{i} 
$$
\State set
\begin{multline*}
\jvecth{s}_{\ell + 1} \varoplus
\jvecth{p}_{\ell + 1} \varoplus \jvecth{t}_{\ell + 1} \varoplus \jvecth{v}_{\ell + 1}\\
 \gets (1 - \step{\ell + 1})
\jvecth{s}_\ell \varoplus
\jvecth{p}_{\ell} \varoplus \jvecth{t}_\ell \varoplus \jvecth{v}_\ell 
\\ + 
\frac{\step{\ell + 1}}{\normconst{\ell + 1} \n[\ell]}
\sstatest(\param[\ell]) \varoplus
\pstatest{}{{\param[\ell]}} \varoplus \tstatest(\param[\ell]) \varoplus \vstatest{}(\param[\ell])
\end{multline*}
\State set $\param[\ell + 1] \gets \globmax(\jvecth{s}_{\ell + 1}, \jvecth{p}_{\ell + 1})$
\Comment{Here $\mixbetavec$ is updated according to $\mixbetavec[\ell + 1] \eqdef \mixbetavec[\ell] - \jvecth{v}^{-1}_{\ell + 1} \jvecth{t}_{\ell + 1}$.}
\EndFor
\end{algorithmic}
\end{algorithmhere}

\subsection{Some practical guidelines to choose the algorithm's parameters}

The algorithm outlined above involves parameters, such as the number $d$ of mixture components and Monte Carlo sample sizes $\{ N_\ell \}_{\ell = 0}^{L - 1}$, that need to be set. Fortunately, these parameters are easy to tune. Moreover, we may expect a significant improvement of the SMC proposal distribution also with a suboptimal choice of the same. Without constituting a definite answer on how to tune the algorithmic parameters, the following guidelines emerge from the simulations in Section~6. 
\begin{itemize}[leftmargin=1em]
\item Concerning the number $d$ of experts, a too small number of strata averts good approximation of strongly nonlinear kernels while a too large $d$ leads to computational overhead. 
For instance, in the simulations in Section~6 we use up to $d = 8$ strata to approximate well a kernel with a ring-shaped, 2-dimensional support. 
\item The initial values of the mixture parameters could be chosen based on prior knowledge of the model by, e.g., fitting, in the case of optimal filtering with informative or non-informative observations, the local likelihood or the prior kernel, respectively. In our simulations, we initialise the algorithm with uniform logistic weights (i.e. with $\mixbetavec$ being a null matrix), regressions being null except for the intercept (implying independence of the ancestors), and covariances making the strata widely overspread on the support of the target.
\item For simplicity, the importance sampling sizes can be taken constant $\n[0] = \n[1] = \ldots = \n[L - 1]$ over all iterations. This is well in line with the theory of stochastic approximation \citep[see e.g.][]{duflo:1997,benveniste:metivier:priouret:1990}, which in general guarantees, if the step size sequence is chosen correctly, convergence of the algorithm as long as the variance of the observation noise is not increasing with the iteration index.  Moreover, the importance sampling size can be chosen relatively to the computational budget, by using, say,  a proportion $\alpha = 20\%$ or  $\alpha = 50\%$ of the total number $N$ of particles that would be used for a plain SMC algorithm for the adaptation step. More specifically, this is achieved by using $N_\ell = \alpha \n / L$ samples per iteration in the adaptation step, and propagating $(1 - \alpha) \n$ particles through the SMC updating step (based on the adapted proposal). In order assure reasonably fast convergence, the proportion $\alpha$ should be large enough to assure a decent sampling size $N_\ell$, typically not less than a hundred particles. It can be helpful to let the sample size $N_0$ be twice or three times larger than the size used at a typical iteration, to recover from a potentially bad initial choice. 
\item The number $L$ of iterations can typically be very small compared to typical stochastic approximation EM runs: as mentioned at the beginning of Section~5, the main gain in KLD typically occurs in the first dozen iterations. 
\item The least precise guideline concerns the step sizes $\{ \lambda_\ell \}_{\ell = 0}^{L - 1}$. Since we not search an exact optimum but rather aim at fast convergence towards the region of interest, we use a slowly decreasing step size. A rate matching this could be, say,  letting $\lambda_{\ell} = \ell^{- 0.6}$ \citep[which was used by e.g.][Section~11.1.5, within the framework of stochastic approximation EM methods]{cappe:moulines:ryden:2005}, which satisfies \eqref{eq:step:size:criterion}. However, in our simulations we use, for simplicity, a constant step size. This is well motivated by the fact that the adaptation algorithm is run for only a few iterations. 
\end{itemize}

%% file: sections/examples_rev.tex
\section{Applications to nonlinear SSMs}
\label{sec:examples}

As mentioned, SMC methods can be successfully applied to optimal filtering in SSMs. An SSM is a bivariate process $\{ (\jvect{X}_k, \jvect{Y}_k) \}_{k \geq 0}$, where $\jvect{X} \eqdef \{ \jvect{X}_k \}_{k \geq 0}$ is an unobserved Markov chain on some state space $\stsp \subseteq \R^x$ and $\{ \jvect{Y}_k \}_{k \geq 0}$ is an observation process taking values in some space $\obsp \subseteq \R^y$. We denote by $\hk$ and $\prior$ the transition density (in the following sometimes referred to as the \emph{prior kernel}) and initial distribution of $\jvect{X}$, respectively. Conditionally on $\jvect{X}$, the observations are assumed to be conditionally independent with the conditional distribution $G$ of a particular $\jvect{Y}_k$ depending on the corresponding $\jvect{X}_k$ only. We will assume that $\hk$ and $G$ have densities $\hkdens$ and $\olik$, respectively, with respect to Lebesgue measure, \ie\
\begin{equation} \label{eq:state:eqn}
	\mathbb{P}(\jvect{X}_{k + 1} \in \set{A} | \jvect{X}_k) = \hk(\jvect{X}_k, \set{A}) = \int_{\set{A}} \hkdens(\jvect{X}_k, \jvect{x}) \, \rmd \jvect{x}
\end{equation}
and
\begin{equation} \label{eq:obs:eqn}
	\mathbb{P}(\jvect{Y}_k \in \set{B} | \jvect{X}_k) = G(\jvect{X}_k, \set{B}) = \int_{\set{B}} \olik(\jvect{X}_k, \jvect{y}) \, \rmd \jvect{y} \eqsp.
\end{equation}
For simplicity, we assume that the Markov chain $\jvect{X}$ is
time-homogeneous and that the distribution $G$ does not depend on $k$;
however, all developments that follow may be extended straightforwardly to
the time-inhomogeneous case. The optimal filtering problem consists of
computing, given a fixed record $\jvect{Y}_{0:n} \eqdef (\jvect{Y}_0, \ldots, \jvect{Y}_n)$ of observations, the \emph{filter distributions}
\begin{equation} \label{eq:def:filter}
	\filtpost{k}(\set{A}) \eqdef \mathbb{P}(\jvect{X}_k \in \set{A} | \jvect{Y}_{0:k}) 
\end{equation}
for $k = 0, 1, \ldots, n$. As the sequence of filter distributions satisfies the nonlinear recursion
\begin{multline}
    \filtpost{k}(\jvect{x}_{k + 1}) =\\  \frac{
    \int \olik(\jvect{x}_{k + 1}, \jvect{Y}_{k + 1}) 
    \hkdens(\jvect{x}_k, \jvect{x}_{k + 1}) \filtpost{k}(\jvect{x}_k) \, \ud \jvect{x}_k}{\iint \olik(\jvect{x}_{k + 1}, \jvect{Y}_{k + 1})  \hkdens(\jvect{x}_k, \jvect{x}_{k + 1}) \filtpost{k}(\jvect{x}_k) \, \ud \jvect{x}_k \, \ud \jvect{x}_{k + 1}} \eqsp, 
\label{eq:filteringRecursion}
\end{multline}
the optimal filtering problem can be perfectly cast into the sequential
importance sampling framework \eqref{eq:SISrecursion} with $\psp = \psptd
\equiv \stsp$, $\jinit \equiv \filtpost{k}$, $\jtarg \equiv \filtpost{k + 1}$,
and $\ukdens(\ppa, \ppr) \equiv g(\ppr, \jvect{Y}_{k + 1}) \hkdens(\ppa, \ppr)$.
Consequently, a particle sample $\{ (\parti{i}, \wgt{i}) \}_{i = 1}^N$
(here and in the following we have omitted, for brevity, the time index from the particles and the associated weights) approximating $\filtpost{k}$ can be transformed into a sample $\{ (\partitd{i}, \wgttd{i}) \}_{i = 1}^N$ approximating the filter $\filtpost{k + 1}$ at the subsequent time step by executing the following steps.
\begin{algorithmhere}
	\caption{}
	\label{alg:particle:filter}
	\begin{algorithmic}[1]
		\Require $\{ (\parti{i}, \wgt{i}) \}_{i = 1}^N$
		\For{$i = 1 \to N$}
			\State draw $\ind{i} \sim \{ \adj{\ell} \wgt{\ell} \}_{\ell = 1}^N$
			\State draw $\partitd{i} \sim \propdens(\parti{\ind{i}}, \cdot)$
			\State set $\wgttd{i} \gets \frac{\olik(\partitd[]{i}, \jvect{y}_{k + 1}) \hkdens(\parti{\ind[]{i}}, \partitd{i}) }{\adj{\ind[]{i}} \propdens(\parti{\ind{i}},\partitd{i})}$
		\EndFor \\
		\Return $\{ (\partitd{i}, \wgttd{i}) \}_{i = 1}^N$
	\end{algorithmic}
\end{algorithmhere}
Here $\{ \adj{i} \}_{i = 1}^N$ is a set of nonnegative adjustment multipliers and the proposal $\propdens$ is a Markov transition density. In the following
examples we illustrate how the proposal $\propdens$ can be designed
adaptively using Algorithm~\ref{alg:main:2} described in the previous section. In Sections~\ref{sec:lgm} and \ref{sec:bessel} we adapt mixtures of Gaussian strata (following Example~\ref{ex:gaussian:strata}) only, while Section~\ref{sec:tobit} provides also a comparison with Student's $t$-distributed strata. 

\subsection{Multivariate linear Gaussian model}
\label{sec:lgm}
We start by considering a simple multivariate linear Gaussian model. In this toy example, the optimal kernel $\uknormdens$ and the optimal adjustment weight function $\adjfuncopt$ (defined in \eqref{eq:def:optimal:kernel} and \eqref{eq:def:optimal:weight}, respectively) are available in closed form; here we use the term ``optimal''---which is standard in the literature---as a proposal distribution $\pi$ based these quantities minimise the KLD \eqref{eq:KLD:define} (which of course then vanishes as $\pi = \auxtarg$ in that case). This makes it possible to compare our algorithm to an exact reference. The optimal kernel does not belong to our family of mixture of experts. In the model under consideration:
\begin{itemize}[leftmargin=1em]
	\item Each $\jvect{Y}_k$, taking values in $\obsp = \R^2$, is a noisy observation of the corresponding hidden state $\jvect{X}_k$ with local likelihood $g(\ppr, \jvect{y}) = \normdens[2]{\jvect{y}}{\ppr}{\jvect{\Sigma}_{\jvect{Y}}}$, where $\jvect{\Sigma}_{\jvect{Y}} = 0.1 \times \jvect{I}_2$.
	\item The prior kernel density is a mixture of multivariate Gaussian
	distributions $\hkdens(\ppa, \ppr) = (\normdens[2]{\ppr}{\jvect{\Amat}_1 \extd{\ppa}}{\jvect{\Sigma}} + \normdens[2]{\ppr}{\jvect{\Amat}_2 \extd{\ppa}}{\jvect{\Sigma}})/2$  with regression matrices
	\begin{align*}
		\jvect{\Amat}_1 &= \left(\begin{array}{cccc}1 & 0 & 1 \\ 0 & 1 & 1 \end{array}\right) \eqsp, &
		\jvect{\Amat}_2 &= \left(\begin{array}{cccc}1 & 0 & 1\\ 0 & 1 & -1 \end{array}\right) \eqsp.
	\end{align*}
	\item The filter $\filtpost{k}$ at time $k$ is assumed to be a bimodal distribution $\stsp = \R^2$ with density $(\mathcal{N}_2(\ppa; (0, 1)^\trans, \jvect{\Sigma}) + \normdens[2]{\ppa}{(0, -1)^\trans}{\jvect{\Sigma}})/2$, where $\jvect{\Sigma} = 0.1 \times \jvect{I}_2$. The two modes are hence well separated.
\end{itemize}
In this first example, we consider \emph{one single updating step of the particle filter}, with initial particles $\{\parti{i}\}_{i=1}^N$ sampled exactly from the filter $\filtpost{k}$. For this single step we study the performance of Algorithm~\ref{alg:main:2}. The observation $\jvect{Y}_{k + 1}$ is chosen to be $\jvect{Y}_{k + 1} = (1, 0)^\trans = \jvect{\Amat}_2 (0, 1)^\trans = \jvect{\Amat}_1 (0, -1)^\trans$. The likelihood of $\jvect{X}_k$ given $\jvect{Y}_{k + 1}$ is thus symmetric around the $x$-axis, giving equal weight to each of the two components of $\filtpost{k}$. Even though the optimal kernel $\uknormdens$ does not belong to our family of experts, it is still a mixture whose weights are highly dependent of location of the ancestor, and we can expect our algorithm to adapt accordingly.

The histogram in Figure~\ref{fig:mixtlgssm:histogram} of the importance weights of the particle swarm produced using the prior kernel shows that $50\%$ of these weights are nearly equal to zero, and the remaining ones are spread over a large range of values. This is confirmed by looking at the corresponding curve of proportions in Figure~\ref{fig:mixtlgssm:allproportions}, showing that $80\%$ of the total mass is carried by only $25\%$ of the particles and $99\%$ of the total mass by $40\%$ of the particles.

\begin{figurehere}
	\centering
	\begin{subfigure}[t]{1\columnwidth}
		\includegraphics[width=1\textwidth]{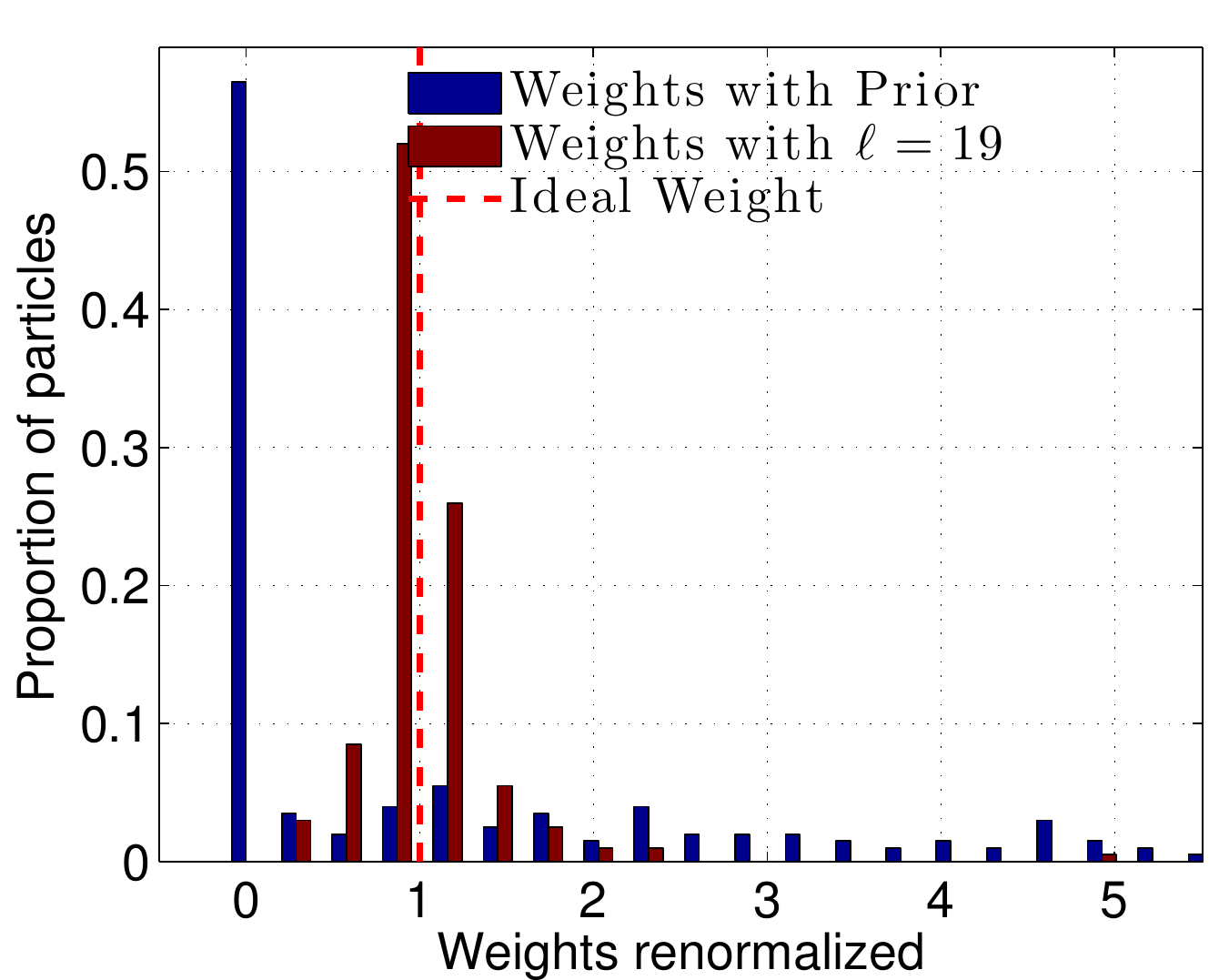}
	\caption{Histogram of the importance weights obtained using the prior kernel and the final adapted kernel in the linear Gaussian model. Weights are re-normalised and multiplied by the size of the sample. In the fully adapted case, all weights would be equal to unity.}
	\label{fig:mixtlgssm:histogram}
	\end{subfigure}\hfill%
	\begin{subfigure}[t]{1\columnwidth}
		\includegraphics[width=1\textwidth]{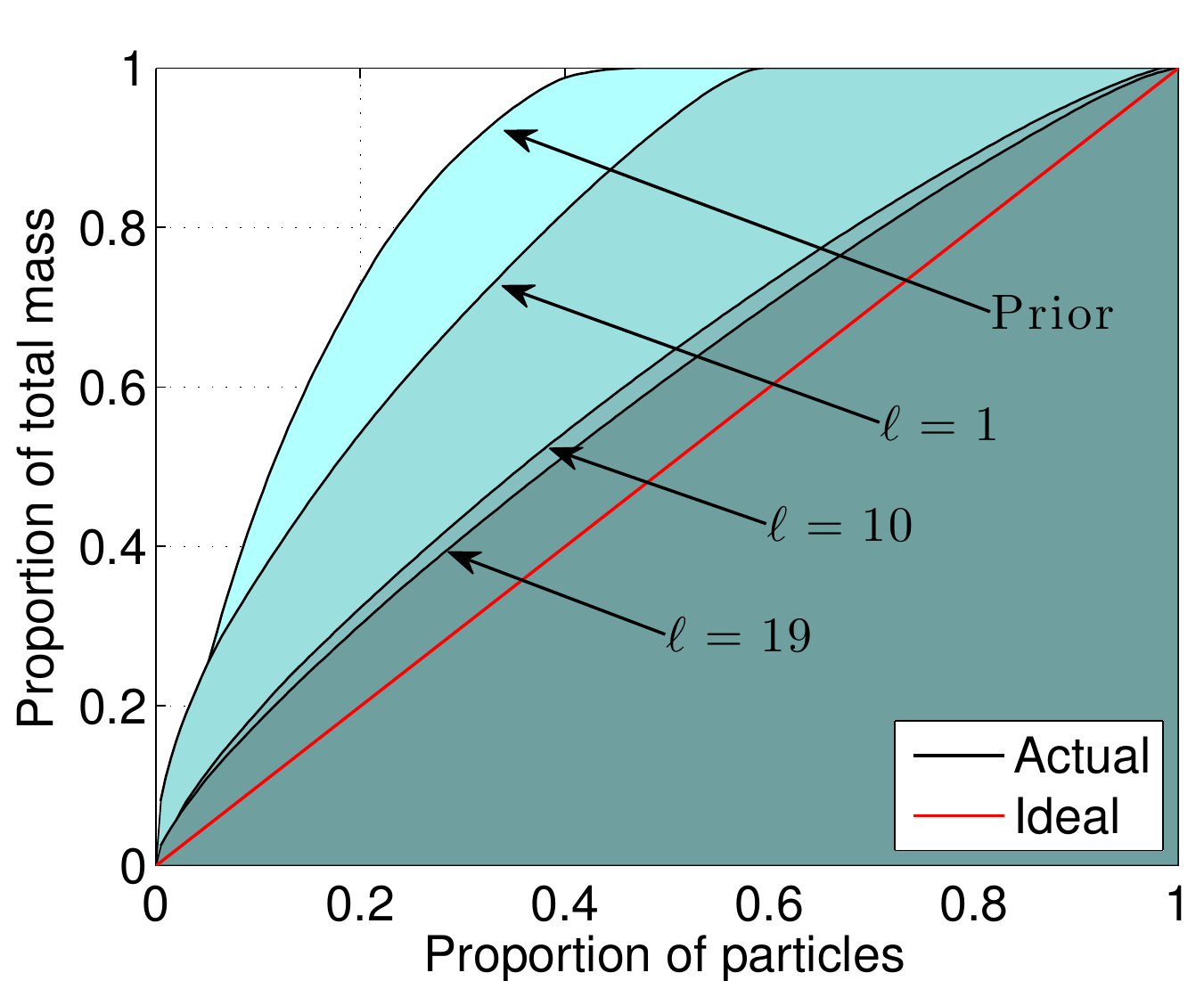}
	\caption{\emph{Curve of proportions}: proportion of particles, sorted by decreasing order of importance weight, against proportion of the total mass, for the prior kernel and for different numbers of iterations of the adaptation algorithm.}
	\label{fig:mixtlgssm:allproportions}
	\end{subfigure}%
	\caption{Evolution of the distribution of the importance weights before and after adaptation.}
	\label{fig:mixtlgssm:evolution}
\end{figurehere}

\begin{figure*}[htbp]
	\centering
	\begin{subfigure}[t]{1\columnwidth}
	\includegraphics[width=1\textwidth]{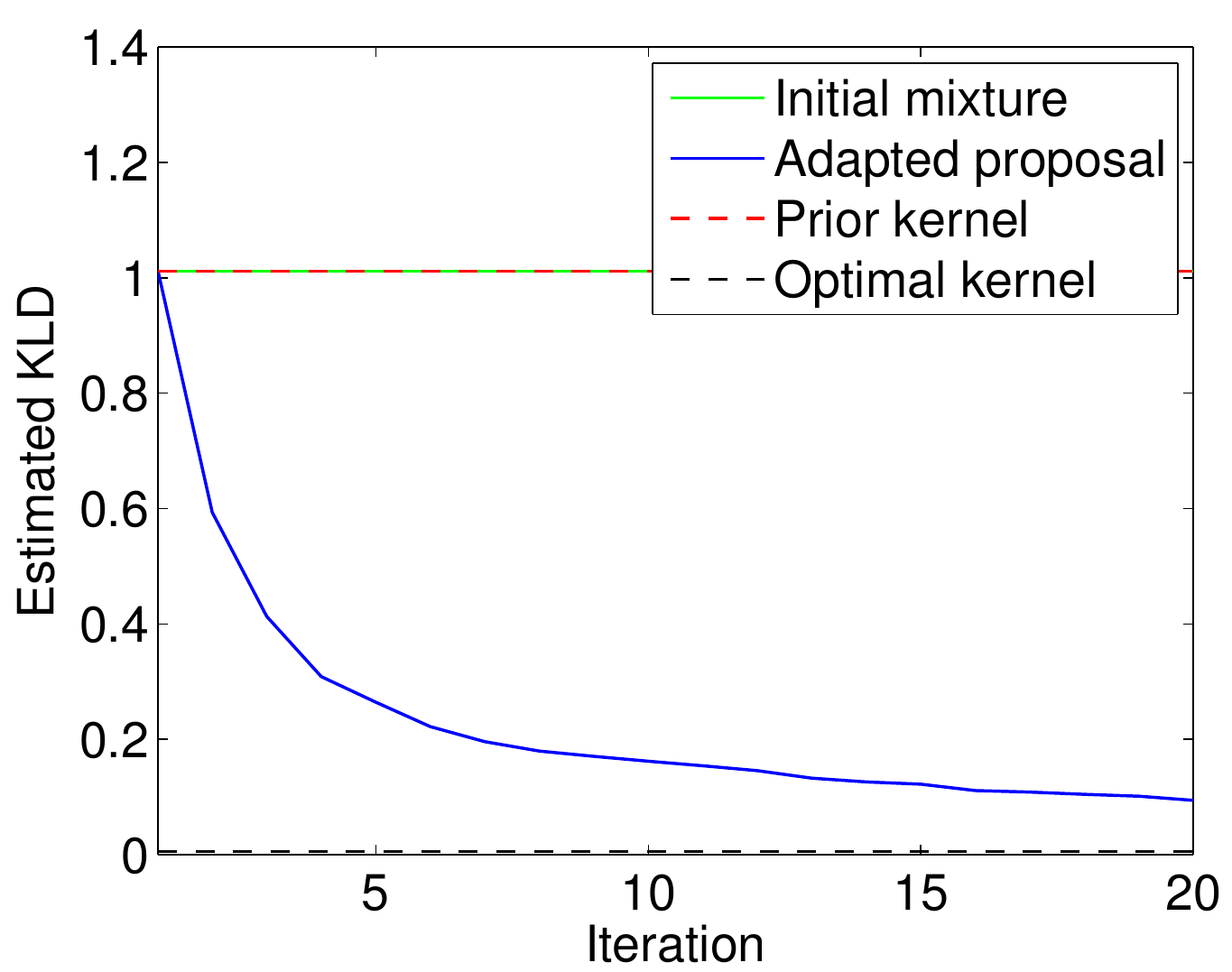}
\caption{Evolution of the KLD for the step size $0.1/\sqrt{L}$.}
	\label{fig:mixtlgssm:kldstep01}\end{subfigure}%
\hfill
	\begin{subfigure}[t]{1\columnwidth}
	\includegraphics[width=1\textwidth]{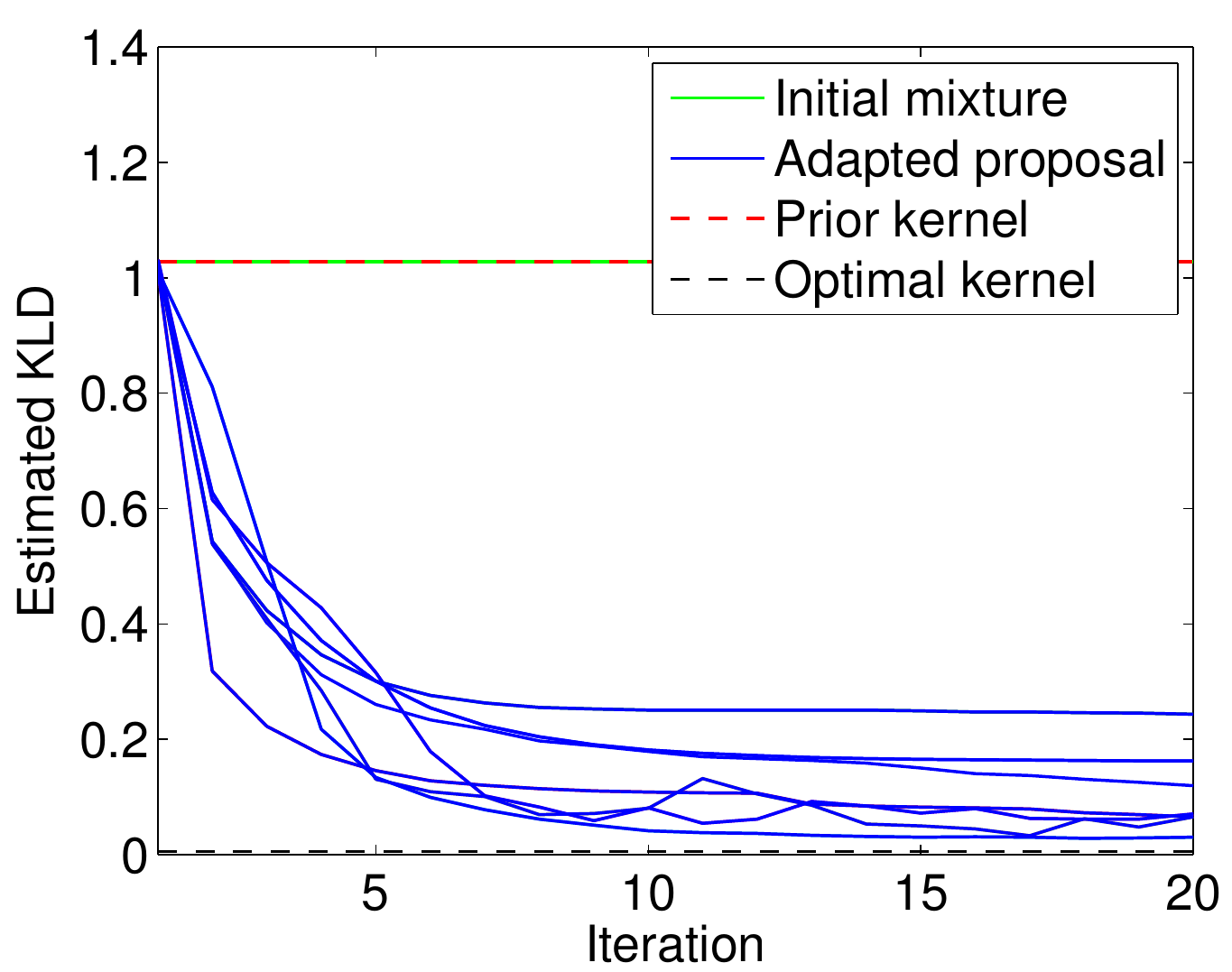}
	\caption{Comparison of evolution of KLDs for step sizes $\{0.001, 0.01, 0.1, 1, 10, 15\}/\sqrt{L}$.}
	\label{fig:mixtlgssm:allklds}\end{subfigure}%
	\caption{Evolution of the KLD over $L = 20$ iterations of adaptation for the linear Gaussian model.}
	\label{fig:mixtlgssm:cmpstepsizes}
\end{figure*}

One single iteration of the adaptation scheme described in
Section~\ref{sec:algorithm:logistic:weights} is sufficient to improve those proportions to $80\%$ of the mass for $40\%$ of the particles and $99\%$ of the mass for $55\%$ of the particles; see the corresponding curve of proportions in Figure~\ref{fig:mixtlgssm:allproportions}. After 10 such iterations, the corresponding curve of proportions (again in Figure~\ref{fig:mixtlgssm:allproportions}) shows that close to maximum improvement has been reached: the first few steps of the optimisation are enough to bring significant improvement. The histogram in Figure~\ref{fig:mixtlgssm:histogram} of the weights obtained at the final iteration shows how the weights are concentrated around unity, the value that corresponds to sampling with the optimal kernel.
We display in Figure~\ref{fig:mixtlgssm:kldstep01} the KLD \eqref{eq:KLD:decouples} between the
fit and the target (estimated by means of a Monte Carlo approximation with a
large sample size), and the same KLD for the proposal based on the prior kernel as well as the proposal based on the optimal kernel with uniform adjustment weights---i.e. all ancestor particles have practically the same optimal adjustment weight: choosing $a \equiv \1_{\stsp}$ makes the second term in \eqref{eq:KLD:decouples} negligible. As mentioned earlier, the KLD decreases very fast, and most of the
improvement is obtained after only a few iterations.
Figure~\ref{fig:mixtlgssm:allklds} compares
the evolution of the KLD for several step sizes covering four orders of magnitude. As the step size is increased, the algorithm converges faster, although less smoothly: this has no impact from a practical point of view as we are only looking for a good proposal kernel in an importance sampling setting, not for the exact optimal one.

\subsection{Bessel process observed in noise}
\label{sec:bessel}
\renewcommand{\ds}{2}

\begin{figure*}[htbp]
	\begin{center}
		\begin{subfigure}[t]{.02\textwidth}
		\end{subfigure}%
		\begin{subfigure}[c]{.31\textwidth}
			\centering \small Ancestors
		\end{subfigure}%
		\hfill%
		\begin{subfigure}[c]{.31\textwidth}
			\centering \small Particles
		\end{subfigure}
		\hfill%
		\begin{subfigure}[c]{.31\textwidth}
			\centering \small Importance weights
		\end{subfigure}\\
		\begin{subfigure}[b]{.02\textwidth}
			\centering \rotatebox[origin=tl]{90}{\small Prior kernel}
		\end{subfigure}%
		\begin{subfigure}[t]{.31\textwidth}
			\includegraphics[width=1\textwidth]{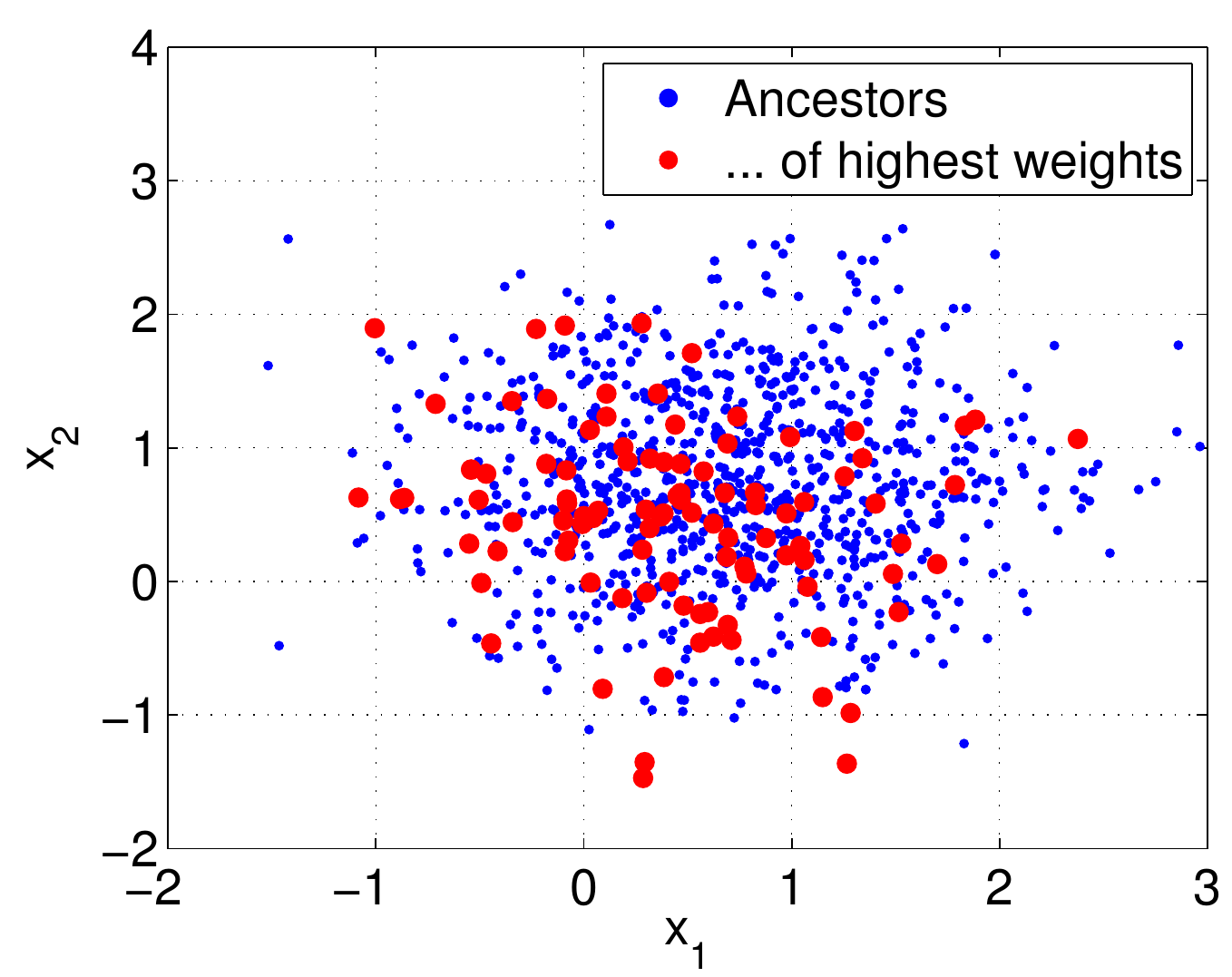}
		\end{subfigure}%
		\hfill%
		\begin{subfigure}[t]{.31\textwidth}
			\includegraphics[width=1\textwidth]{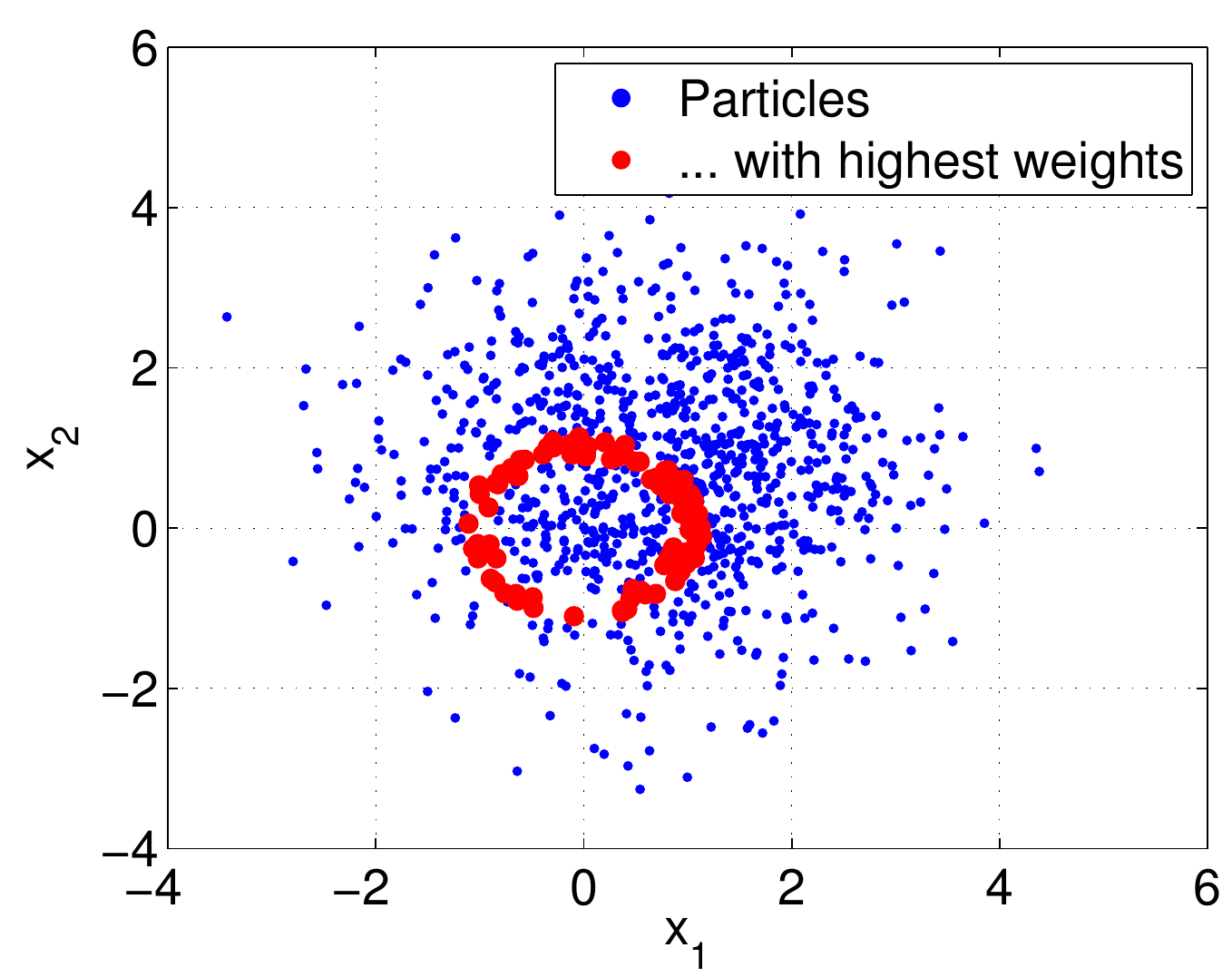}
		\end{subfigure}%
		\hfill%
		\begin{subfigure}[t]{.31\textwidth}
			\includegraphics[width=1\textwidth]{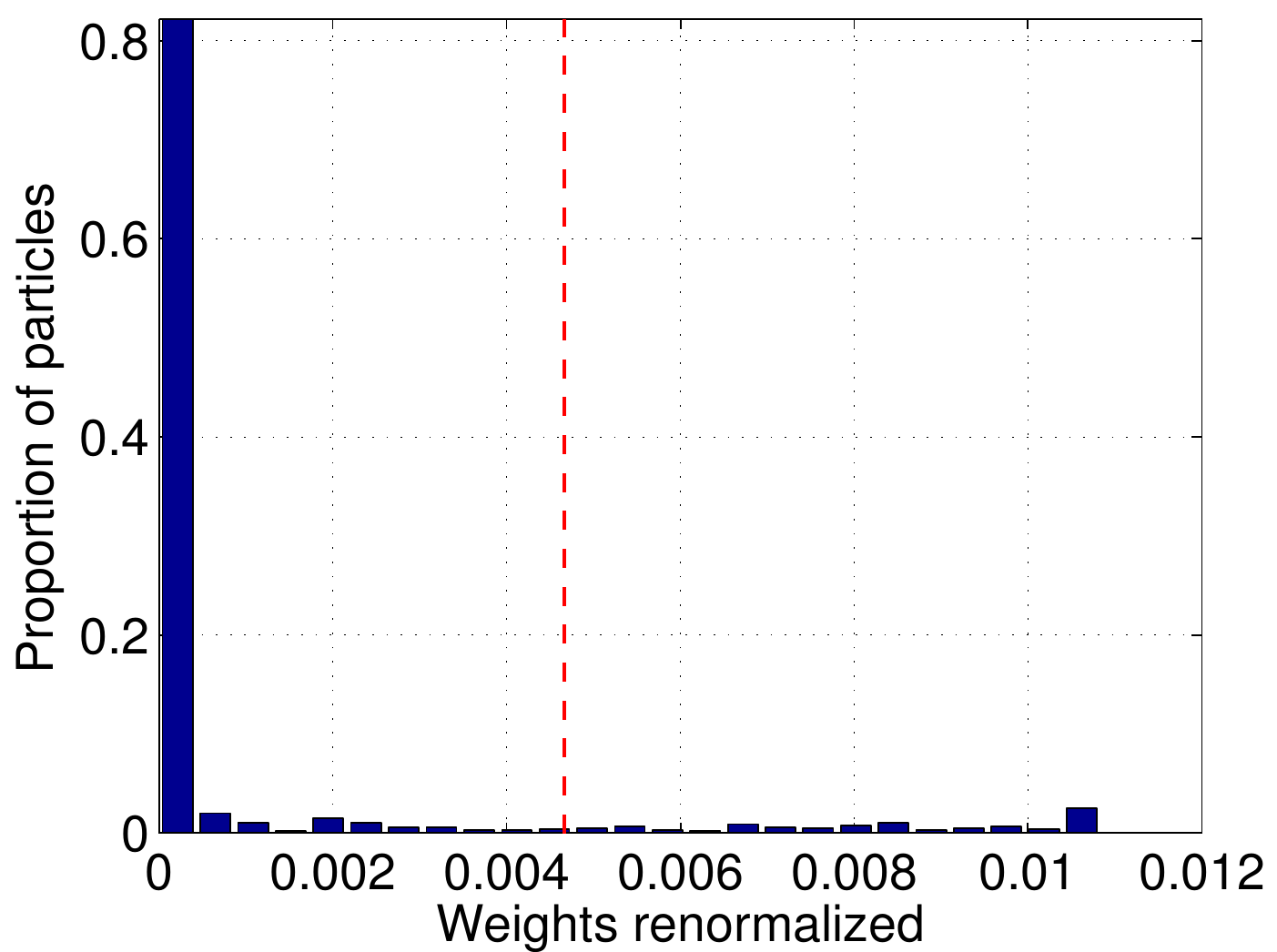}
		\end{subfigure}\\
		\begin{subfigure}[b]{.02\textwidth}
			\centering \rotatebox[origin=tl]{90}{\small Adapted kernel}
		\end{subfigure}%
		\begin{subfigure}[t]{.31\textwidth}
			\includegraphics[width=1\textwidth]{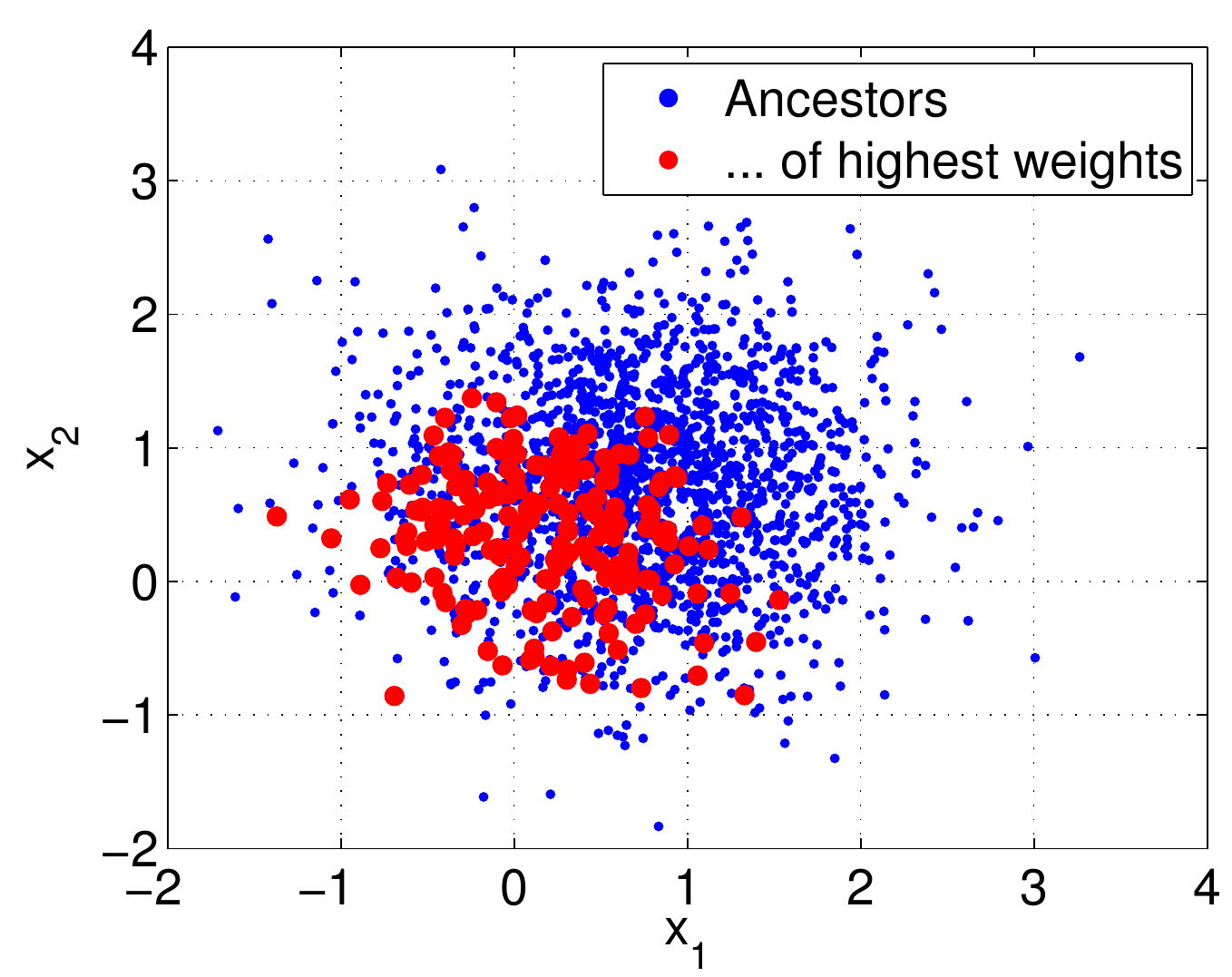}
		\end{subfigure}%
		\hfill%
		\begin{subfigure}[t]{.31\textwidth}
			\includegraphics[width=1\textwidth]{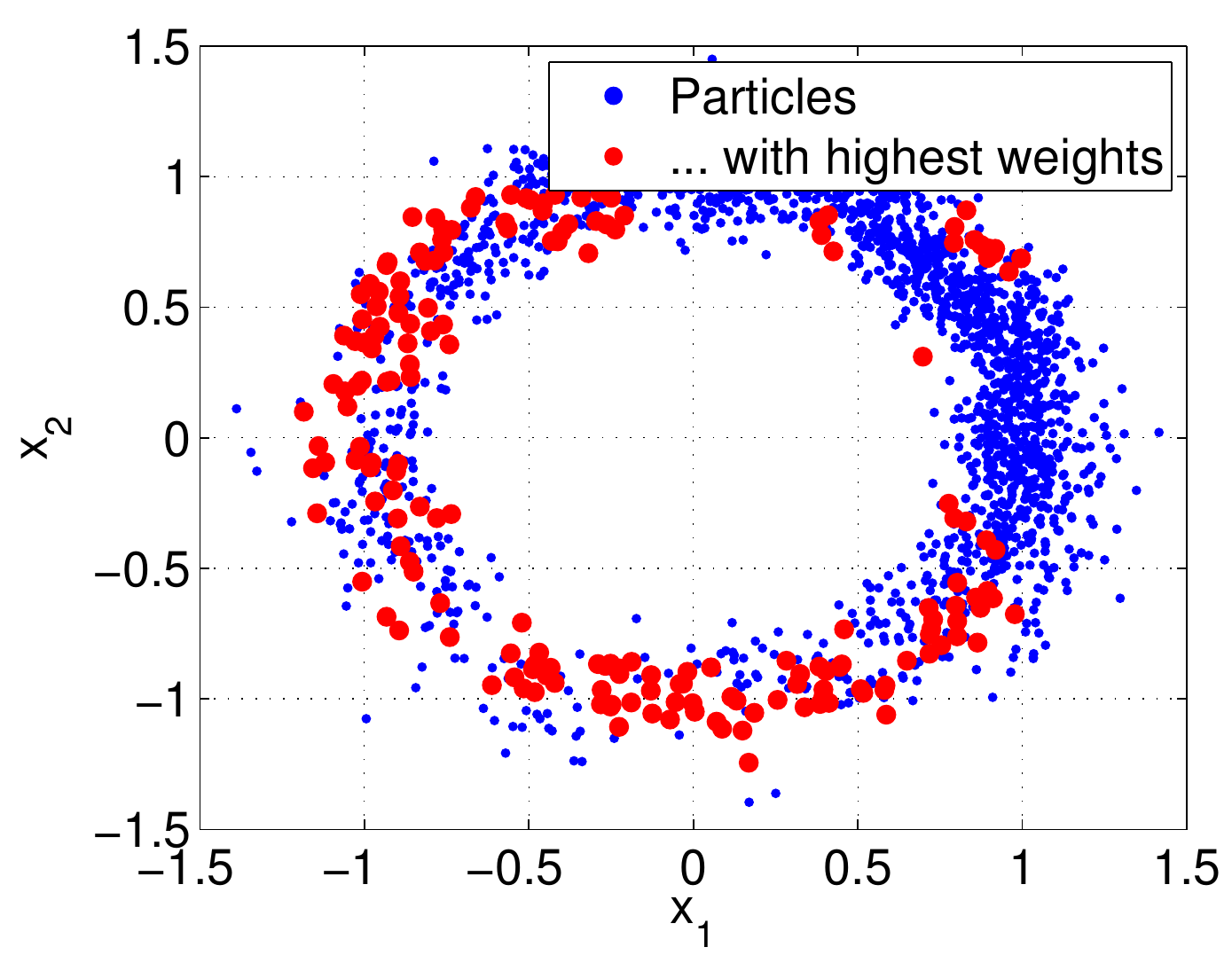}
		\end{subfigure}%
		\hfill%
		\begin{subfigure}[t]{.31\textwidth}
			\includegraphics[width=1\textwidth]{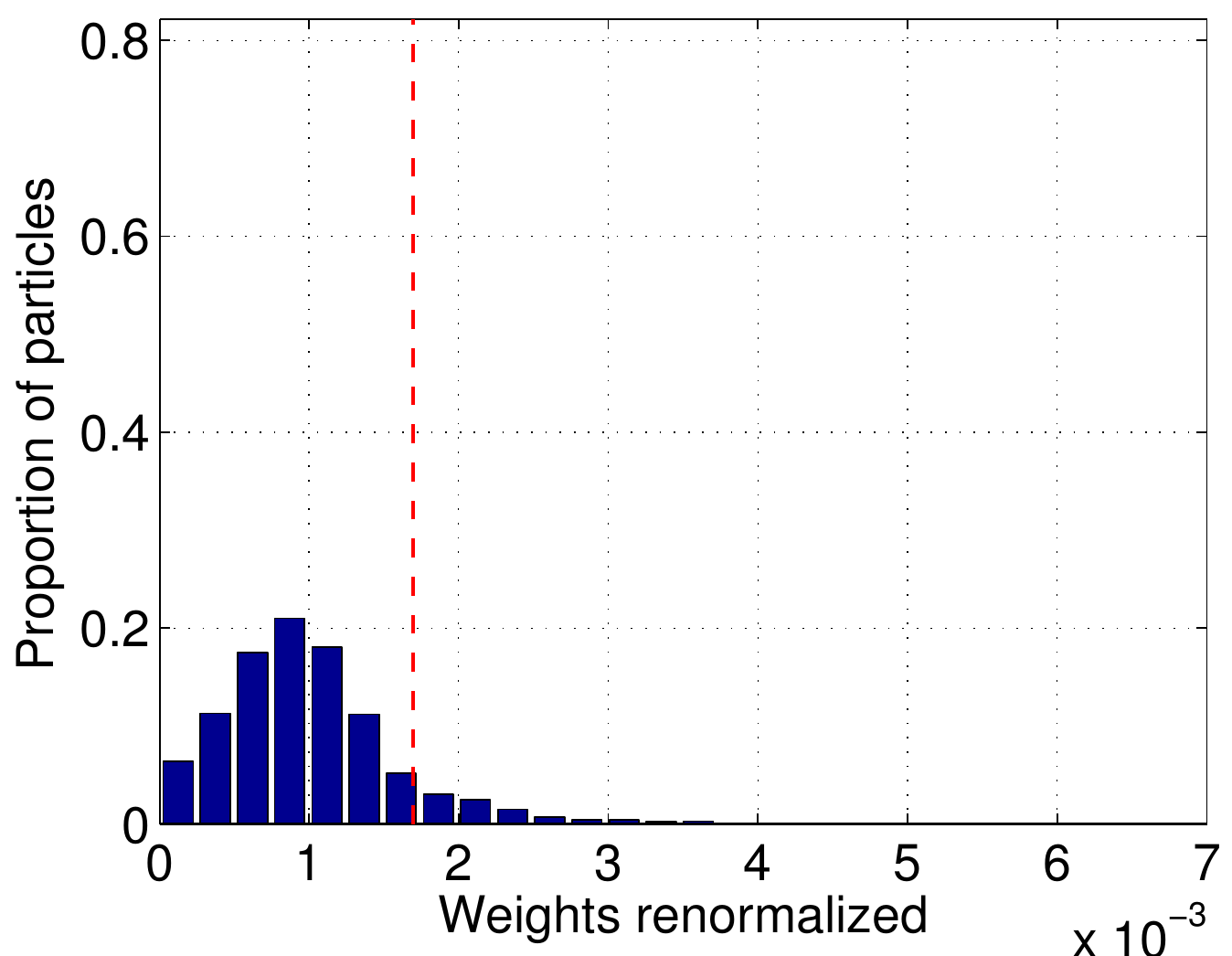}
		\end{subfigure}
	\end{center}
	\caption{Comparison of $1,\!000$ and $200$ particles from the prior and final adapted kernels, respectively, with the 10\% largest weights plotted in red, along with the corresponding ancestors and the distribution of the correspoding importance weights. For the prior kernel, the support of the proposal distribution is over-spread and as much as $80\%$ of particles have null weight. On the other hand, the adaptation algorithm concentrates the particles on the support of the target distribution, thus narrowing the span of the importance weights and balancing the contributions of all particles.}
	\label{fig:bessel:gaussian:priorbest}
\end{figure*}

\begin{figure*}[htbp]
	\begin{center}
		\begin{subfigure}[t]{.02\textwidth}
			\rotatebox[origin=tl]{90}{ }
		\end{subfigure}%
		\begin{subfigure}[c]{.31\textwidth}
			\centering \small Ancestor $(-0.7, -0.7)$
		\end{subfigure}%
		\hfill%
		\begin{subfigure}[c]{.31\textwidth}
			\centering \small Ancestor $(0, 0)$
		\end{subfigure}
		\hfill%
		\begin{subfigure}[c]{.31\textwidth}
			\centering \small Ancestor $(1.4, 1.4)$
		\end{subfigure} \\
		\begin{subfigure}[b]{.02\textwidth}
			\centering \rotatebox[origin=tl]{90}{\small Prior kernel}
		\end{subfigure}%
		\begin{subfigure}[t]{.31\textwidth}
			\includegraphics[width=1\textwidth]{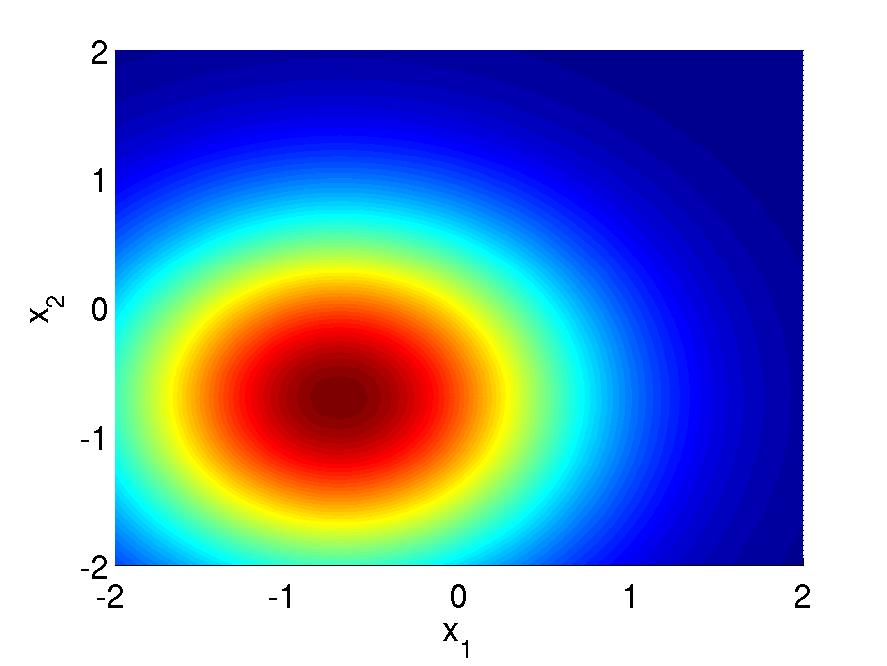}
		\end{subfigure}%
		\hfill%
		\begin{subfigure}[t]{.31\textwidth}
			\includegraphics[width=1\textwidth]{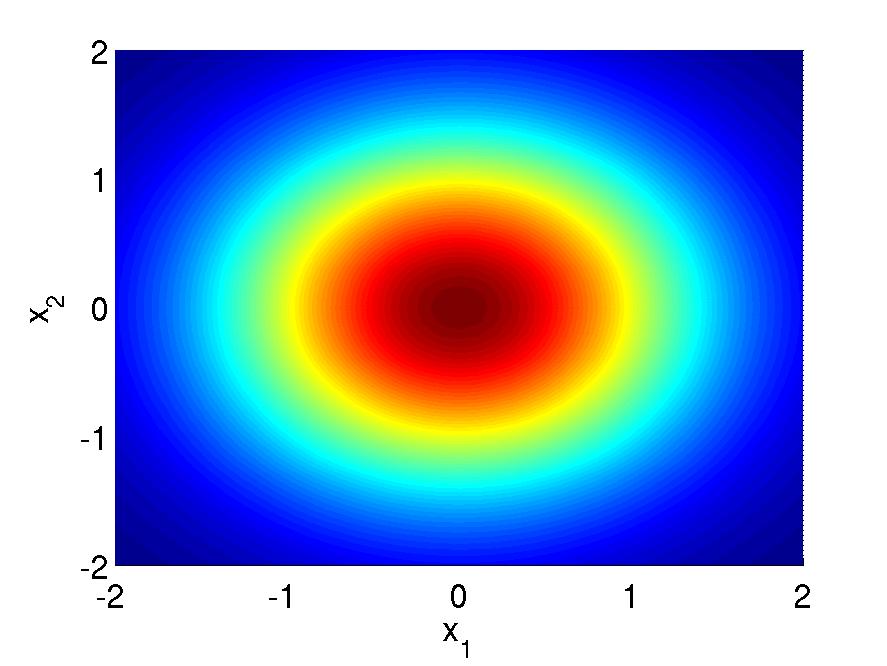}
		\end{subfigure}
		\hfill%
		\begin{subfigure}[t]{.31\textwidth}
			\includegraphics[width=1\textwidth]{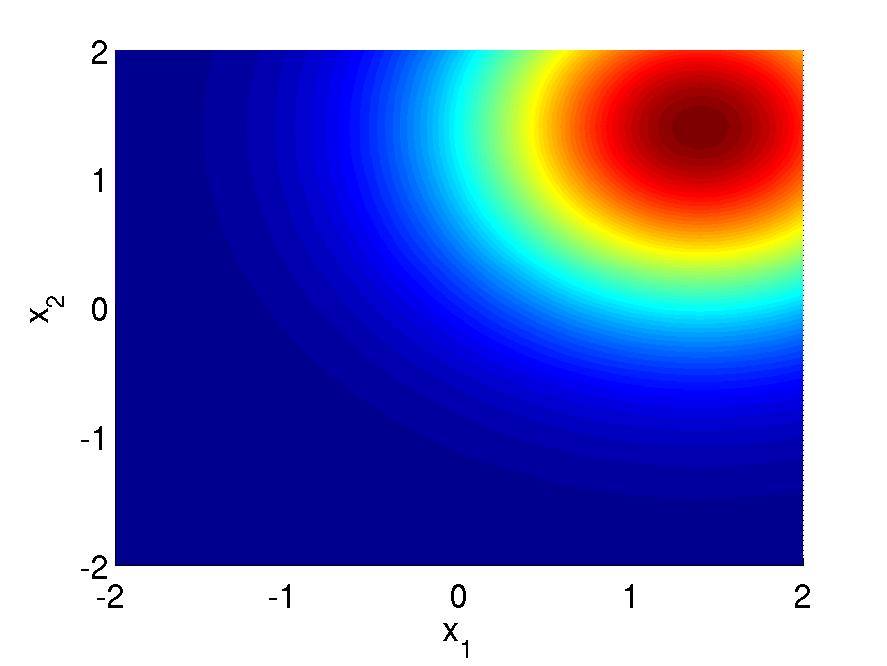}
		\end{subfigure}\\[-1em]
		\begin{subfigure}[b]{.02\textwidth}
			\centering \rotatebox[origin=tl]{90}{\small Local Likelihood}
		\end{subfigure}%
		\begin{subfigure}[t]{.31\textwidth}
			\includegraphics[width=1\textwidth]{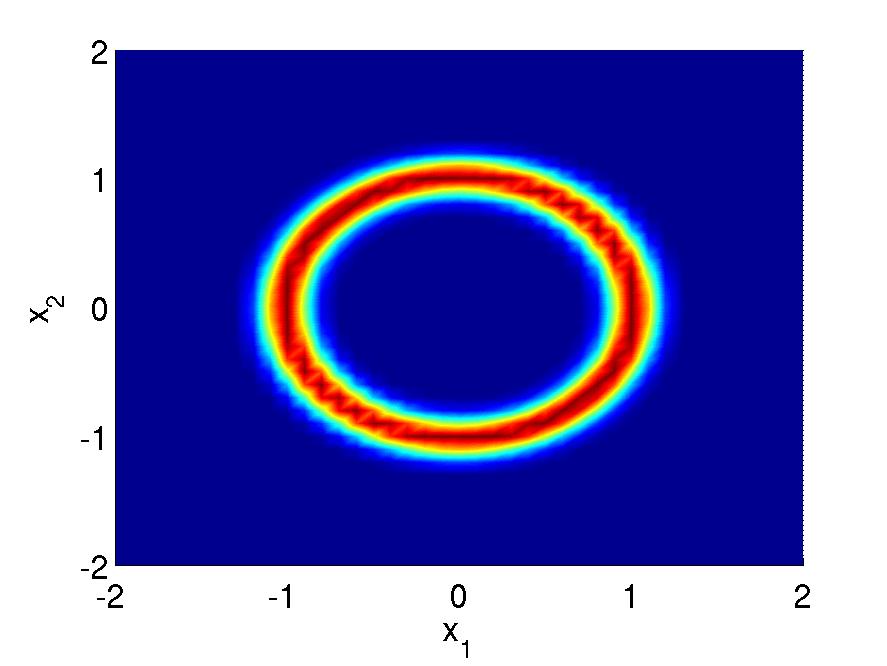}
		\end{subfigure}%
		\hfill%
		\begin{subfigure}[t]{.31\textwidth}
			\includegraphics[width=1\textwidth]{OneStep_size_18_Loclike}
		\end{subfigure}
		\hfill%
		\begin{subfigure}[t]{.31\textwidth}
			\includegraphics[width=1\textwidth]{OneStep_size_18_Loclike}
		\end{subfigure}\\[-1em]
		\begin{subfigure}[b]{.02\textwidth}
			\centering \rotatebox[origin=tl]{90}{\small Optimal kernel}
		\end{subfigure}%
		\begin{subfigure}[t]{.31\textwidth}
			\includegraphics[width=1\textwidth]{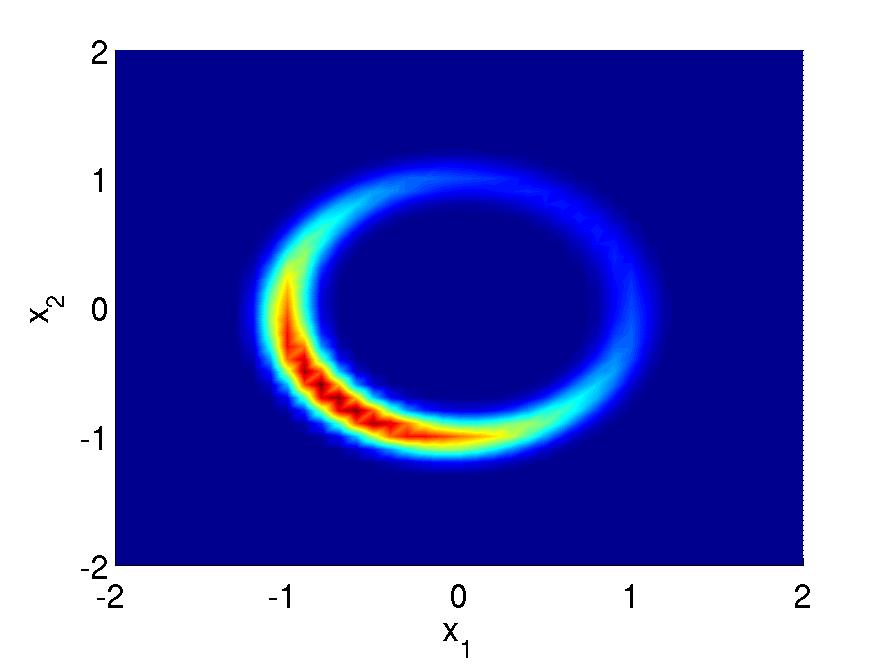}
		\end{subfigure}%
		\hfill%
		\begin{subfigure}[t]{.31\textwidth}
			\includegraphics[width=1\textwidth]{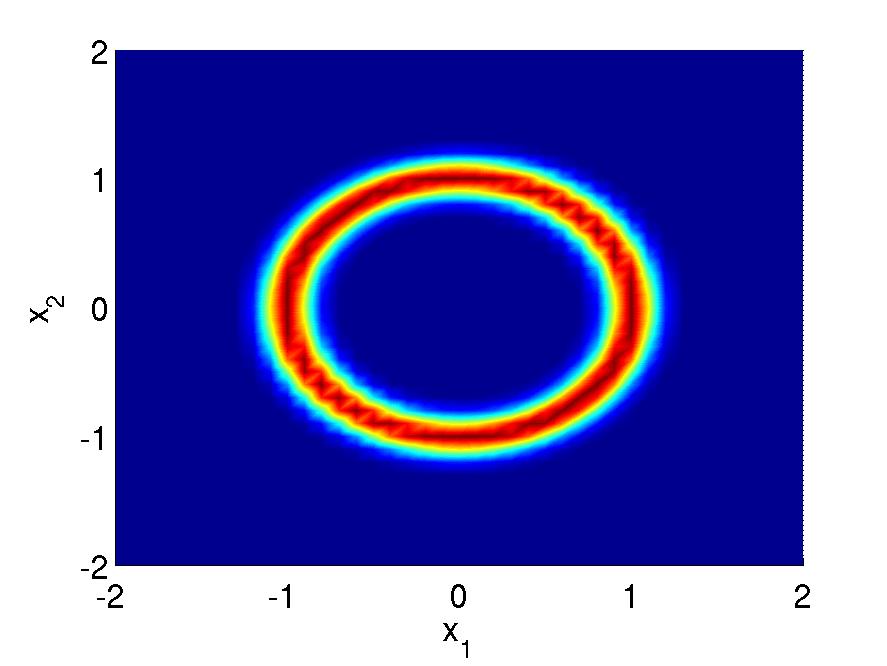}
		\end{subfigure}
		\hfill%
		\begin{subfigure}[t]{.31\textwidth}
			\includegraphics[width=1\textwidth]{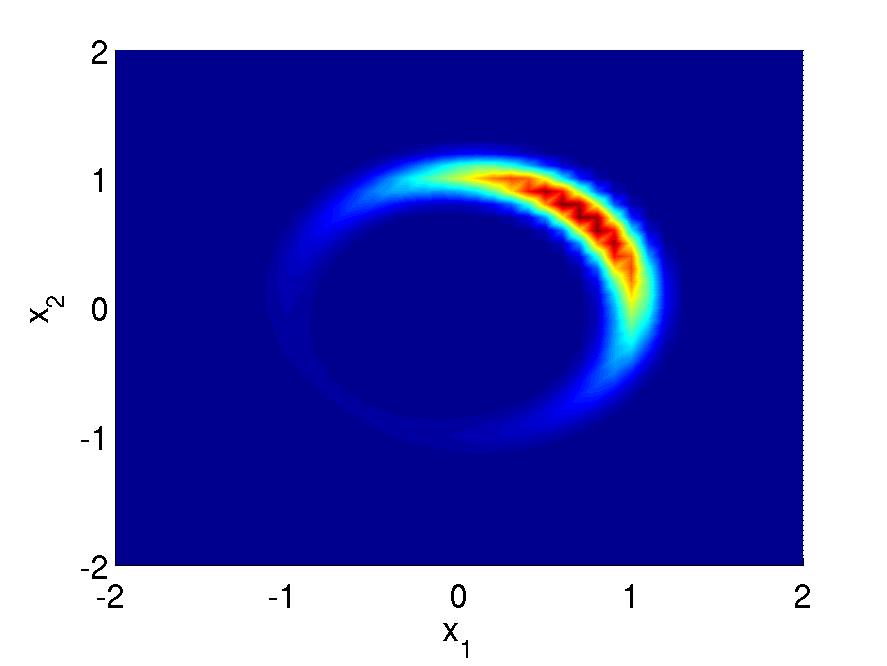}
		\end{subfigure}\\[-1em]
		\begin{subfigure}[b]{.02\textwidth}
			\centering \rotatebox[origin=tl]{90}{\small Initial fit}
		\end{subfigure}%
		\begin{subfigure}[t]{.31\textwidth}
			\includegraphics[width=1\textwidth]{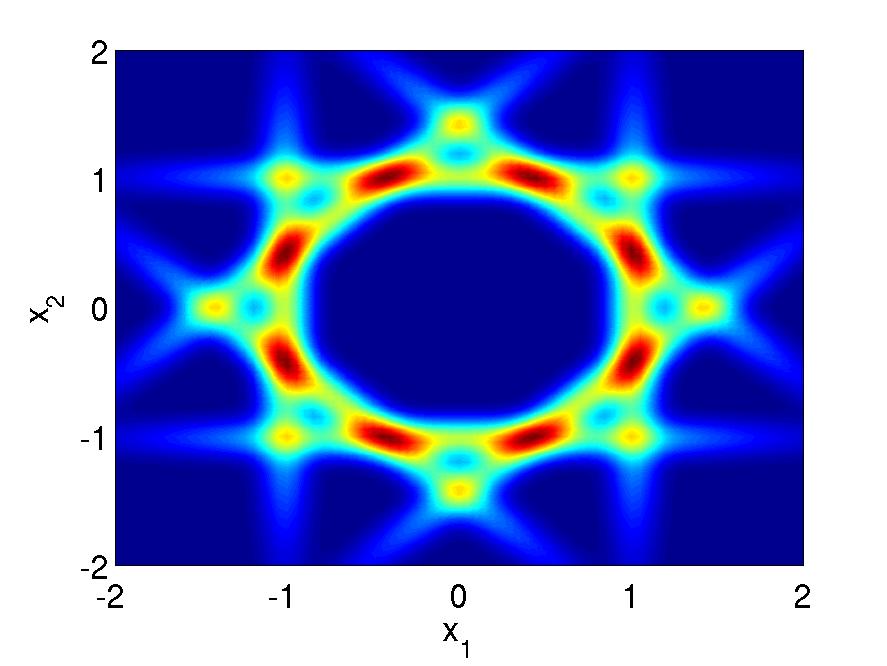}
		\end{subfigure}%
		\hfill%
		\begin{subfigure}[t]{.31\textwidth}
			\includegraphics[width=1\textwidth]{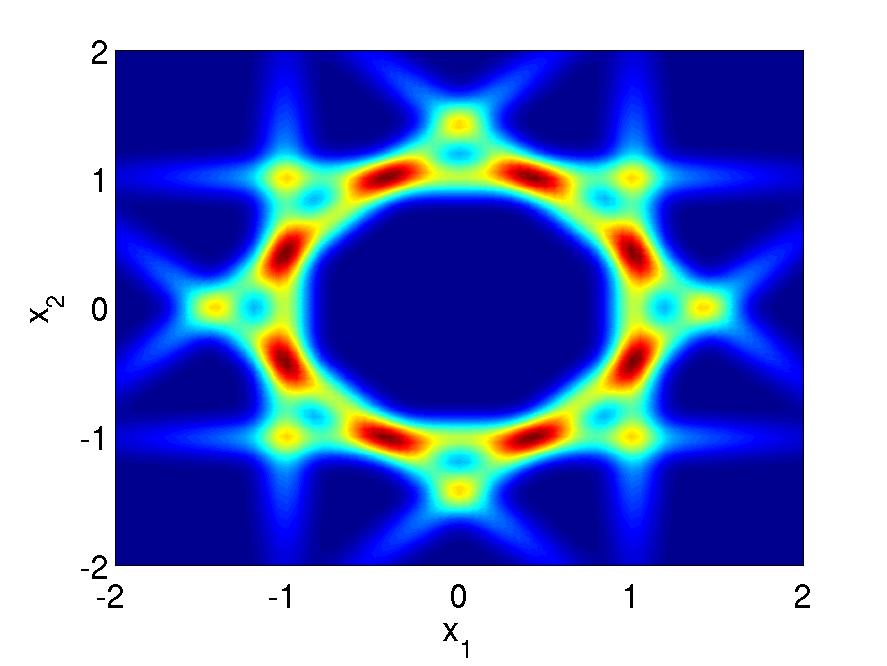}
		\end{subfigure}
		\hfill%
		\begin{subfigure}[t]{.31\textwidth}
			\includegraphics[width=1\textwidth]{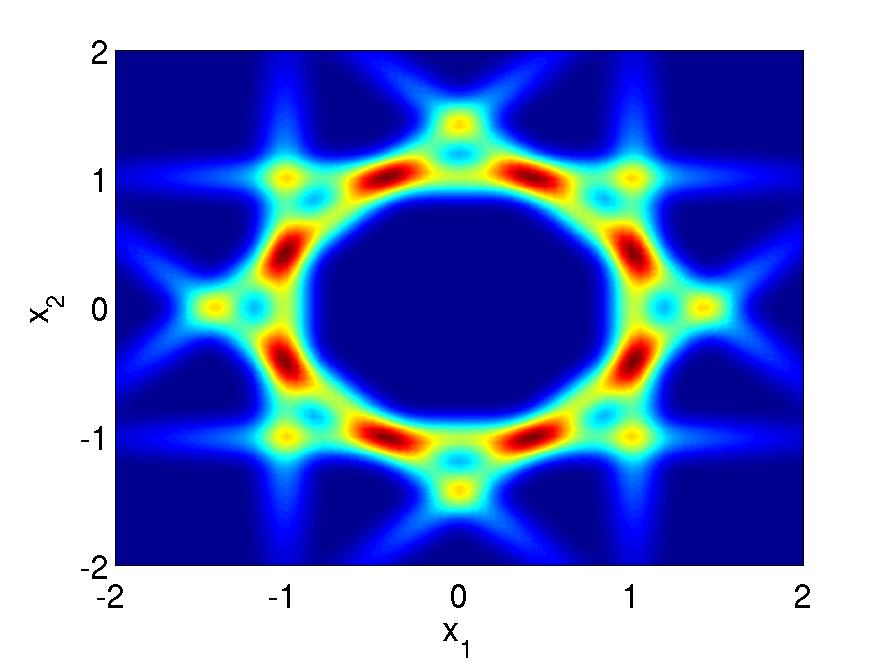}
		\end{subfigure}\\[-1em]
		\begin{subfigure}[b]{.02\textwidth}
			\centering \rotatebox[origin=tl]{90}{\small Iteration $1$}
		\end{subfigure}%
		\begin{subfigure}[t]{.31\textwidth}
			\includegraphics[width=1\textwidth]{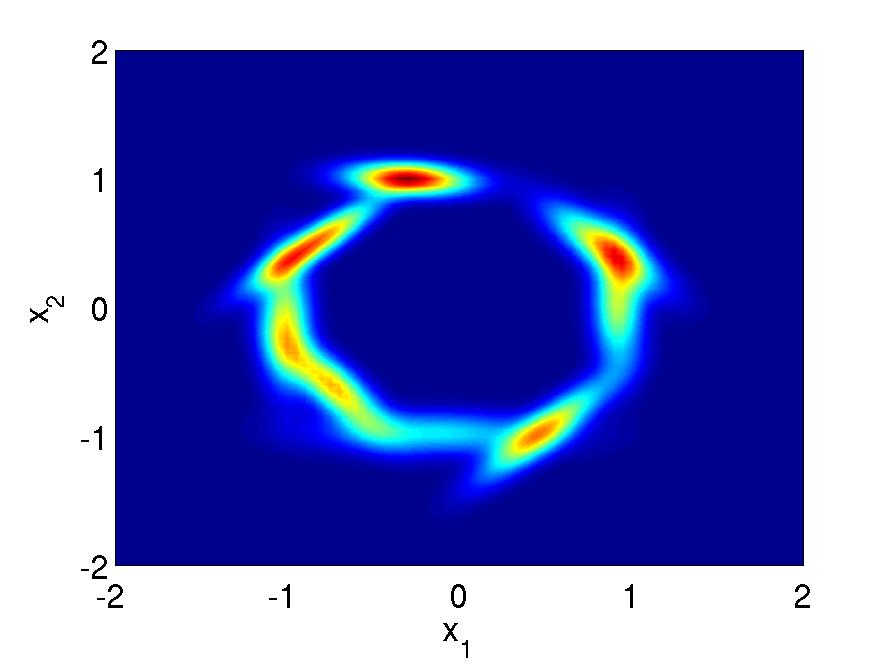}
		\end{subfigure}%
		\hfill%
		\begin{subfigure}[t]{.31\textwidth}
			\includegraphics[width=1\textwidth]{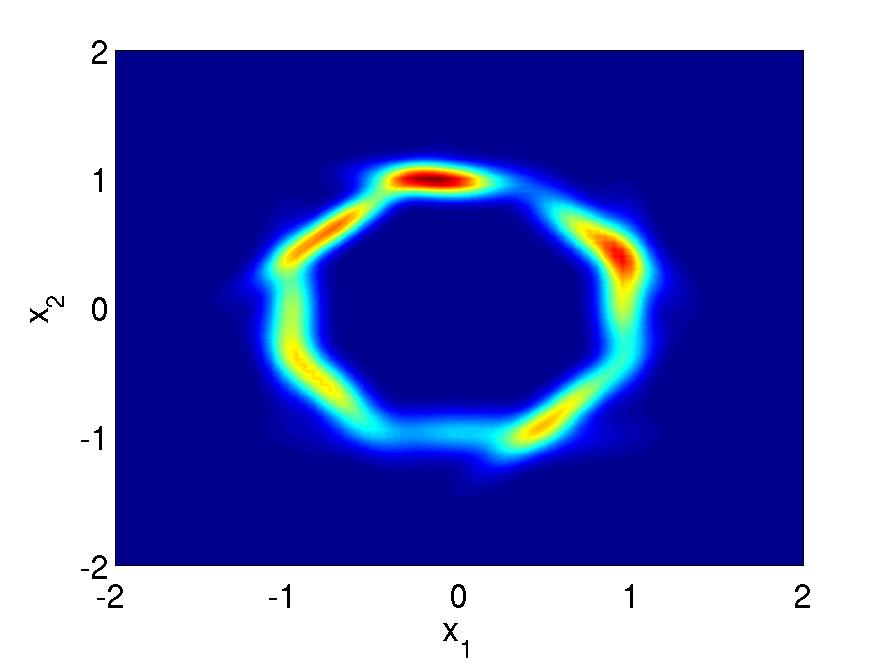}
		\end{subfigure}
		\hfill%
		\begin{subfigure}[t]{.31\textwidth}
			\includegraphics[width=1\textwidth]{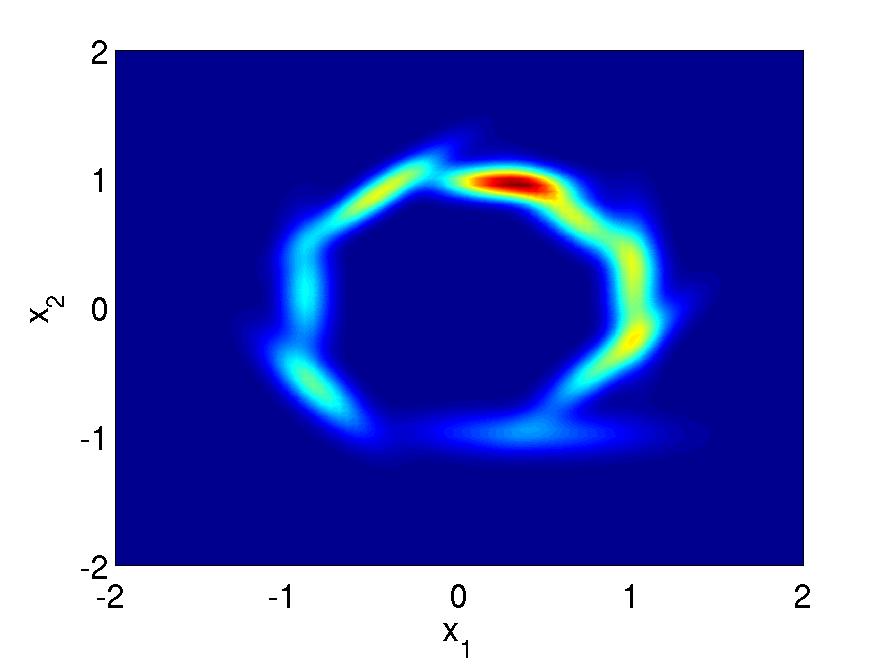}
		\end{subfigure}\\[-1em]
		\begin{subfigure}[b]{.02\textwidth}
			\centering \rotatebox[origin=tl]{90}{\small Iteration $30$}
		\end{subfigure}%
		\begin{subfigure}[t]{.31\textwidth}
			\includegraphics[width=1\textwidth]{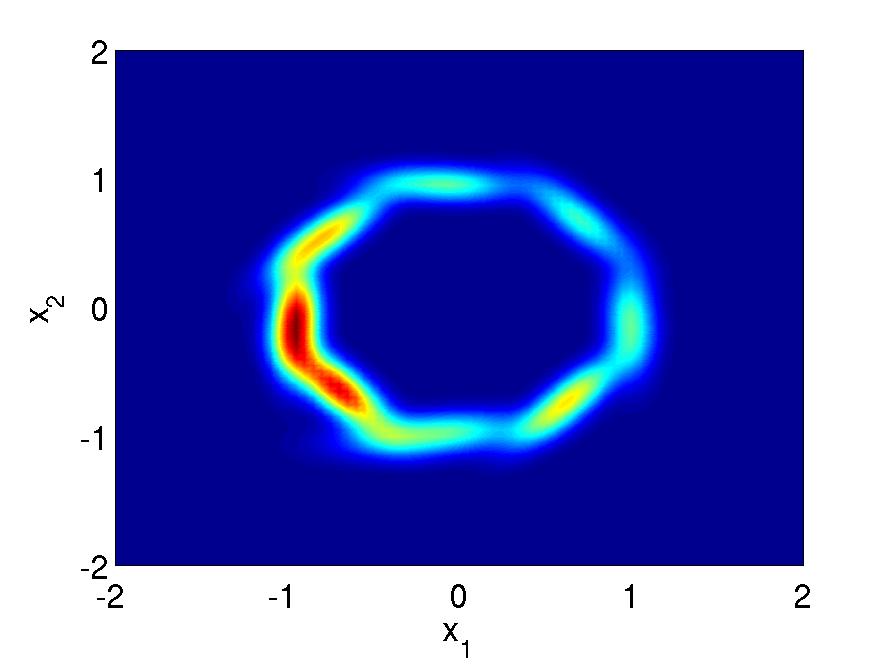}
		\end{subfigure}%
		\hfill%
		\begin{subfigure}[t]{.31\textwidth}
			\includegraphics[width=1\textwidth]{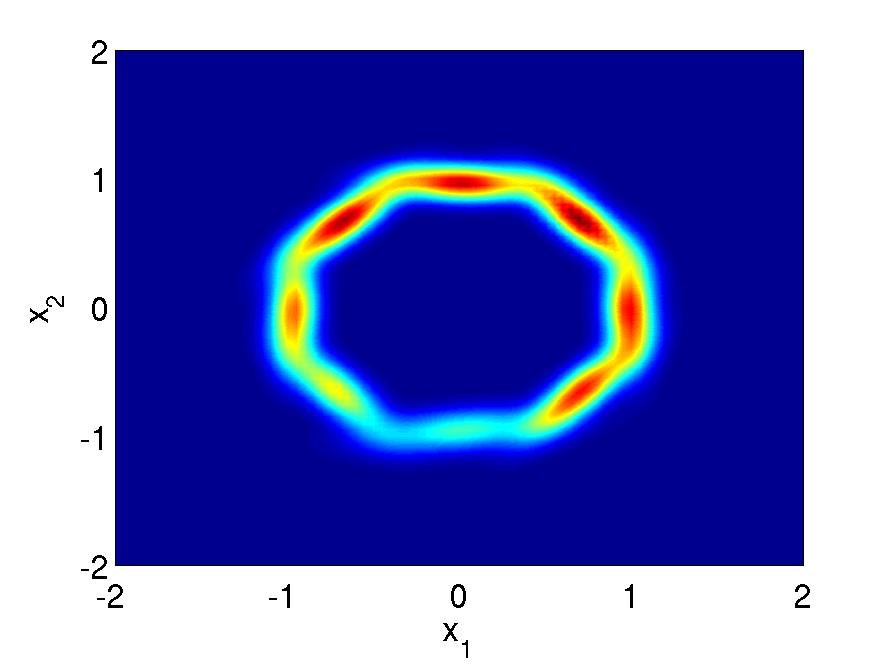}
		\end{subfigure}
		\hfill%
		\begin{subfigure}[t]{.31\textwidth}
			\includegraphics[width=1\textwidth]{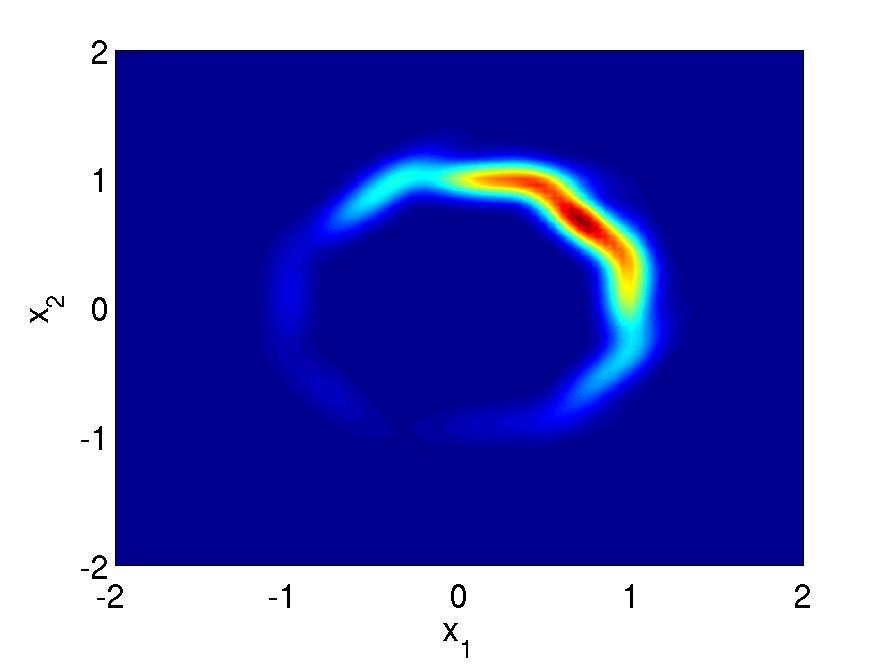}
		\end{subfigure}
\end{center}
\caption{Evolution of the optimal and adapted kernels for the Bessel model for 3 distinct ancestors in different regions and for the observation $Y_{k + 1} = 1.0$. The last three rows show the evolution of the adapted proposal kernel after the initial fit, one iteration, and $30$ iterations, respectively. The adaptation algorithm used a constant sample size $N_\ell = 200$ for all $\ell \in \{0, 2, \ldots, 30\}$. After $30$ iterations (the last row), the adapted kernel is visually very close to the optimal kernel.}
		\label{fig:bessel:gaussian:evolution}
\end{figure*}

As a more challenging example we consider optimal filtering of
the Brownian motion underlying a Bessel process observed in noise, also known as \emph{range-only filtering} of a Gaussian random walk. We let the state process evolve in the plane for visualisation purposes. More specifically, the SSM is given by
\[
	\begin{split}
		\jvect{X}_{k+1} &= \jvect{X}_k + \jvect{V}_k \eqsp, \\
		Y_k &= \vectornorm{\jvect{X}_k}_2 + W_k \eqsp,
	\end{split}
\]
where $\{ \jvect{V}_k\}_{k \geq 0}$ and $\{ W_k\}_{k \geq 0}$ are sequences of mutually independent $\mathcal{N}_2(\jvect{0}, \jvect{\Sigma}_{\jvect{X}})$-distributed and $\mathcal{N}_2(0, \sigma_Y^2)$-distributed, respectively, noise variables and $\vectornorm{\jvect{X}}_2$ is the Euclidean $\mathcal{L}_2$ norm on the state space $\stsp \eqdef \R^2$. Note that the observations $\{ Y_k \}_{k \geq 0}$ are real-valued in this case. With the notation in~\eqref{eq:state:eqn} and \eqref{eq:obs:eqn}, it holds that $\hkdens(\ppa, \ppr) \eqdef \normdens[\ds]{\ppr}{\ppa}{\jvect{\Sigma}_\jvect{X}}$
the density of the prior kernel and by $\olik(\ppr, y) \eqdef
\mathcal{N}(y, \|\ppr\|_2, \sigma_Y^2)$ the local likelihood.
We ensure a diffuse prior kernel and informative observations by setting $\jvect{\Sigma}_{\jvect{X}} = \jvect{I}_2$ and $\sigma_Y^2 = 0.01$.
As the hidden state is observed range-only, the state equation provides most of the information concerning the
bearing while the local likelihood is invariant under rotation of the state around the origin. This induces a variety of nonlinear shapes of the optimal kernel depending on the location of the ancestor---see the three top rows of Figure~\ref{fig:bessel:gaussian:evolution}.

To start with, we consider again \emph{a single step} of the particle filter, updating a weighted particle sample $\{(\parti{i},\wgt{i})\}_{i=1}^N$ approximating the filter $\filtpost{k}$ at some time index $k$ to a weighted sample $\{(\partitd{i},\wgttd{i})\}_{i=1}^N$ approximating the filter $\filtpost{k + 1}$ at the next time step. We assume that $\filtpost{k} = \normlaw[\ds]{(0.7, 0.7)^\trans}{0.5 \times \jvect{I}_2}$.
In our experiment, the original weighted sample $\{(\parti{i},\wgt{i})\}_{i=1}^N$ consists of a sample of $N = 20,\!000$ \iid~particles drawn exactly from the filter distribution $\filtpost{k}$ (hence, the particles have uniform weights). The observation at time $k + 1$ is $Y_{k + 1} = 1.0$. We initialise the algorithm with a sample $\{(\partitd[]{i}, \wgttd[]{i})\}_{i=1}^{N_0}$ of size $N_0 = 1,\!000$ using the prior kernel. The resulting cloud, along with the $100$ particles with largest weights and the corresponding ancestors, is plotted in
Figure~\ref{fig:bessel:gaussian:priorbest} (top row).
The support of the proposal distribution is over-spread:
most of the particles have negligible weights, and only a few
particles have comparatively large weights; indeed, $80\%$ of particles
have null weight. This is confirmed by the curve of proportions in  Figure~\ref{fig:bessel:gaussian:proportions} (left): only $20\%$ of the proposed particles carry the total mass. Adaptation of the proposal kernel is
thus highly relevant. Here again, adaptation of the adjustment weights is not required, as the ancestors of the particles with highest importance weights are not located in any specific
region of the state space.

Based on these $1,\!000$ particles, the first iteration of the adaptation is carried through using conditional probabilities from the initial fit displayed in Figure~\ref{fig:bessel:gaussian:evolution} (fourth row), whose components are chosen to be independent of the ancestor, \ie~only the constant term is non-zero. The resulting kernel is plotted in Figure~\ref{fig:bessel:gaussian:evolution} (fifth row). We use it to propose $200$ new particles, serving as an importance sampling approximation of $\auxtarg$ at the second interation. After $30$ such iterations, the adapted kernel visible in Figure~\ref{fig:bessel:gaussian:evolution} (last row) is visually very close to the optimal kernel. Note the impact of the location of the ancestor on the un-normalised transition kernel, whose mass shifts on a circle, and how the adapted kernel shifts accordingly. Figure~\ref{fig:bessel:gaussian:priorbest} (bottom row) shows that the adaptation algorithm concentrates the particles on the support of the target distribution, thus narrowing the span of the importance weights and balancing the contributions of all particles.

\begin{figurehere}
	\begin{center}
		\begin{subfigure}[t]{\columnwidth}
			\includegraphics[width=1\textwidth]{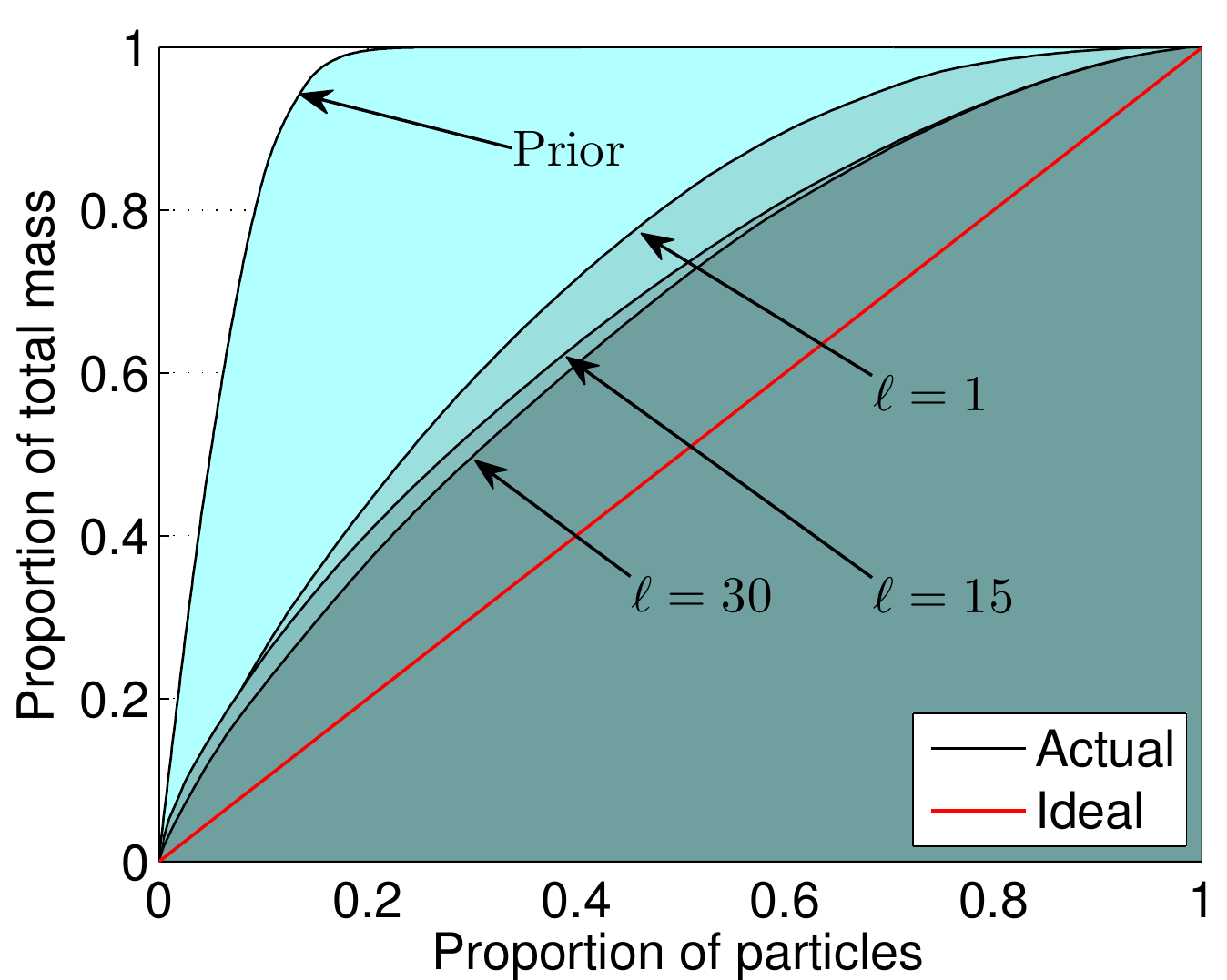}
		\end{subfigure}\hfill
		\begin{subfigure}[t]{\columnwidth}
			\includegraphics[width=1\textwidth]{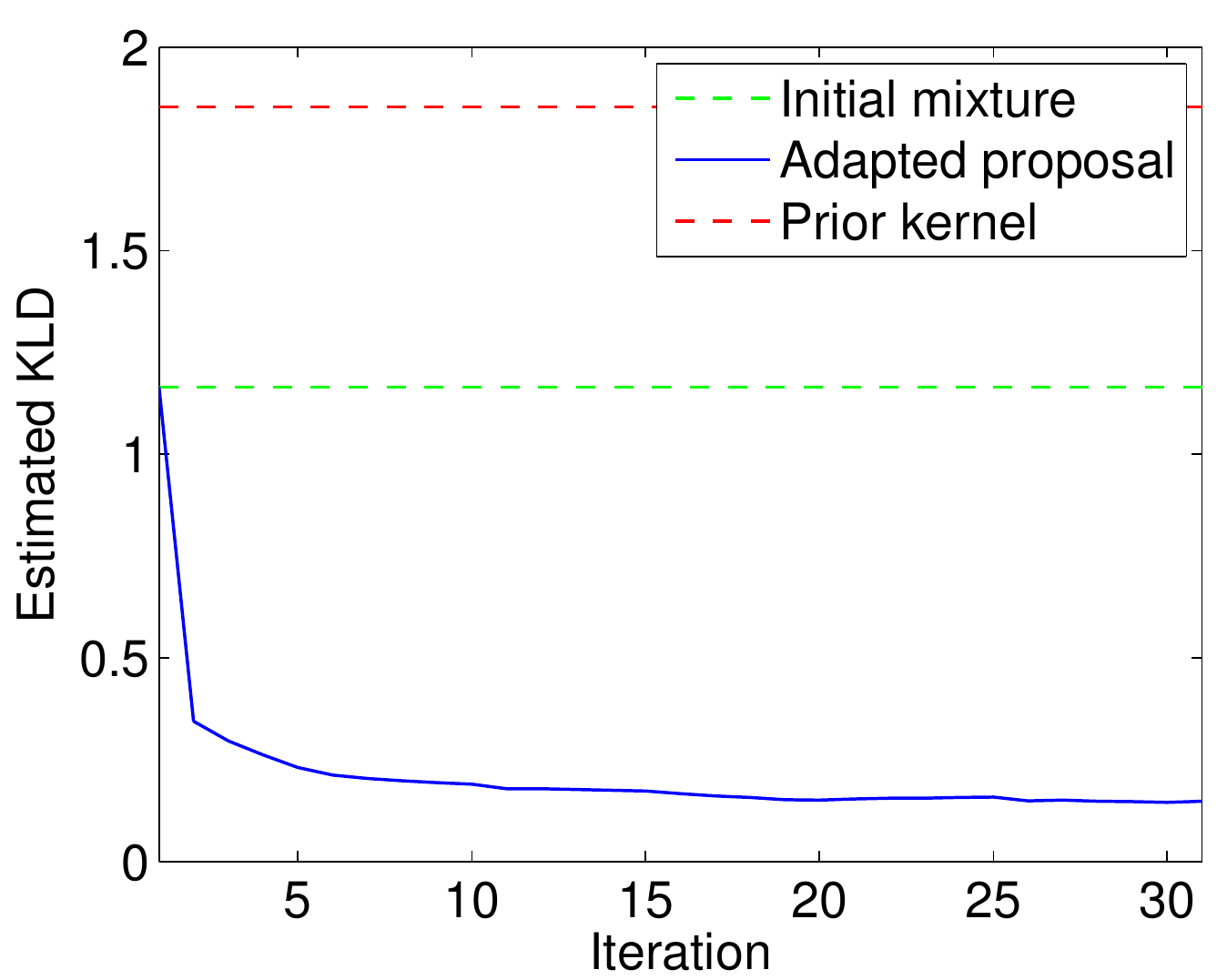}
		\end{subfigure}
	\caption{Improvement of the proposal via adaptation over $30$ iterations. The plot of proportions (left) and the KLD (right) improve dramatically over the very first iterations. While the prior kernel puts $90\%$ of the mass on $15\%$ of the particles, adaptation increases it to $70\%$ of the particles in one step and stabilises to $80\%$ after a very few iterations.}
	\label{fig:bessel:gaussian:proportions}
	\end{center}
\end{figurehere}

Most importantly, a very small number of iterations suffices to achieve significantly more uniformly distributed importance weights, \ie~to significantly lower the KLD: Figure~\ref{fig:bessel:gaussian:proportions} (right) shows how the KLD drops after the first 2 iterations and then stabilises near null, while the curve of proportions in Figure~\ref{fig:bessel:gaussian:proportions} (left) shows that the distribution of the weights is essentially unchanged past the first few iterations.

\begin{figure*}[htbp]
	\begin{center}
		\begin{subfigure}[t]{.31\textwidth}
			\includegraphics[width=1\textwidth]{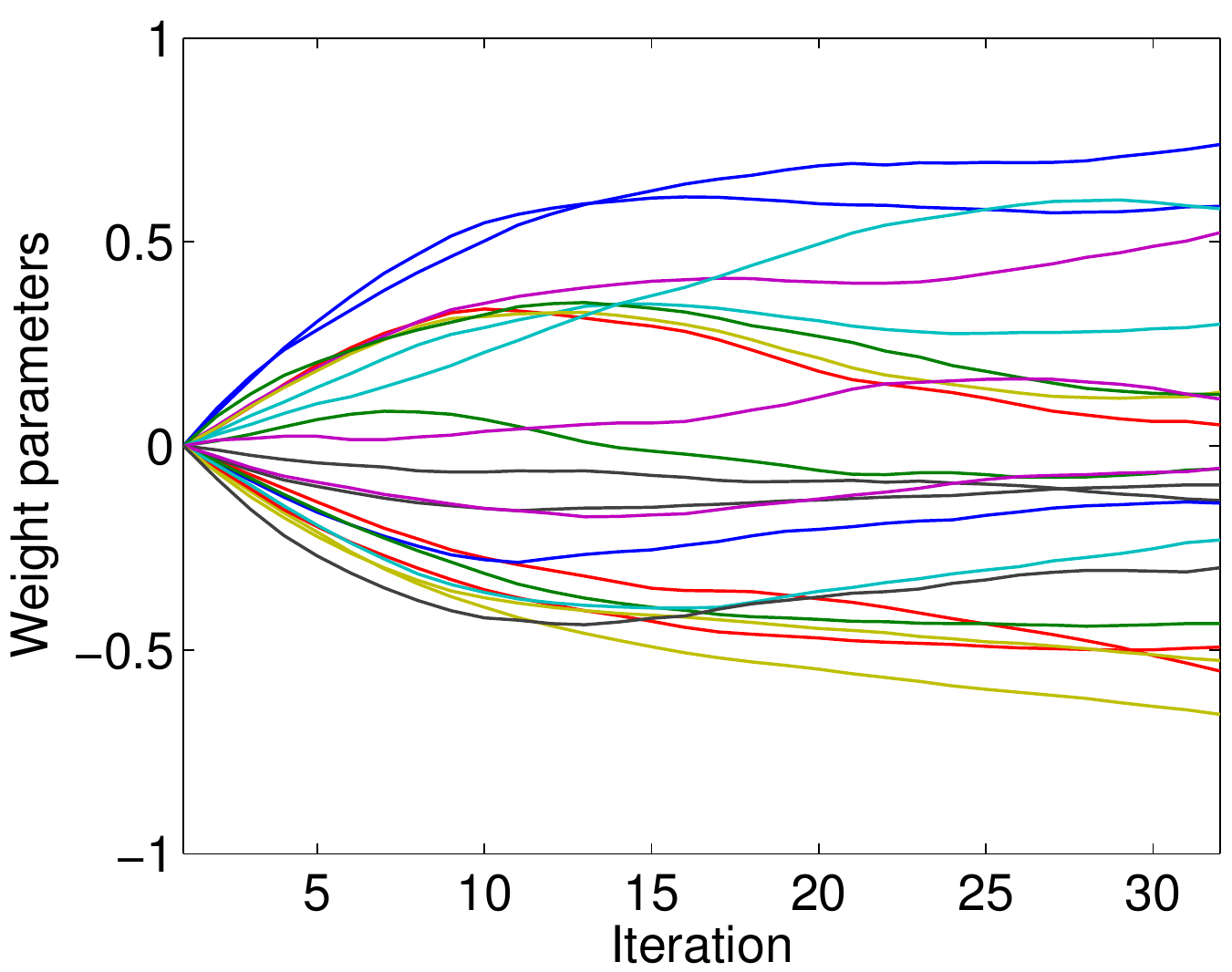}
			\caption{Components of the logistic weight parameters $\mixbetavec^{\ell}$.}
		\end{subfigure}%
		\hfill%
		\begin{subfigure}[t]{.31\textwidth}
			\includegraphics[width=1\textwidth]{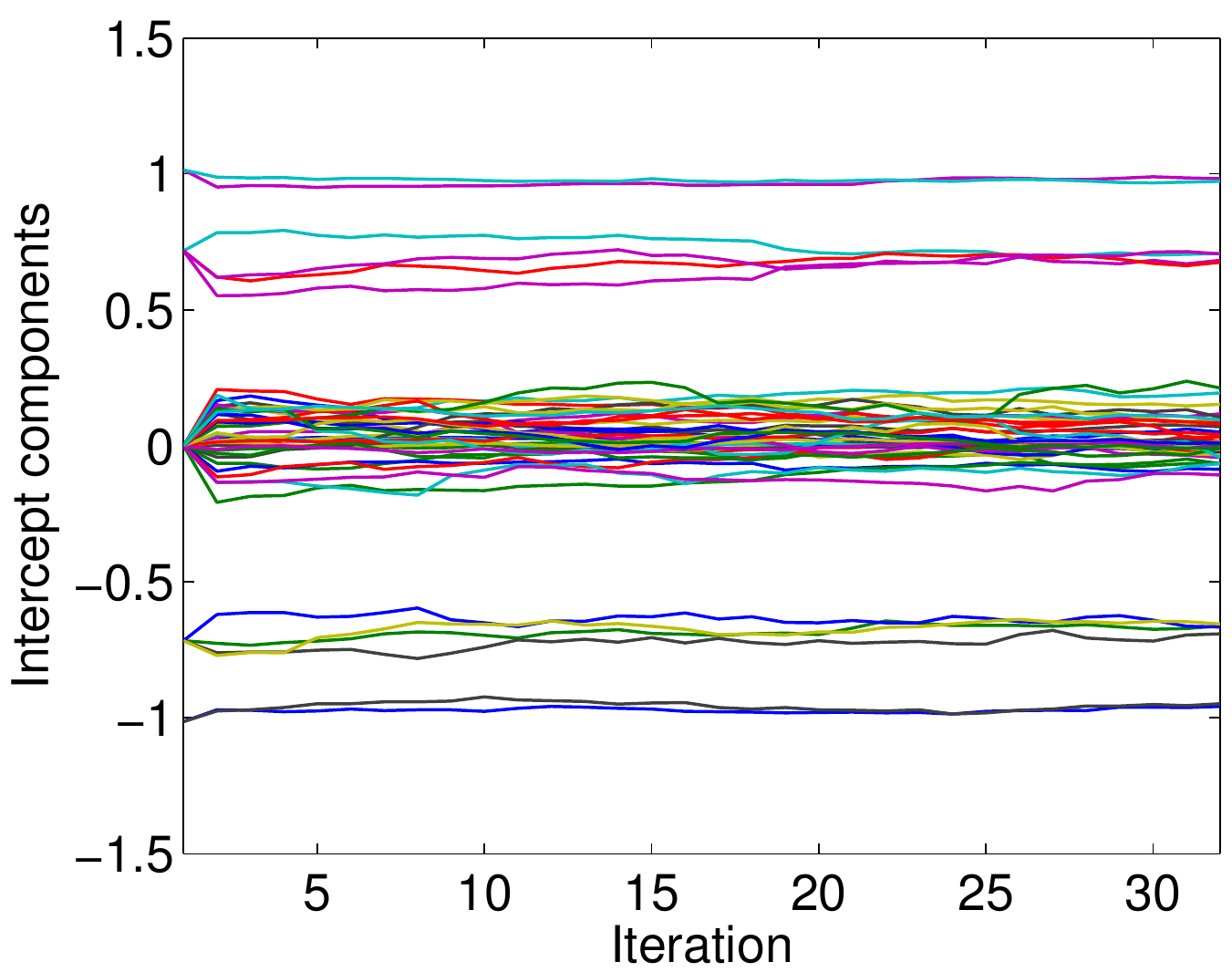}
			\caption{Components of the intercept matrices $\mixmean{\ell}$.}
		\end{subfigure}%
		\hfill%
		\begin{subfigure}[t]{.31\textwidth}
			\includegraphics[width=1\textwidth]{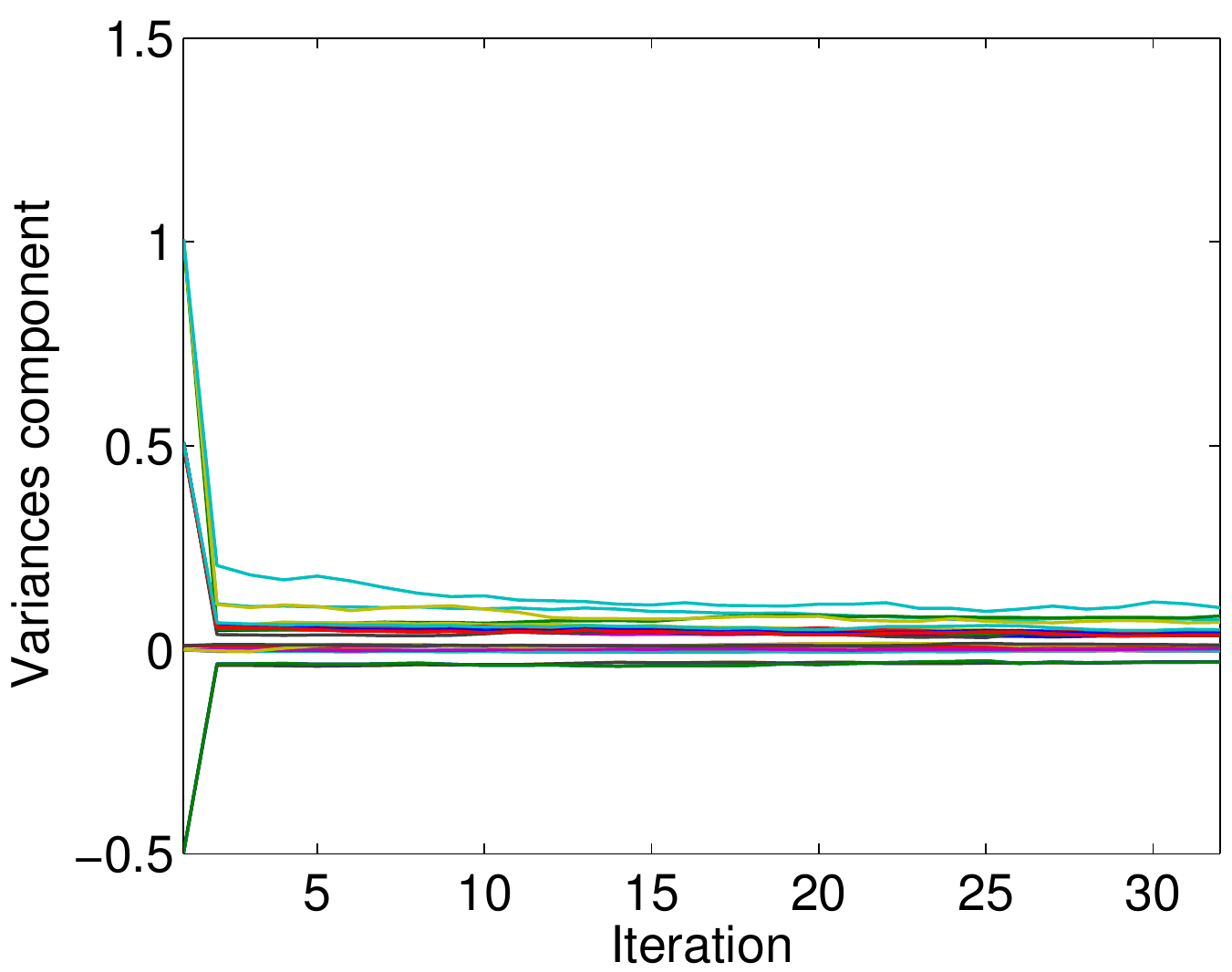}
			\caption{Components of the covariance matrices $\mixvar{\ell}$.}
		\end{subfigure}%
	\end{center}
	\caption{Parameters $\param[\ell]$ of the adapted kernel over 30 iterations of the algorithm: practically, convergence is achieved after a few steps only.}
	\label{fig:bessel:gaussian:plotparameters}
\end{figure*}

As a final look at convergence, Figure~\ref{fig:bessel:gaussian:plotparameters} displays the evolution of all the estimated parameters over the $30$ iterations, confirming that the fit stabilises after a few steps: the result of the very first couple of iterations could serve as a more efficient proposal than the prior kernel.
The parameters of the logistic weights $\mixbetavec[\ell]$ are the slowest to stabilise due to the stochastic gradient used to palliate for the lack of closed-form update as mentioned in Section~\ref{sec:algorithm:logistic:weights}.

\begin{figure*}[htbp]
	\begin{center}
		\begin{subfigure}[t]{1\columnwidth}
			\includegraphics[width=1\textwidth]{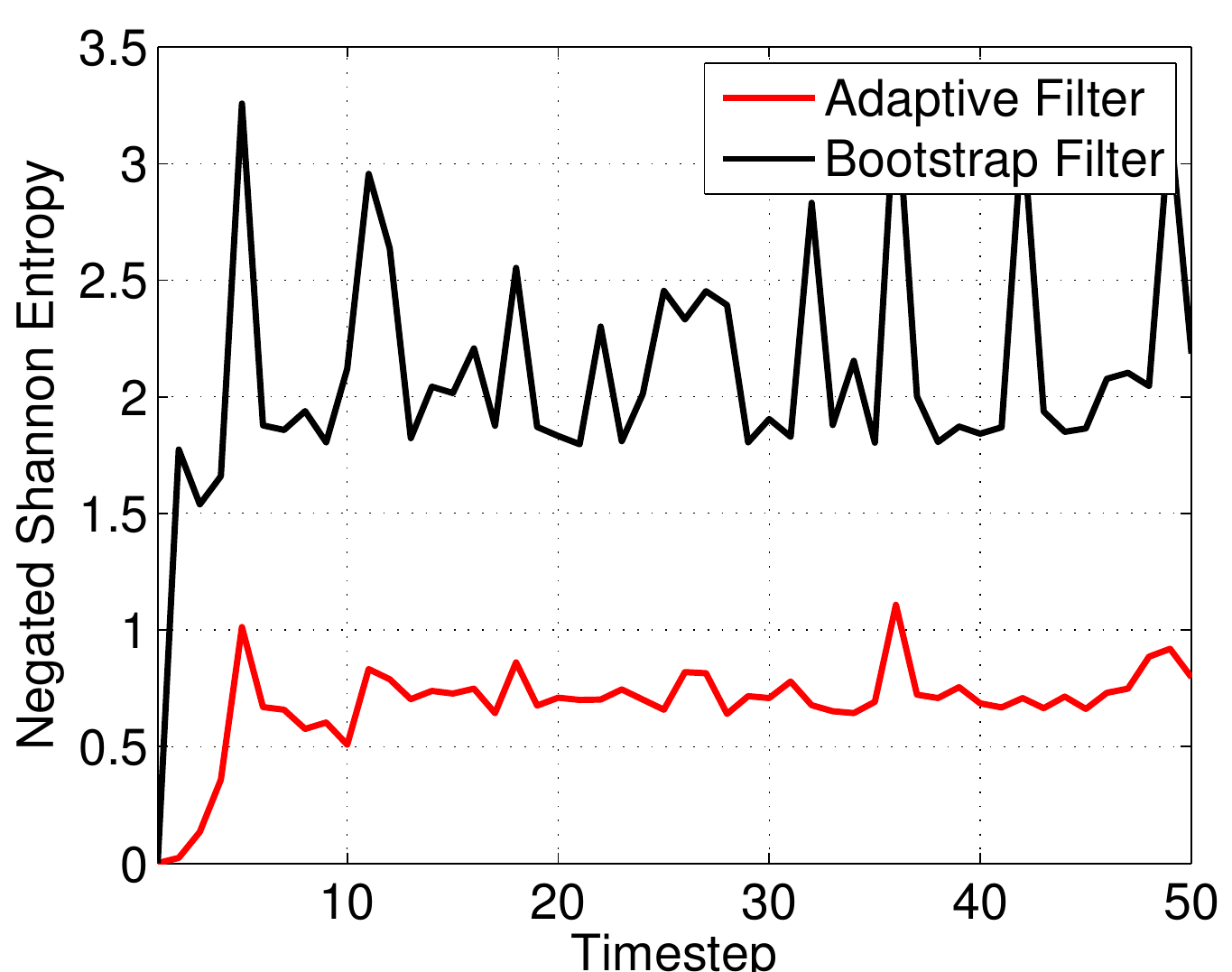}
			\caption{Negated Shannon entropy}
			\label{fig:bessel:multisteps:criteria:entropy}
		\end{subfigure}%
		\hfill%
		\begin{subfigure}[t]{1\columnwidth}
			\includegraphics[width=1\textwidth]{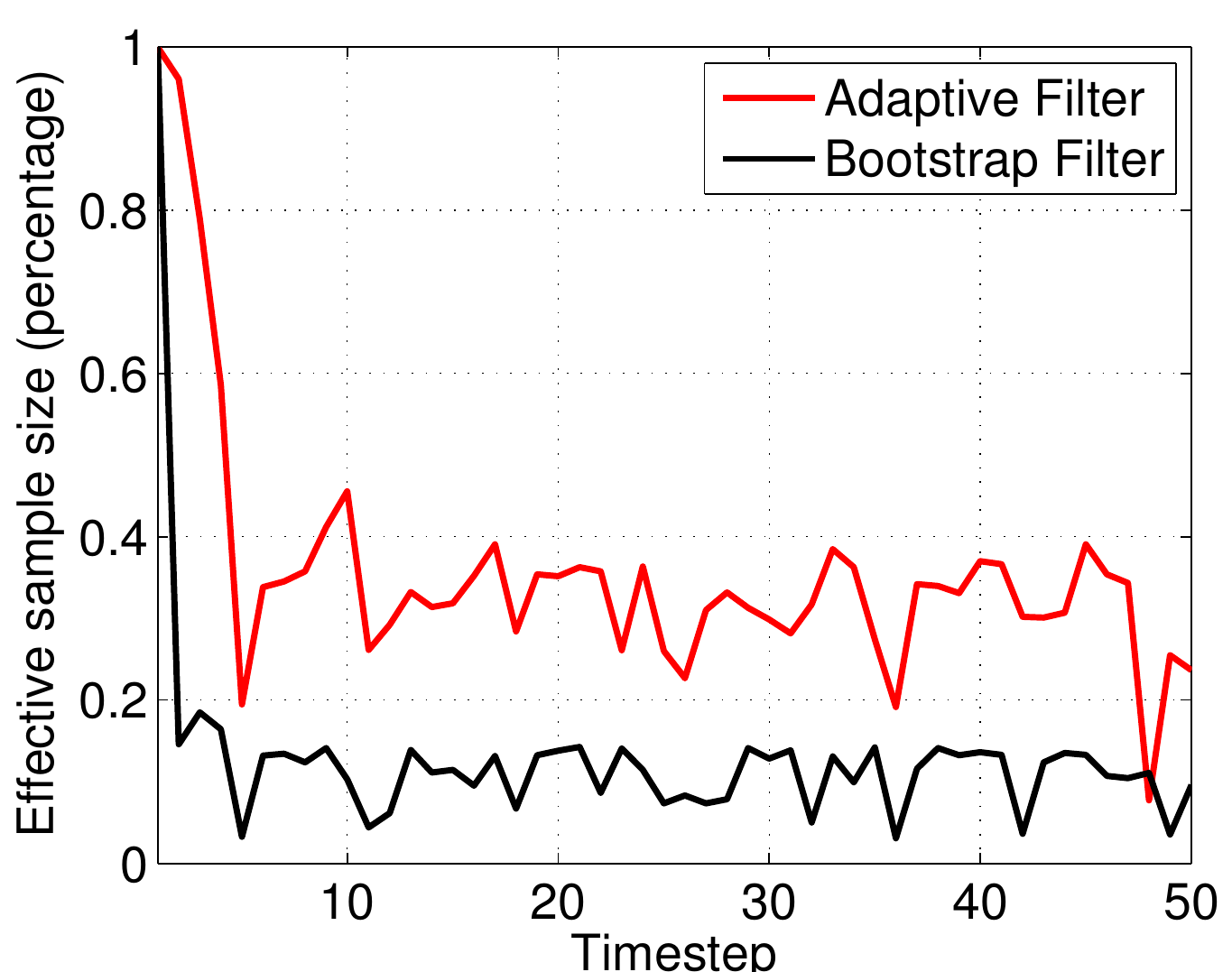}
			\caption{Relative effective sample size}
			\label{fig:bessel:multisteps:criteria:ess}
		\end{subfigure}%
	\caption{Evolution of the effective sample size and the entropy between target and proposal for the bootstrap filter and the adaptive filter. As intended, the adaptive filter leads to higher effective sample size and lower negated Shannon entropy.}
		\label{fig:bessel:multisteps:criteria}
	\end{center}
\end{figure*}

We now turn focus away from a single time step and run the adaptive particle filter in parallel with a plain bootstrap filter for 50 times steps with a simulated observation record $\jvect{Y}_{0:50}$ as input. At each iteration we let the adaptive filter optimise the proposal kernel using Algorithm~\ref{alg:main:2}. The forgetting of the initial distribution of the random walk entails that the filter distribution flow converges to a distribution
symmetric under rotation and supported on a ring centered at $\jvect{0}$.
To connect with widespread measures of efficiency discussed in~\cite{cornebise:moulines:olsson:2008}, we compare in
Figure~\ref{fig:bessel:multisteps:criteria} the negated Shannon
entropy (lower is better) of the importance weights and relative effective sample size (in percentage of the total sample; higher is better) of the two filters. The negated Shannon entropy shows improvement by our adaptive filter, which is not surprising in the light of the convergence results of~\citet[][Theorems~1--2]{cornebise:moulines:olsson:2008}: it is a consistent estimate of the same KLD when the number of particles tends to infinity. In this example also the relative effective sample size also exhibits an improvement, even though it is a consistent estimate of the chi-square divergence between the same distributions rather than the KLD \citep[see again][Theorem~1--2, for an exact formulation of this result]{cornebise:moulines:olsson:2008}. Even though this makes sense to some extent, one should not expect this to hold systematically. 

\subsection{Multivariate tobit model}
\label{sec:tobit}

\begin{figure*}[htbp]
	\begin{center}
		\begin{subfigure}[t]{.02\textwidth}
			\rotatebox[origin=tl]{90}{ }
		\end{subfigure}%
		\begin{subfigure}[c]{.31\textwidth}
			\centering \small Ancestor $(-3, 1)$
		\end{subfigure}%
		\hfill%
		\begin{subfigure}[c]{.31\textwidth}
			\centering \small Ancestor $(1, 1)$
		\end{subfigure}
		\hfill%
		\begin{subfigure}[c]{.31\textwidth}
			\centering \small Ancestor $(9, 5)$
		\end{subfigure} \\
		\begin{subfigure}[b]{.02\textwidth}
			\centering \rotatebox[origin=tl]{90}{\small Prior kernel}
		\end{subfigure}%
		\begin{subfigure}[t]{.31\textwidth}
			\includegraphics[width=1\textwidth]{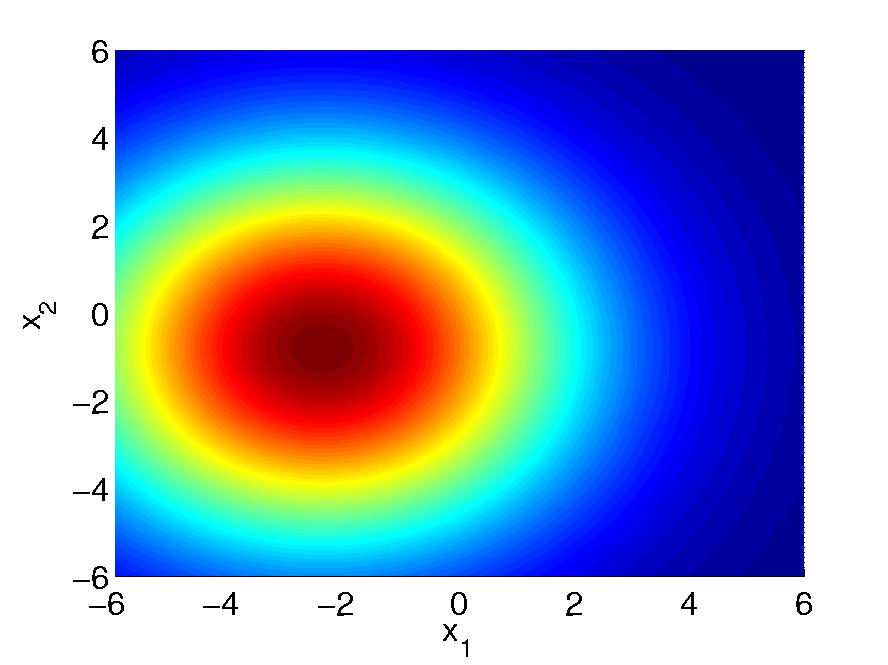}
		\end{subfigure}%
		\hfill%
		\begin{subfigure}[t]{.31\textwidth}
			\includegraphics[width=1\textwidth]{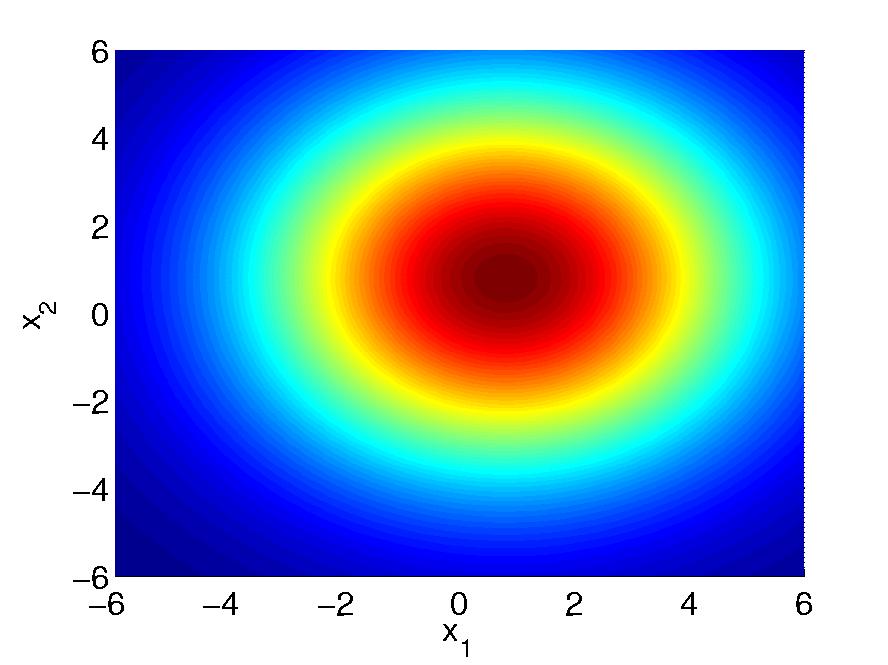}
		\end{subfigure}
		\hfill%
		\begin{subfigure}[t]{.31\textwidth}
			\includegraphics[width=1\textwidth]{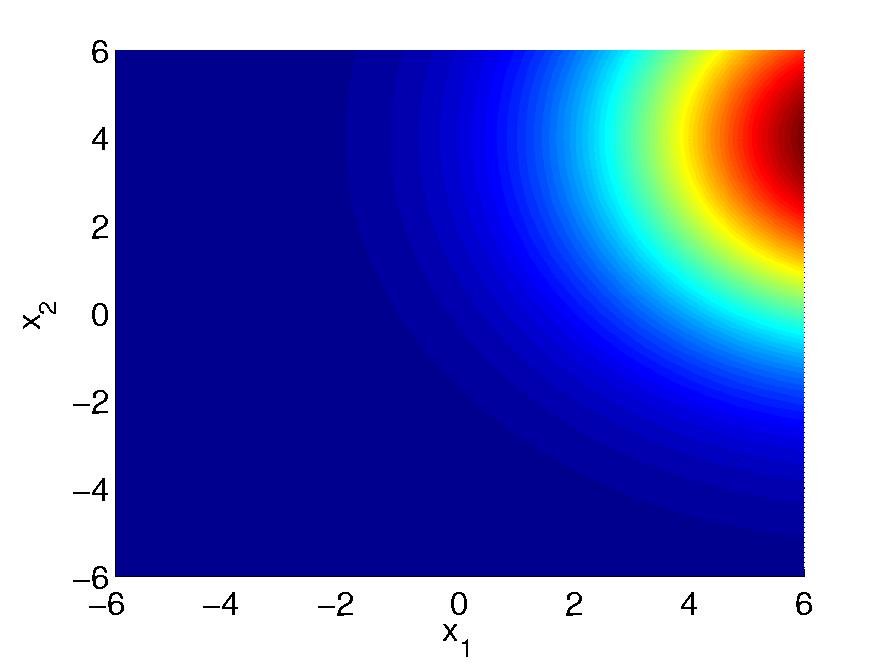}
		\end{subfigure}\\
		\begin{subfigure}[b]{.02\textwidth}
			\centering \rotatebox[origin=tl]{90}{\small Local Likelihood}
		\end{subfigure}%
		\begin{subfigure}[t]{.31\textwidth}
			\includegraphics[width=1\textwidth]{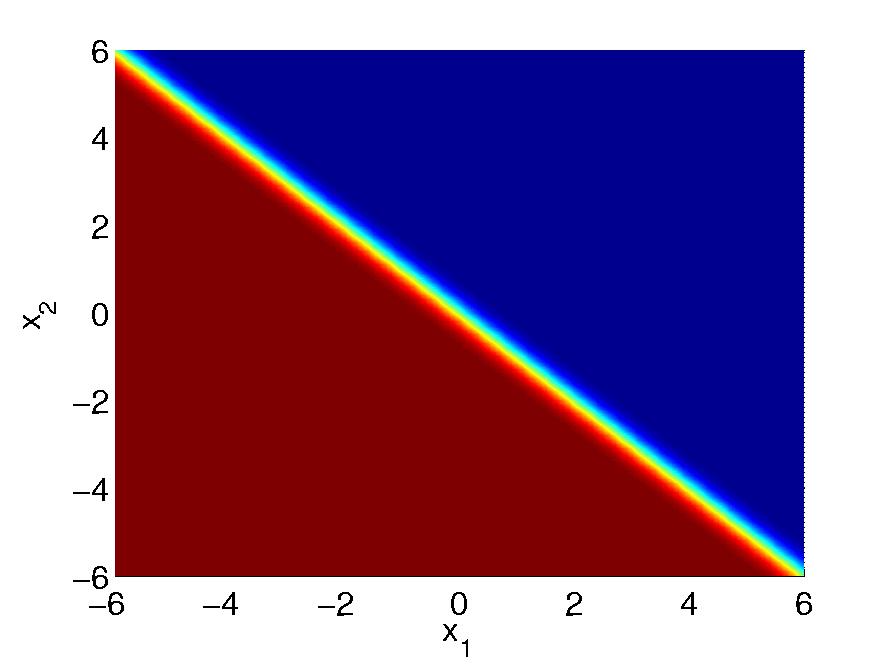}
		\end{subfigure}%
		\hfill%
		\begin{subfigure}[t]{.31\textwidth}
			\includegraphics[width=1\textwidth]{Tobit_Loclike}
		\end{subfigure}
		\hfill%
		\begin{subfigure}[t]{.31\textwidth}
			\includegraphics[width=1\textwidth]{Tobit_Loclike}
		\end{subfigure}
		\begin{subfigure}[b]{.02\textwidth}
			\centering \rotatebox[origin=tl]{90}{\small Optimal kernel}
		\end{subfigure}%
		\begin{subfigure}[t]{.31\textwidth}
			\includegraphics[width=1\textwidth]{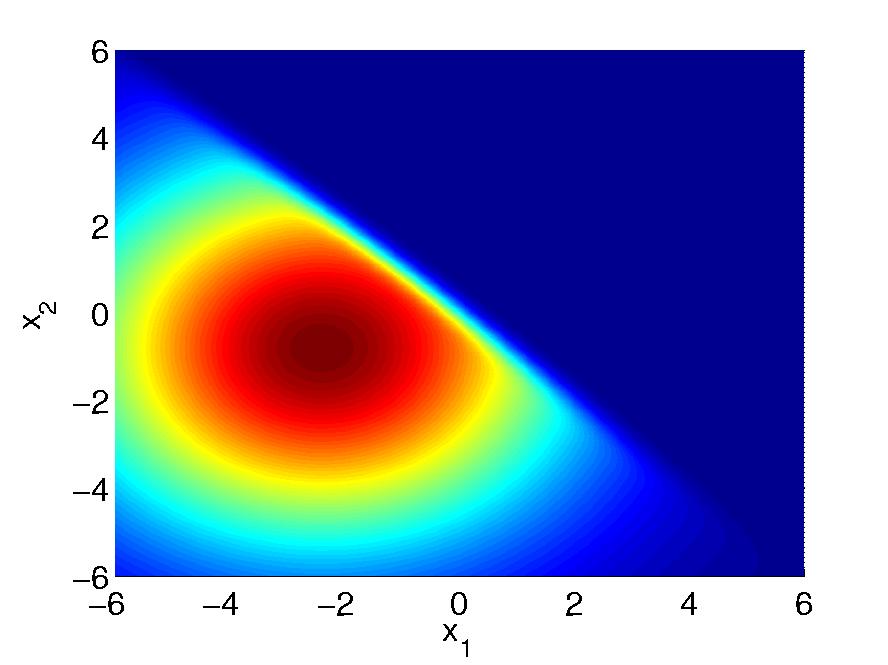}
		\end{subfigure}%
		\hfill%
		\begin{subfigure}[t]{.31\textwidth}
			\includegraphics[width=1\textwidth]{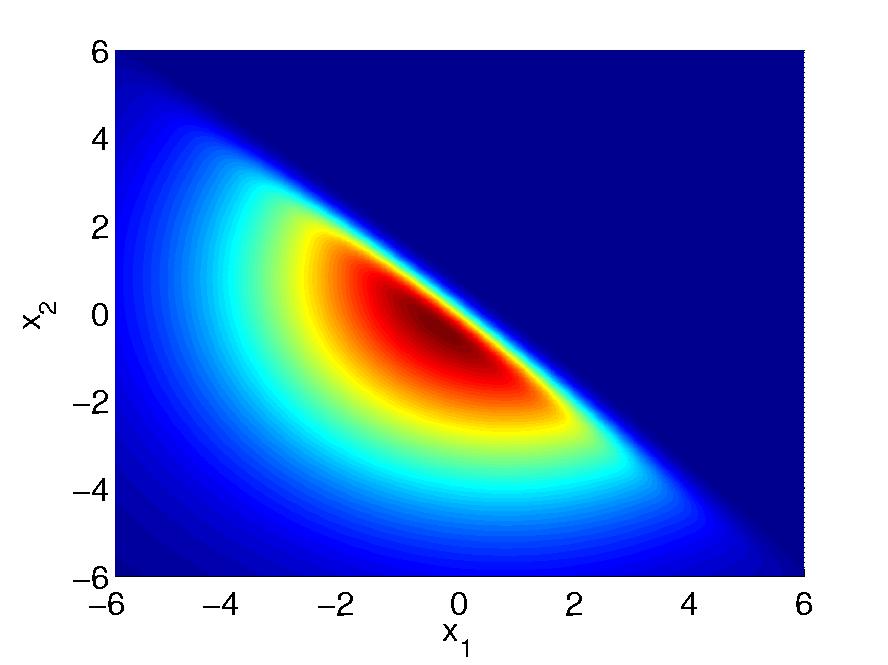}
		\end{subfigure}
		\hfill%
		\begin{subfigure}[t]{.31\textwidth}
			\includegraphics[width=1\textwidth]{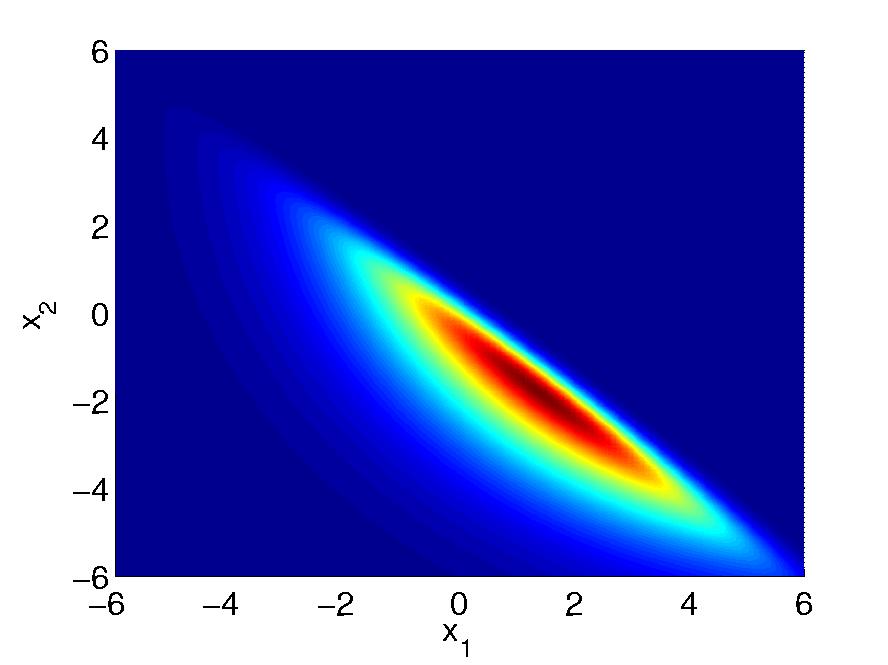}
		\end{subfigure}\\
		\begin{subfigure}[b]{.02\textwidth}
			\centering \rotatebox[origin=tl]{90}{\small Iteration $10$}
		\end{subfigure}%
		\begin{subfigure}[t]{.31\textwidth}
			\includegraphics[width=1\textwidth]{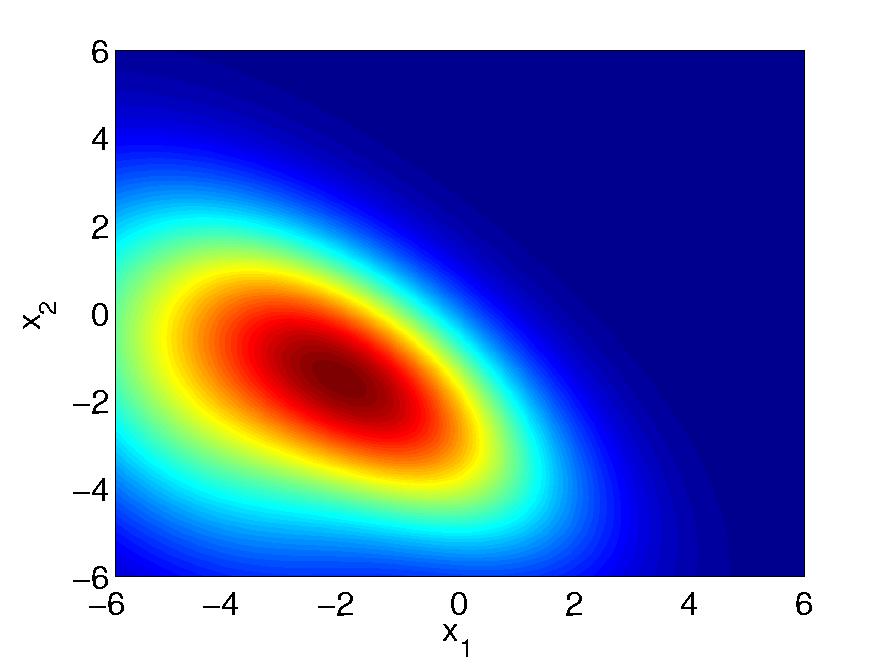}
		\end{subfigure}%
		\hfill%
		\begin{subfigure}[t]{.31\textwidth}
			\includegraphics[width=1\textwidth]{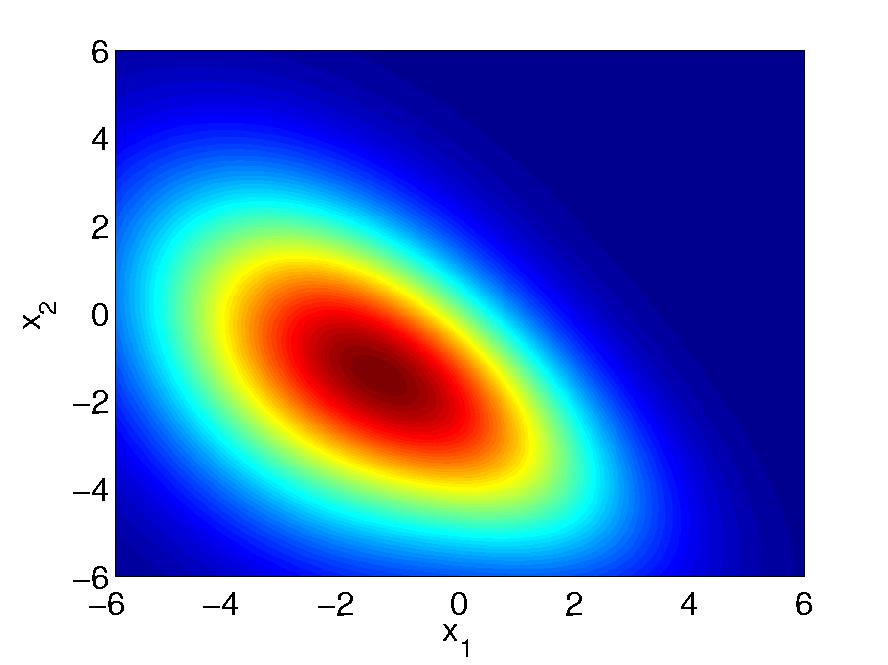}
		\end{subfigure}
		\hfill%
		\begin{subfigure}[t]{.31\textwidth}
			\includegraphics[width=1\textwidth]{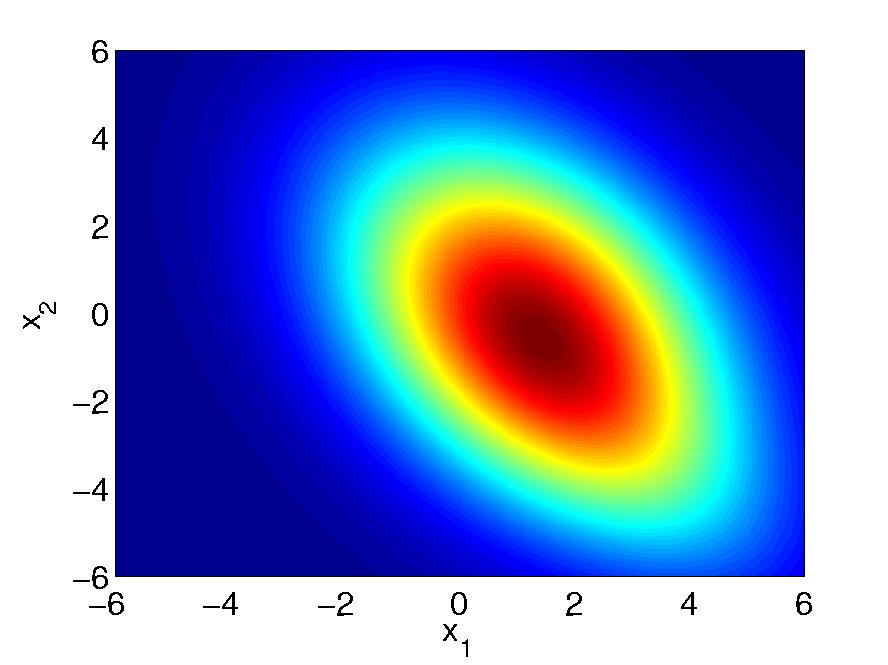}
		\end{subfigure}\\
\end{center}
\caption{Evolution of the adapted kernel, for three ancestors
in different regions for the tobit model. In this example the observation brings, thanks to the censorship, a lot of information concerning the location of the ancestor. Thus, adapting only the proposal kernel will not lead to perfect adaptation, as can be seen on the fit after 10 iterations. In this case it is thus highly relevant to consider adaptation also of the adjustment multiplier weights.}
\label{fig:tobit:3refs}
\end{figure*}

We now briefly illustrate how far adaptation of the proposal kernel leads---and where it stops. Consider a partially observed multivariate dynamic \emph{tobit} model
\[
	\begin{split}
		\jvect{X}_{k + 1}  &= \jvect{A} \jvect{X}_k + \jvect{U}_{k + 1} \eqsp, \\
		Y_k  &= \max \left(\jvect{B}^\trans \jvect{X}_k  + V_k, 0 \right) \eqsp,
	\end{split}
\]
where $\stsp = \R^2$, $\jvect{A} = 0.8 \times \jvect{I}_2$,
and $\jvect{B} = (1, 1)^\trans \in \R^2$, so that $Y_k$ takes values in $\R$. Here $\{ \jvect{U}_k \}_{k \geq 1}$ and $\{ V_k \}_{k \geq 0}$ are independent sequences of mutually independent and identically distributed Gaussian variables with variances $\jvect{\Sigma}_{\jvect{U}} = 2 \times \jvect{I}_2$ and $\sigma_v^2 = 0.1$, respectively. The observation process consists of noisy observations of the sum of its components of the hidden states. In addition, the observations are left-censored. We consider again a single update of the particle swarm for a given time step $k$, where we have $N = 20,\!000$ ancestor particles distributed according to $\normlaw[2]{(1,1)^\trans}{10 \times \jvect{I}_2}$ and set $Y_{k + 1} = 0$. The local likelihood is hence null above the line $\Delta = \{\jvect{x} \in \R^2 : \jvect{B}^\trans \jvect{x} = 0 \}$ and constant below with a narrow transition region, as displayed in Figure~\ref{fig:tobit:3refs} (second row). The prior kernel displayed in Figure~\ref{fig:tobit:3refs} (first row) can have most of its mass out of the high-likelihood regions, depending on the ancestor. Figure~\ref{fig:tobit:3refs} (third row) illustrates the un-normalised optimal transition kernel for three reference ancestors, showing how the match between the supports of the prior kernel and the local likelihood varies depending on the position of the ancestor particle relatively to the line $\jvect{A}^{-1} \Delta$. Hence, half of the original particles have very small adjustment multiplier weights.

Adapting the proposal kernel will hence require at least two components, and
will not lead to perfect adaptation, as can be seen on the fit after 10 iterations displayed in Figure~\ref{fig:tobit:3refs} (last row). The ancestor $(1,1)^\trans$ is the center of the ancestor sample, and the un-normalised optimal kernel is the prior kernel truncated in its middle; $(-3,-1)^\trans$ is the bottom left of the ancestor sample, and the un-normalised optimal kernel almost matches the prior, save for a truncation in the upper right tail; $(9, 5)^\trans$ is the top right of the ancestor sample, and the un-normalised optimal kernel differs widely from the prior kernel, as only the very far tails of the latter match non-null local likelihood.

Finally, without getting into details, we make our point through
Figure~\ref{fig:tobit:KLD:compare} showing again how the KLD drops in the first iteration for families of both Gaussian distributions and Student's $t$-distributions. We nevertheless purposefully keep the algorithm running for a large number of iterations, to illustrate the difference between the Gaussian and the Student's $t$ parametrisations. The Gaussian parametrisation stabilises close to the minimum achieved by the optimal kernel with uniform adjustment weights, which is not negligible. In addition, the Student's $t$-distribution allows for heavier tails at the price of a higher attainable lower bound on the KLD.
\begin{figurehere}
		\begin{center}
	\includegraphics[width=1\columnwidth]{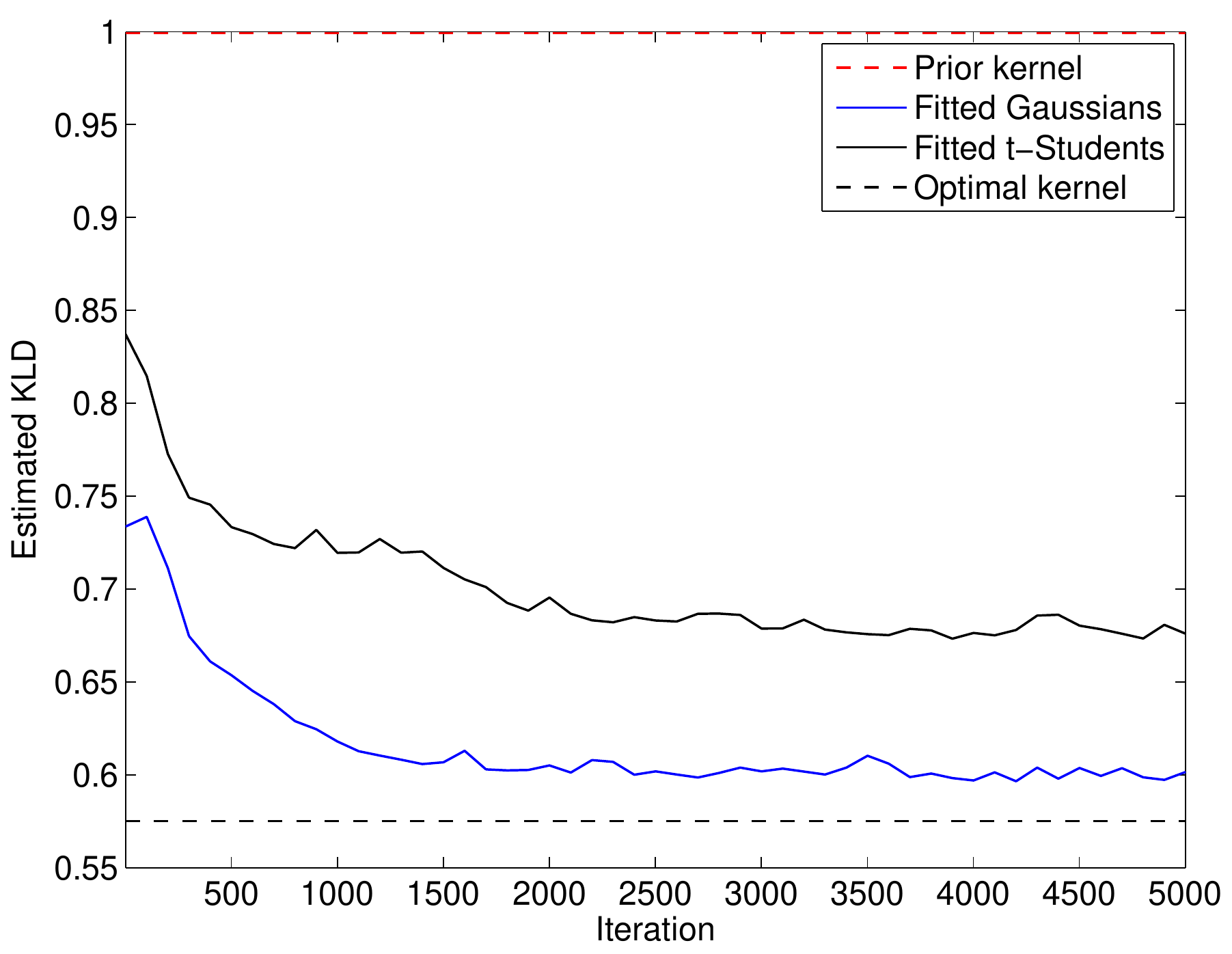}
\caption{Comparison of the evolution of the KLD for the Gaussian
			experts and the Student's $t$-distributed expert over $L = 5,\!000$ iterations of the algorithm.}
	\label{fig:tobit:KLD:compare}
		\end{center}
\end{figurehere}

Finally, Figure~\ref{fig:tobit:KLD:compare} illustrates a limit case for kernel-only adaptation: replacing the prior kernel by the optimal one reduces only the the KLD, which is still far from zero, by half. In this example, chosen for that purpose, the observation brings, thanks to the censorship, a lot of information concerning the location of the ancestor. In this case it is thus highly relevant to consider adaptation also of the adjustment multiplier weights. Naturally, this could, as a logical next step in development of adaptive SMC algorithms, be done by designing a minimisation algorithm for the second term of \eqref{eq:KLD:decouples}, which we leave as an open problem.

%% file: sections/conclusion_rev.tex
\section{Future work and Conclusion}

Relying on the results of \cite{cornebise:moulines:olsson:2008}, we have built
new algorithms approximating the so-called optimal proposal kernel at a given
time step of an auxiliary particle filter by means of minimisation of the KLD
between the auxiliary target and instrumental distributions of the particle
filter. More specifically, the algorithm fits a weighted mixture of integrated
curved exponential distributions with logistic weights to the auxiliary target
distribution by minimising the KLD between the two using a Monte Carlo version of the online EM method proposed by \cite{cappe:moulines:2009}. 

In addition, we have applied successfully this relatively simple algorithm to optimal filtering in SSMs; indeed, running the stochastic
approximation-based adaptation procedure for only a few iterations at every time step as the particles evolve leveled off
significantly the distribution of the total weight mass among the particles
also for models exhibiting very strong nonlinearity. Thus, adding the
adaptation step to an existing particle filter implementation implies only
limited computational demands. 

A proof of convergence, combining the theory developed in Section~\ref{sec:algo} with existing theory of stochastic approximation with Markovian pertubations \citep[treated by, e.g.,][]{duflo:1997,benveniste:metivier:priouret:1990} of the algorithm is currently in progress. In addition, we investigate at present the possibility of extending the approach to comprise adaptation also of the adjustment multipliers.

%% file: sections/proofs_rev.tex
\section{Proof of Proposition~\ref{prop:key:result}}
\label{sec:proofs}
The proof follows the lines the proof of \citet[Proposition~1]{cappe:moulines:2009}. For vector-valued differentiable functions $h = (h_1, \ldots, h_m)^\trans$ from $\Theta$ to $\rset^m$ we denote by $\nablapar h^\trans$ the $|\Theta| \times m$ matrix having the gradient $\nablapar h_j$ as $j$th column, this is, the inverse of the Jacobian of the same mapping. 

Let $(\jvect{s}^*, \jvect{p}^*)$ be a zero of $\jvect{h}$ and set $\mixparamvec^* \eqdef\globmax(\jvect{s}^\ast, \jvect{p}^\ast)$. Since for all  $\curvpar$ and $\jvect{s} \in \rset^{\sdim{1} \times \sdim{2}}$, 
\begin{equation} \label{eq:trace:gradient}
\nablapar \trace(\cB(\curvpar)^\trans \jvect{s}) = \sum_{\ell = 1}^{\sdim{2}} \nablapar \colv{\cB}{\ell}(\curvpar)^\trans \colv{\jvect{s}}{\ell} \eqsp, 
\end{equation}
it holds, by definition, 
\begin{equation} \label{eq:nabla:l}
\begin{split} 
\lefteqn{\nablapar  l(\jvect{s}^*, \jvect{p}^* ; \mixparamvec) \big|_{\mixparamvec = \mixparamvec^*}} \\
&= - \left( \nablapar \cA(\curvpar_1) \cdots  \nablapar \cA(\curvpar_d) \right) \big |_{\curvparvec = \curvparvec^*} \jvect{p}^* \\
&\hphantom{=}+ \left( \nablapar \log \beta_1 \cdots \nablapar \log \beta_d \right) \big |_{\underline{\boldsymbol{\beta}} = \underline{\boldsymbol{\beta}}^*} \jvect{p}^* 
\\
&\hphantom{=}+ \sum_{j = 1}^d \sum_{\ell = 1}^{\sdim{2}} \nablapar \colv{\cB}{\ell}(\curvpar_j)^\trans \big |_{\curvparvec = \curvparvec^*} \jvect{s}^*_{j|\ell} \\
&= \mathbf{0} \eqsp. 
\end{split}
\end{equation}
On the other hand, by Fisher's identity,
\begin{equation*} 
\begin{split}
\lefteqn{\nablapar \log \auxinstrparam{\mixparamvec}(\indv, \ppr)}\\ 
&=\E_{\completelike{\mixparamvec}} \left[ \left. \nablapar \log \completelike{\mixparamvec}(\indva, \compva, \pprva, \uuva) \right| \pprva = \ppr, \indva = \indv \right]\\
&=  - \sum_{j = 1}^d \mixwgtcond{\mixparamvec}{j}{i}{\ppr} \nablapar \cA(\curvpar_j)\\
&\hphantom{=}+ \sum_{j = 1}^d \mixwgtcond{\mixparamvec}{j}{i}{\ppr} \nablapar \log \beta_j\\
&\hphantom{=}+ \sum_{j = 1}^d  \mixwgtcond{\mixparamvec}{j}{i}{\ppr} \int \nablapar \trace(\cB(\curvpar_j)^\trans \suffstat(\parti{i}, \ppr, \uu))\\
&\hspace{15mm} \times \bar{\rho}(\uu | \parti{i}, \ppr ; \curvparj{j}) \, \ud \uu \eqsp,
\end{split}
\end{equation*}
and, by \eqref{eq:trace:gradient},  
\begin{equation*} 
\begin{split}
\lefteqn{\nablapar \log \auxinstrparam{\mixparamvec}(\indv, \ppr)}\\ 
&=  - \sum_{j = 1}^d \mixwgtcond{\mixparamvec}{j}{i}{\ppr} \nablapar \cA(\curvpar_j)\\
&\hphantom{=}+ \sum_{j = 1}^d \mixwgtcond{\mixparamvec}{j}{i}{\ppr} \nablapar \log \beta_j\\
&\hphantom{=}+ \sum_{j = 1}^d \sum_{\ell = 1}^{\sdim{2}} \mixwgtcond{\mixparamvec}{j}{i}{\ppr} \nablapar \colv{\cB}{\ell}(\curvpar_j)^\trans \\
&\hspace{15mm} \times \int \colv{\suffstat}{\ell}(\parti{i}, \ppr, \uu) \bar{\rho}(\uu | \parti{i}, \ppr ; \curvparj{j}) \, \ud \uu \eqsp. 
\end{split}
\end{equation*}
Consequently, 
\begin{equation} \label{eq:nabla:KL}
\begin{split}
\lefteqn{\nablapar \KL(\auxtarg \| \auxinstrparam{\mixparamvec})} \\
&= \E_{\auxtarg} \left[ \nablapar \log \auxinstrparam{\mixparamvec}(\indva, \pprva) \right] \\
&= - \left( \nablapar \cA(\curvpar_1) \cdots \nablapar \cA(\curvpar_d) \right) \pstat{}{\mixparamvec}\\
&\hphantom{=}+ \left(\nablapar \log \beta_1 \cdots \nablapar \log \beta_d \right) \pstat{}{\mixparamvec} \\
&\hphantom{=} + \sum_{j = 1}^d \sum_{\ell = 1}^{\sdim{2}} \nablapar \colv{\cB}{\ell}(\curvpar_j)^\trans \curvexpestat{j|\ell}{\mixparamvec} \eqsp. 
\end{split}
\end{equation}
Thus, as 
\begin{multline*}
\meanfd{\jvect{s}^*, \jvect{p}^*} = \jvect{0} \Rightarrow\\ \curvexpestat{}{\globmax(\jvect{s}^*, \jvect{p}^*)} \varoplus \pstat{}{\globmax(\jvect{s}^*, \jvect{p}^*)} = \jvect{s}^* \varoplus \jvect{p} \eqsp, 
\end{multline*}
we obtain 
\begin{equation} \label{eq:stat:KL}
\nablapar \KL(\auxtarg \| \auxinstrparam{\mixparamvec}) \big|_{\mixparamvec = \mixparamvec^*} = \jvect{0} \eqsp. 
\end{equation}

Conversely, let $\mixparamvec^*$ be a stationary point of $\mixparamvec \mapsto \KL(\auxtarg \| \auxinstrparam{\mixparamvec})$ (i.e. \eqref{eq:stat:KL} holds) and let $(\jvect{s}^*, \jvect{p})$ be given by \eqref{eq:def:root:candidate}. Then we conclude, via \eqref{eq:nabla:l} and \eqref{eq:nabla:KL}, that $\mixparamvec^*$ is a stationary point of $\mixparamvec \mapsto l(\jvect{s}^*, \jvect{p}^* ; \mixparamvec)$ as well. However, by Assumption~\ref{ass:existence:maximum}, this point is unique and equal to $\globmax(\jvect{s}^*, \jvect{p}^*)$; thus, 
\begin{multline*}
\jvect{s}^* \varoplus \jvect{p}^*
= \curvexpestat{}{\globmax(\jvect{s}^*, \jvect{p}^*)} \varoplus
\pstat{}{\globmax(\jvect{s}^*, \jvect{p}^*)} \\ 
\Rightarrow \meanfd{\jvect{s}^*, \jvect{p}^*} = \jvect{0} \eqsp,
\end{multline*}
which completes the proof.